# Review of superconducting properties of $MgB_2$

**Cristina Buzea [1, 2] and Tsutomu Yamashita [2, 3]**

[1] Research Institute of Electrical Communication, Tohoku University, Sendai 980-8577, Japan
[2] New Industry Creation Hatchery Center, Tohoku University, Sendai 980-8579, Japan
[3] CREST Japan Science and Technology Corporation (JST)

**Abstract**
This review paper illustrates the main normal and superconducting state properties of magnesium diboride, a material known since early 1950's, but recently discovered to be superconductive at a remarkably high critical temperature $T_c$=40K for a binary compound. What makes $MgB_2$ so special? Its high $T_c$, simple crystal structure, large coherence lengths, high critical current densities and fields, transparency of grain boundaries to current promises that $MgB_2$ will be a good material for both large scale applications and electronic devices. During the last seven month, $MgB_2$ has been fabricated in various forms, bulk, single crystals, thin films, tapes and wires. The largest critical current densities >10MA/cm$^2$ and critical fields 40T are achieved for thin films. The anisotropy ratio inferred from upper critical field measurements is still to be resolved, a wide range of values being reported, $\gamma = 1.2 \div 9$. Also there is no consensus about the existence of a single anisotropic or double energy gap. One central issue is whether or not $MgB_2$ represents a new class of superconductors, being the tip of an iceberg who awaits to be discovered. Up to date $MgB_2$ holds the record of the highest $T_c$ among simple binary compounds. However, the discovery of superconductivity in $MgB_2$ revived the interest in non-oxides and initiated a search for superconductivity in related materials, several compounds being already announced to become superconductive: $TaB_2$, $BeB_{2.75}$, C-S composites, and the elemental B under pressure.

(Some figures in this article are in colour only in the electronic version; see www.iop.org)

---

**Contents**

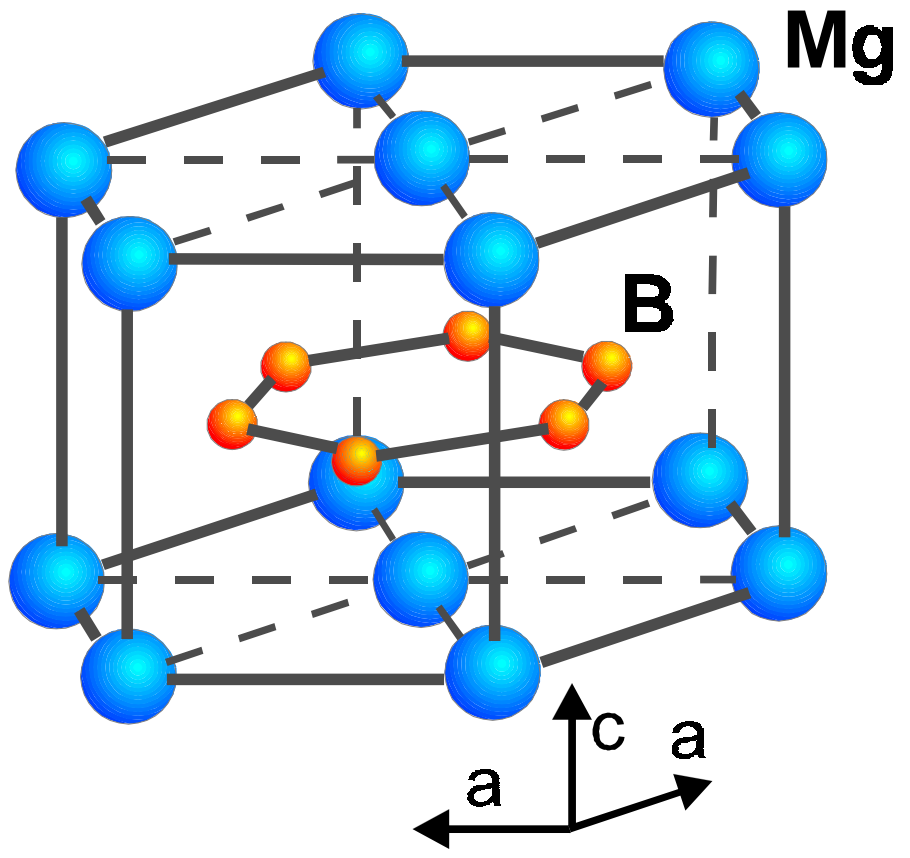

**Figure 1.** The structure of $MgB_2$ containing graphite-type B layers separated by hexagonal close-packed layers of Mg.

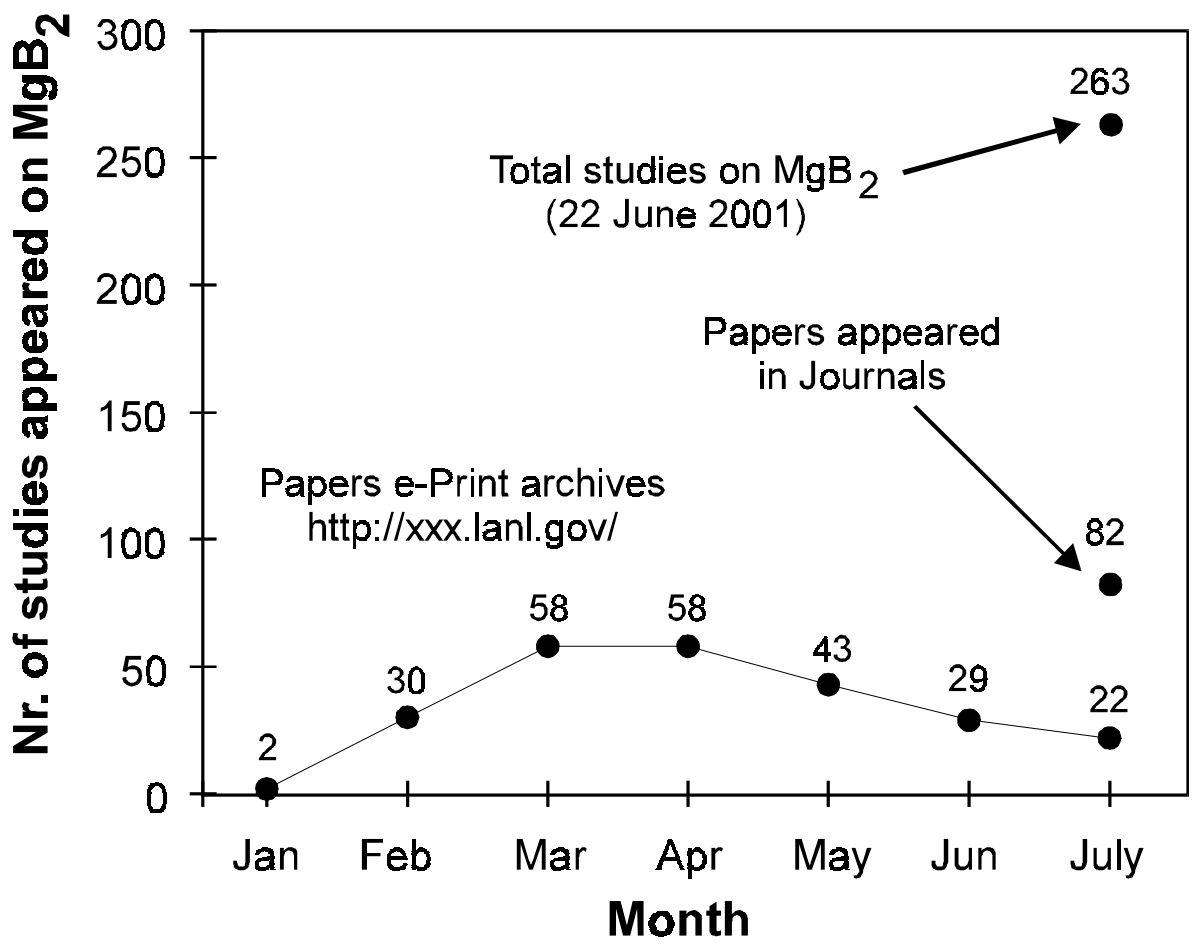

**Figure 2.** Studies appeared about magnesium diboride between January and July 2001

## 1. Introduction

$MgB_2$ is an "old" material, known since early 1950's, but only recently discovered to be superconductor [Akimitsu], [Nagamatsu] at a remarkably high critical temperature - about 40K - for its simple hexagonal structure (Fig. 1).

Since 1994 there has been a renewed interest in intermetallic superconductors which incorporate light elements, such as boron, due to the discovery of the new class of borocarbides RE-$TM_2B_2C$, where RE = Y, Lu, Er, Dy or other rare earths, and TM = Ni or Pd [Nagarajan], [Cava]. The main characteristics of these compounds are very high $T_c$ among intermetallics ($T_c$ = 23K in $YPd_2B_2C$), the anisotropic layered structure (unique for intermetallics) and a strong interplay between magnetism and superconductivity [Eisaki].

In the framework of BCS theory [Bardeen] the low mass elements result in higher frequency phonon modes which may lead to enhanced transition temperatures. The highest superconducting temperature is predicted for the lightest element, hydrogen [Ascroft], [Richardson] under high pressure. In 1986 investigations of the electrical resistance of Li under pressure up to 410 Kbar showed a sudden electrical resistance drop at around 7 K between 220 and 230Kbar, suggesting a possible superconducting transition [Lin]. Extremely pure beryllium superconducts at ordinary pressure with a $T_c$ of 0.026 K [Falge]. Its critical temperature can be increased to about 9-10K for amorphous films [Lazarev], [Takei]. Finally, the recent discovery of superconductivity in $MgB_2$ confirms the predictions of higher $T_c$ in compounds containing light elements, being believed that the metallic B layers play a crucial role in the superconductivity of $MgB_2$ [Kortus].

The discovery of superconductivity in $MgB_2$ certainly revived the interest in the field of superconductivity, especially non-oxides, and initiated a search for superconductivity in related boron compounds [Felner], [Young], [Gasparov], [Kaczorowski (a)], [Strukova]. Its high critical temperature gives hopes for obtaining even higher $T_c$'s for simple compounds.

$MgB_2$ superconductivity announcement was the catalyst for the discovery of several superconductors, some related to magnesium diboride, $TaB_2$ with $T_c$ = 9.5 K [Kaczorowski (a)], $BeB_{2.75}$ $T_c$ = 0.7 K [Young], graphite-sulphur composites $T_c$ = 35 K [da Silva], and another not related, but "inspired" by it, $MgCNi_3$ with $T_c$ = 8 K [He]. Probably the most impressive is the recent report related to superconductivity under pressure of B, with a very high critical temperature for a simple element of $T_c$ = 11.2 K [Eremets]. One have to mention that graphite-sulfur composites, C-S, are similar materials with $MgB_2$ electronically and crystallographically.

Its critical temperature of about 40K is close to or above the theoretical value predicted from the BCS theory [McMillan]. This may be a strong argument to consider $MgB_2$ as a non-conventional superconductor.

Akimitsu's group reported the superconductivity of $MgB_2$ on January 10th 2001 at a Conference in Sendai, Japan [Akimitsu]. Since then until the end of July 2001, have appeared more than 260 studies about this superconductor, i. e. in average of 1.3 papers/day. From this 260 studies, 80 have already appeared in Journals (as of the end of July 2001). Most of the 260 studies related to $MgB_2$ have been posted in electronic format on e-print archives of Los Alamos server at http://xxx.lanl.gov/.

In Fig.2 are shown the number of papers posted in electronic format since January until July, with a maximum of 58 in March and April. This coincided with the 12th March American Physical Society Meeting from Seattle, where more than 50 post-deadline contributions have been presented in a late-night session.

After April, the number of studies about $MgB_2$ decrease to 43 in May, 29 June and 22 in July. Anyway, the decrease in the number of $MgB_2$ studies does not reflect the lost of interest in this compound, but the proximity of summer holidays and the attendance of summer conferences.

The topics of these 260 studies cover a wide area of subjects, such as: preparation in the form of bulk, thin films, wires, tapes; the effect of substitution with various elements on $T_c$, isotope and Hall effect measurements, thermodynamic studies, critical currents and fields dependencies, to microwave and tunneling properties.

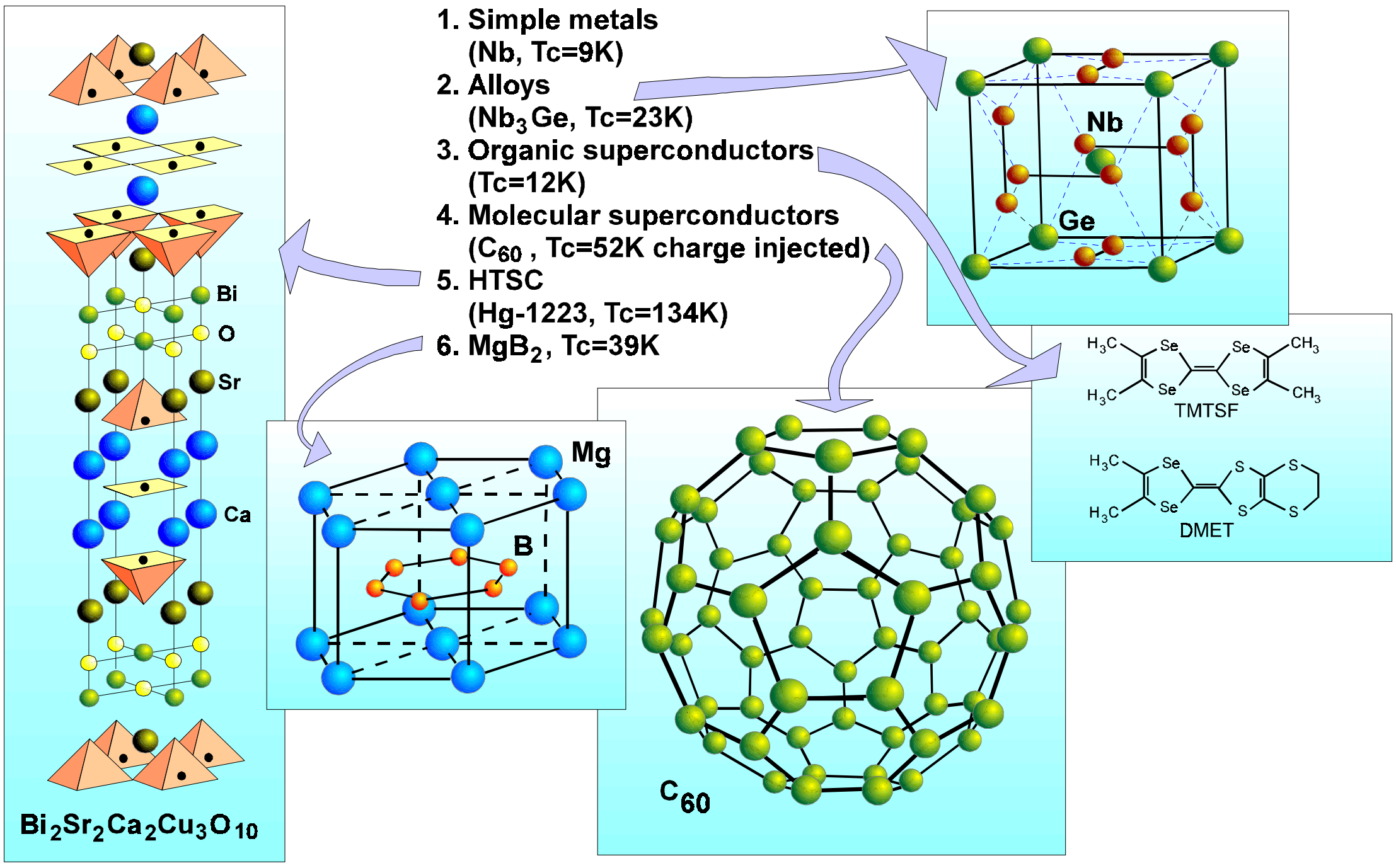

**Figure 3.** Comparison between the structures of different classes of superconductors

Since January of this year a considerable amount of effort has expended in order to understand the origin of superconductivity in this compound. Several theories have been already proposed [Baskaran], [Hirsch (a)], [Hirsch (b)], [Hirsch (c)], [Imada], [Voelker], [Haas], [Alexandrov], [Furukawa], [Cappelluti]. However, the superconductivity mechanism in $MgB_2$ is still to be decided. Recent calculations try to theoretically forecast the electronic properties of this material or similar compounds [An], [Antropov], [Bascones], [Bohnen], [Haas], [Kortus], [Kobayashi], [Kohmoto], [Lampakis], [Manske], [Medvedeva (a)], [Medvedeva (b)], [Medvedeva (c)], [Mehl], [Papaconstantopoulos], [Park], [Ravindran], [Satta], [Singh (a)], [Singh (b)], [Vajeeston], [Wan], [Yamaji], [Yildirim].

Why such a large interest in $MgB_2$ from the physics community one may ask? After all its critical temperature is only 40 K, more than three times lower than 134 K attained by the mercury-based high-$T_c$ superconducting (HTSC) cuprates. Besides, we already have wires made of high-$T_c$ copper oxides which operate above liquid nitrogen temperature (77 K). One important reason is the cost - HTSC wires are 70% silver [Grant], therefore expensive. Unlike the cuprates, $MgB_2$ has a lower anisotropy, larger coherence lengths, transparency of the grain boundaries to current flow, which makes it a good candidate for applications. $MgB_2$ promises a higher operating temperature and higher device speed than the present electronics based on Nb. Moreover, high critical current densities, $J_c$, can be achieved in magnetic fields by oxygen alloying [Eom], and irradiation shows an increase of $J_c$ values [Bugoslavski (b)].

$MgB_2$ possesses the simple hexagonal $AlB_2$-type structure (space group P6/mmm), which is common among borides. The $MgB_2$ structure is shown in Figs. 1, 3. It contains graphite-type boron layers which are separated by hexagonal close-packed layers of magnesium. The magnesium atoms are located at the center of hexagons formed by borons and donate their electrons to the boron planes. Similar to graphite, $MgB_2$ exhibits a strong anisotropy in the B-B lengths: the distance between the boron planes is significantly longer than in-plane B-B distance. Its transition temperature is almost twice as high as the highest $T_c$ in binary superconductors, $Nb_3Ge$ $T_c$=23K. Making a comparison to other types of superconductors (Fig. 3), one can see that $MgB_2$ may be the "ultimate" low-$T_c$ superconductor with the highest critical temperature.

According to initial findings, $MgB_2$ seemed to be a low-$T_c$ superconductor with a remarkably high critical temperature, its properties resembling that of conventional superconductors rather than of high-$T_c$ cuprates. This include isotope effect [Hinks], [Bud'ko (b)], a linear T-dependence of the upper critical field with a positive curvature near $T_c$ (similar to borocarbides) [Bud'ko (a)], a shift to lower temperatures of both $T_c$(onset) and $T_c$(end) at increasing magnetic fields as observed in resistivity R(T) measurements [Lee (a)], [Xu].

On the other hand, the quadratic T-dependence of the penetration depth $\lambda$(T) [Panagopoulos], [Pronin], [Klein], as well as the sign reversal of the Hall coefficient near $T_c$ [Jin (a)] indicates unconventional superconductivity similar to cuprates. One should also pay more attention to the layered structure of $MgB_2$, which may be the key of a higher $T_c$, as in cuprates and borocarbides.

## 2. Other diborides

After the announcement of $MgB_2$ superconductivity everybody hoped that this material will be the tip of a much hotter “iceberg”, being the first in a series of diborides with much higher $T_c$. However, up to date $MgB_2$ holds the record of $T_c$ among borides, as can be seen in Table 1.

The search for superconductivity in borides dates long time ago, in 1949 Kiessling founding a $T_c$ of 4K in TaB [Kiessling]. In 1970 Cooper et al. and in 1979 Leyarovska and Leyarovski looked for superconductivity in various borides (see Table 1).

Since the discovery of superconductivity in $MgB_2$ [Nagamatsu] there have been several theoretical studies to search for the potential high $T_c$ binary and ternary borides in isoelectronic systems such as $BeB_2$, $CaB_2$, transition metal (TM) diborides $TMB_2$, hole doped systems $Mg_{1-x}Li_xB_2$, $Mg_{1-x}Na_xB_2$, $Mg_{1-x}Cu_xB_2$, noble metal diborides $AgB_2$ and $AuB_2$, $CuB_2$ and related compounds [Satta], [Neaton], [Medvedeva (a)], [Medvedeva (b)], [Medvedeva (c)], [Ravindran], [Kwon], [Mehl].

Also there are further attempts to prepare new superconducting borides. The reports are still controversial, some authors reporting superconductivity in one compound and other authors finding the material normal. This is the case of $TaB_2$ found non-superconductive in earlier experiments [Leyaroska] and recently discovered to have a transition temperature of $T_c$=9.5 K [Kaczorowski (a)]. Similar situations apply for $ZrB_2$ found non-superconductive by Kaczorowski et al. [Kaczorowski (a)] and superconducting at 5.5K by Gasparov et al. [Gasparov], as well as $BeB_2$ found nonsuperconductive in stoichiometric form [Felner] but superconductive at 0.7K for the composition $BeB_{2.75}$ [Young].

The fact that some borides have been found superconducting by some authors while other authors didn't found traces of superconductivity in the same materials, suggests that nonstoichiometry may be an important fact in the superconductivity of this family. Extrapolating, in the case of $MgB_2$ it is also possible that the composition for which the critical temperature is a maximum to be slightly nonstoichiometric. The nonstoichiometry requirement for best superconducting properties is frequently seen in low-$T_c$ as well in high-$T_c$ superconductors.

Therefore, in the search for new superconducting borides one should take into account several items.

First, one should try several compositions, as the superconductivity may arise only in nonstoichiometric compounds.

Secondly, the contamination by non-reacted simple elements or other phases has to be ruled out by comparing the critical temperature of the new compound with the $T_c$ of the simple elements contained in the composition, and with other possible phases.

In order to make easier the search for new superconducting borides, in Table 1 we present a list of critical temperature for binary and ternary borides. Also, for the superconducting temperature of simple elements we show an updated picture in Fig. 4 [Yamashita].

## 3. Preparation

One of the advantages of $MgB_2$ fabrication is that magnesium diboride is already available from chemical suppliers, as it is synthesised since early 1950’s. However, sometimes the quality of $MgB_2$ powder commercially available is not as high as desirable. For example, $MgB_2$ commercial powders have wider transition in superconductive state [Tsindlekht] and slightly lower $T_c$ than the materials prepared in laboratory from stoichiometric Mg and B powders.

During the last seven month, $MgB_2$ has been synthesised in various forms: bulk (pollycrystals), thin films, powders, wires, tapes, as well as single crystals.

In Fig. 5 is a schematic picture of the fabrication methods used up to date for $MgB_2$ thin films, powders, single crystals, wires and tapes.

Typical methods of film fabrications used up to date are: pulsed laser deposition (PLD), co-evaporation, deposition from suspension, Mg diffusion, and magnetron sputtering. Please note that some authors refer to their preparation method as PLD, but they use in fact the Mg diffusion method for B films prepared by PLD. Different substrates have already been used for the deposition of $MgB_2$ thin films: SiC [Blank]; Si [Brinkman (a)], [Blank], [Zhai (a)], [Zhai (b)], [Plecenik (b)]; $LaAlO_3$ [Christen], [Zhai (b)]; $SrTiO_3$ [Eom], [Blank]; MgO [Grassano], [Moon], [Blank], [Ferdeghini]; $Al_2O_3$ [Grassano], [Christen], [Kang (d)], [Paranthaman (a)], [Wang (a)], [Zeng], [Zhai (b)], [Plecenik (b)], [Kim (a)], [Berenov], [Ferdeghini], [Ermolov]; stainless steel (SS) [Li (a)].

In Table 2, 3, 4 are shown the preparation conditions together with the critical temperatures of films prepared on various substrates by pulsed laser deposition, co-deposition, and Mg diffusion, respectively.

In the case of film fabrication, Mg volatility reflects the need for unheated substrates and Mg enriched targets. Due to magnesium volatility, an essential problem is to establish the minimum deposition and growth temperature at which the film crystallizes into the hexagonal structure, but at which Mg is not lost from the film. A recent report has used thermodynamics to predict the conditions under which $MgB_2$ synthesis would be possible under vacuum conditions [Liu (d)]. Important information on thermal stability of $MgB_2$ can be found in an experimental study which measures the $MgB_2$ decomposition rate [Fan].

In Fig. 6 is presented the critical temperature of films prepared by different methods on different substrates: $Al_2O_3$, $SrTiO_3$, Si, SiC, MgO and stainless steel (SS). The reports using sapphire show the highest $T_c$’s and sharpest transitions by Mg diffusion method [Zhai (b)], [Kim (a)], [Paranthaman (a)], [Kang (d)], [Plecenik (b)], [Wang (a)], [Zhai (b)].

For the same substrate, $Al_2O_3$, the thin films prepared by PLD have lower $T_c$ and usually wider transitions [Zeng], [Grassano], [Christen], [Zhai (b)] than the films prepared by Mg diffusion method. In order to prepare better quality films by PLD, the fabrication procedure must be optimised.

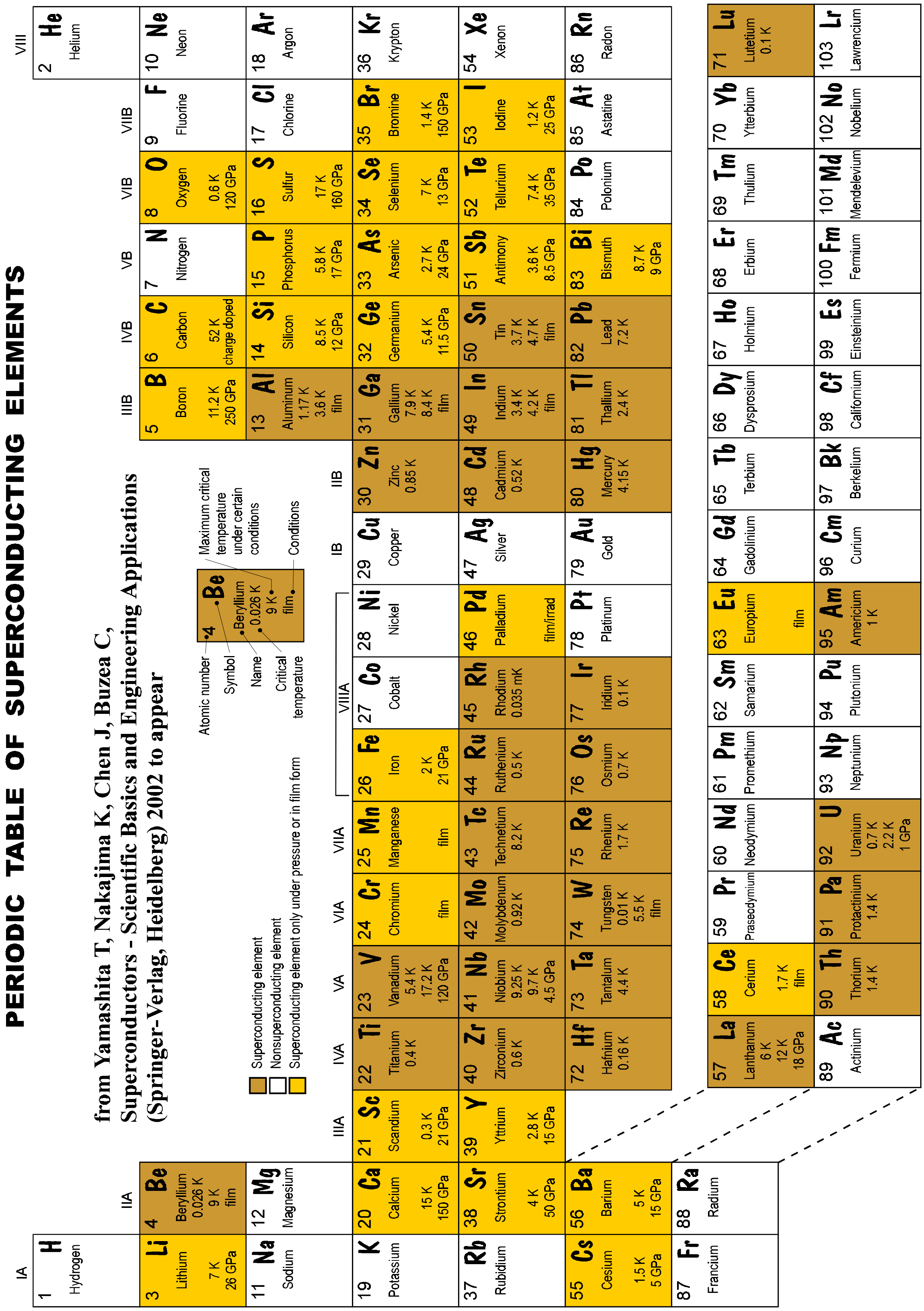

**Figure 4.** Periodic Table of elements with critical temperature at normal pressure, and maximum critical temperature attained under certain conditions (pressure, film form or charge injected) [Yamashita].

**Table 1.** List of binary, ternary, quaternary borides and borocarbides, their critical temperature $T_c$, and structure type. References: 1 [Kiessling]; 2 [Shulishova]; 3 [Savitskii]; 4 [Nowotny]; 5 [Matthias (a)]; 6 [Nagamatsu]; 7 [Cooper]; 8 [Gasparov]; 9 [Leyaroska]; 10 [Kaczorowski (a)]; 11 [Felner]; 12 [Young]; 13 [Strukova]; 14 [Havinga]; 15 [Hulm]; 16 [Matthias (b)]; 17 [Ku (a)]; 18 [Shelton (b)]; 19 [Lejay (b)]; 20 [Lejay (a)]; 21 [Rogl]; 22 [Sakai]; 23 [Shelton (a)]; 24 [Ku (d)]; 25 [Vandenberg]; 26[Ku (c)]; 27[Yvon]; 28 [Ku (b)]; 29[Johnston]; 30 [Watanabe]; 31 [Hsu]; 32 [Poole].

| Compound | $T_c$ (K) | Structure | Ref. | Compound | $T_c$ (K) | Structure | Ref. |
|---|---|---|---|---|---|---|---|
| TaB | 4 | $\alpha$TlI | 1, 2, 3 | $YB_2C_2$ | 3.6 | $YB_2C_2$ | 22 |
| NbB | 8.25 | $\alpha$TlI | 4, 5, 3 | $LuB_2C_2$ | 2.4 | $YB_2C_2$ | 22 |
| ZrB | 2.8-3.4 | | 3 | $Ca_{0.67}Pt_3B_2$ | 1.57 | $Ba_{0.67}Pt_3B_2$ | 23 |
| HfB | 3.1 | | 3 | $Sr_{0.67}Pt_3B_2$ | 2.78 | $Ba_{0.67}Pt_3B_2$ | 23 |
| MoB | 0.5 | | 3 | $Ba_{0.67}Pt_3B_2$ | 5.6 | $Ba_{0.67}Pt_3B_2$ | 23 |
| **$MgB_2$** | 40 | $AlB_2$ | **6** | $LaRh_3B_2$ | 2.82 | $CeCo_3B_2$ | 24 |
| $NbB_2$ | - | $AlB_2$ | 7, 8 | $LaIr_3B_2$ | 1.65 | $CeCo_3B_2$ | 24 |
| | 0.62 | | 9 | $LuOs_3B_2$ | 4.67 | $CeCo_3B_2$ | 24 |
| $NbB_{2.5}$ | 6.4 | $AlB_2$ | 7 | $ThRu_3B_2$ | 1.79 | $CeCo_3B_2$ | 24 |
| $Nb_{0.95}Y_{0.05}B_{2.5}$ | 9.3 | $AlB_2$ | 7 | $ThIr_3B_2$ | 2.09 | $CeCo_3B_2$ | 24 |
| $Nb_{0.9}Th_{0.1}B_{2.5}$ | 7 | $AlB_2$ | 7 | $Sc_{0.65}Th_{0.35}Rh_4B_4$ | 8.74 | $CeCo_4B_4$ | 25 |
| $MoB_2$ | - | $AlB_2$ | 7, 9 | $YRh_4B_4$ | 11.34 | $CeCo_4B_4$ | 25 |
| $MoB_{2.5}$ | 8.1 | $AlB_2$ | 7 | $NdRh_4B_4$ | 5.36 | $CeCo_4B_4$ | 25 |
| $Mo_{0.9}Sc_{0.1}B_{2.5}$ | 9 | $AlB_2$ | 7 | $SmRh_4B_4$ | 2.51 | $CeCo_4B_4$ | 25 |
| $Mo_{0.95}Y_{0.05}B_{2.5}$ | 8.6 | $AlB_2$ | 7 | $ErRh_4B_4$ | 8.55 | $CeCo_4B_4$ | 25 |
| $Mo_{0.85}Zr_{015}B_{2.5}$ | 11.2 | $AlB_2$ | 7 | $TmRh_4B_4$ | 9.86 | $CeCo_4B_4$ | 25 |
| $Mo_{0.9}Hf_{0.1}B_{2.5}$ | 8.7 | $AlB_2$ | 7 | $LuRh_4B_4$ | 11.76 | $CeCo_4B_4$ | 25 |
| $Mo_{0.85}Nb_{0.15}B_{2.5}$ | 8.5 | $AlB_2$ | 7 | $ThRh_4B_4$ | 4.34 | $CeCo_4B_4$ | 25 |
| $TaB_2$ | - | | 8, 9 | $DyRh_2Ir_2B_4$ | 4.64 | $CeCo_4B_4$ | 26 |
| | 9.5 | | 10 | $HoRh_2Ir_2B_4$ | 6.41 | $CeCo_4B_4$ | 26 |
| $BeB_2$ | - | $AlB_2$ | 11 | $Y_{0.5}Lu_{0.5}Ir_4B_4$ | 3.21 | $CeCo_4B_4$ | 26 |
| $BeB_{2.75}$ | 0.7 | not $AlB_2$ | 12 | $Ho_{0.1}Ir_4B_{3.6}$ | 2.12 | $CeCo_4B_4$ | 26 |
| $ZrB_2$ | - | | 9, 10 | $ErIr_4B_4$ | 2.34 | $CeCo_4B_4$ | 26 |
| | 5.5 | | 8 | $TmIr_4B_4$ | 1.75 | $CeCo_4B_4$ | 26 |
| $ReB_{1.8-2}$ | 4.5-6.3 | | 13 | $ErRh_4B_4$ | 4.3 | $LuRh_4B_4$ | 27 |
| $TiB_2$ | - | | 9, 10 | $TmRh_4B_4$ | 5.4 | $LuRh_4B_4$ | 27 |
| $HfB_2$ | - | | 9, 10 | $LuRh_4B_4$ | 6.2 | $LuRh_4B_4$ | 27 |
| $VB_2$ | - | | 9, 10 | $ScRu_4B_4$ | 7.23 | $LuRu_4B_4$ | 28 |
| $CrB_2$ | - | | 9 | $YRu_4B_4$ | 1.4 | $LuRu_4B_4$ | 29 |
| $Mo_2B$ | 5.07 | $\theta$-$CuAl_2$ | 14 | $LuRu_4B_4$ | 2.06 | $LuRu_4B_4$ | 29 |
| | 4.74 | | 3 | $YRh_4B_4$ | 10 | $LuRu_4B_4$ | 29 |
| $W_2B$ | 3.22 | $\theta$-$CuAl_2$ | 14 | $Y(Rh_{0.85}Ru_{0.15})_4B_4$ | 9.56 | $LuRu_4B_4$ | 29 |
| | 3.1 | | 3 | $Pr(Rh_{0.85}Ru_{0.15})_4B_4$ | 2.41 | $LuRu_4B_4$ | 29 |
| $Ta_2B$ | 3.12 | | 3 | $Eu(Rh_{0.85}Ru_{0.15})_4B_4$ | 2 | $LuRu_4B_4$ | 29 |
| $Re_2B$ | 2.8 | | 3, 15 | $Dy(Rh_{0.85}Ru_{0.15})_4B_4$ | 4.08 | $LuRu_4B_4$ | 29 |
| $Re_3B$ | 4.7 | | 13 | $Ho(Rh_{0.85}Ru_{0.15})_4B_4$ | 6.45 | $LuRu_4B_4$ | 29 |
| $Ru_7B_3$ | 2.58 | | 3 | $ErRh_4B_4$ | 7.8 | $LuRu_4B_4$ | 30 |
| $YB_6$ | 7.1 | $CaB_6$ | 16 | $Er(Rh_{0.85}Ru_{0.15})_4B_4$ | 8.02 | $LuRu_4B_4$ | 29 |
| $LaB_6$ | 5.7 | $CaB_6$ | 16 | $Tm(Rh_{0.85}Ru_{0.15})_4B_4$ | 8.38 | $LuRu_4B_4$ | 29 |
| $ThB_6$ | 0.74 | $CaB_6$ | 16 | $Lu(Rh_{0.85}Ru_{0.15})_4B_4$ | 9.16 | $LuRu_4B_4$ | 29 |
| $NdB_6$ | 3 | | 3 | $YRu_2B_2C$ | 9.7 | $LuNi_2B_2C$ | 31 |
| $ScB_{12}$ | 0.39 | $UB_{12}$ | 16 | $DyNi_2B_2C$ | 6.2 | $LuNi_2B_2C$ | 32 |
| $YB_{12}$ | 4.7 | $UB_{12}$ | 16 | $HoNi_2B_2C$ | 8.7 | $LuNi_2B_2C$ | 32 |
| $LuB_{12}$ | 0.48 | $UB_{12}$ | 16 | $ErNi_2B_2C$ | 10.5 | $LuNi_2B_2C$ | 32 |
| $ZrB_{12}$ | 5.82 | $UB_{12}$ | 16 | $TmNi_2B_2C$ | 11 | $LuNi_2B_2C$ | 32 |
| $YRuB_2$ | 7.8 | $LuRuB_2$ | 17, 18 | $LuNi_2B_2C$ | 16.1 | $LuNi_2B_2C$ | 32 |
| $Y_{0.8}Sc_{0.2}RuB_2$ | 8.1 | $LuRuB_2$ | 17 | $YNi_2B_2C$ | 15.6 | $LuNi_2B_2C$ | 32 |
| $LuRuB_2$ | 9.99 | $LuRuB_2$ | 17, 18 | $ScNi_2B_2C$ | 15.6 | $LuNi_2B_2C$ | 32 |
| $ScOsB_2$ | 1.34 | $LuRuB_2$ | 17, 18 | $ThNi_2B_2C$ | 8 | $LuNi_2B_2C$ | 32 |
| $YOsB_2$ | 2.22 | $LuRuB_2$ | 17, 18 | **$YPd_2B_2C$** | **23** | $LuNi_2B_2C$ | 32 |
| $LuOsB_2$ | 2.66 | $LuRuB_2$ | 17, 18 | $YPd_2B_2C$ | 14.5 | $LuNi_2B_2C$ | 32 |
| $Mo_2BC$ | 7.5 | $Mo_2BC$ | 19 | $YPt_2B_2C$ | 10 | $LuNi_2B_2C$ | 32 |
| $Mo_{1.18}Rh_{0.2}BC$ | 9 | $Mo_2BC$ | 20 | $LaPt_2B_2C$ | 10 | $LuNi_2B_2C$ | 32 |
| $Nb_2BN_{0.98}$ | 2.5 | $Mo_2BC$ | 21 | $ThPt_2B_2C$ | 6.5 | $LuNi_2B_2C$ | 32 |
| | | | | $PrPt_2B_2C$ | 6 | $LuNi_2B_2C$ | 32 |

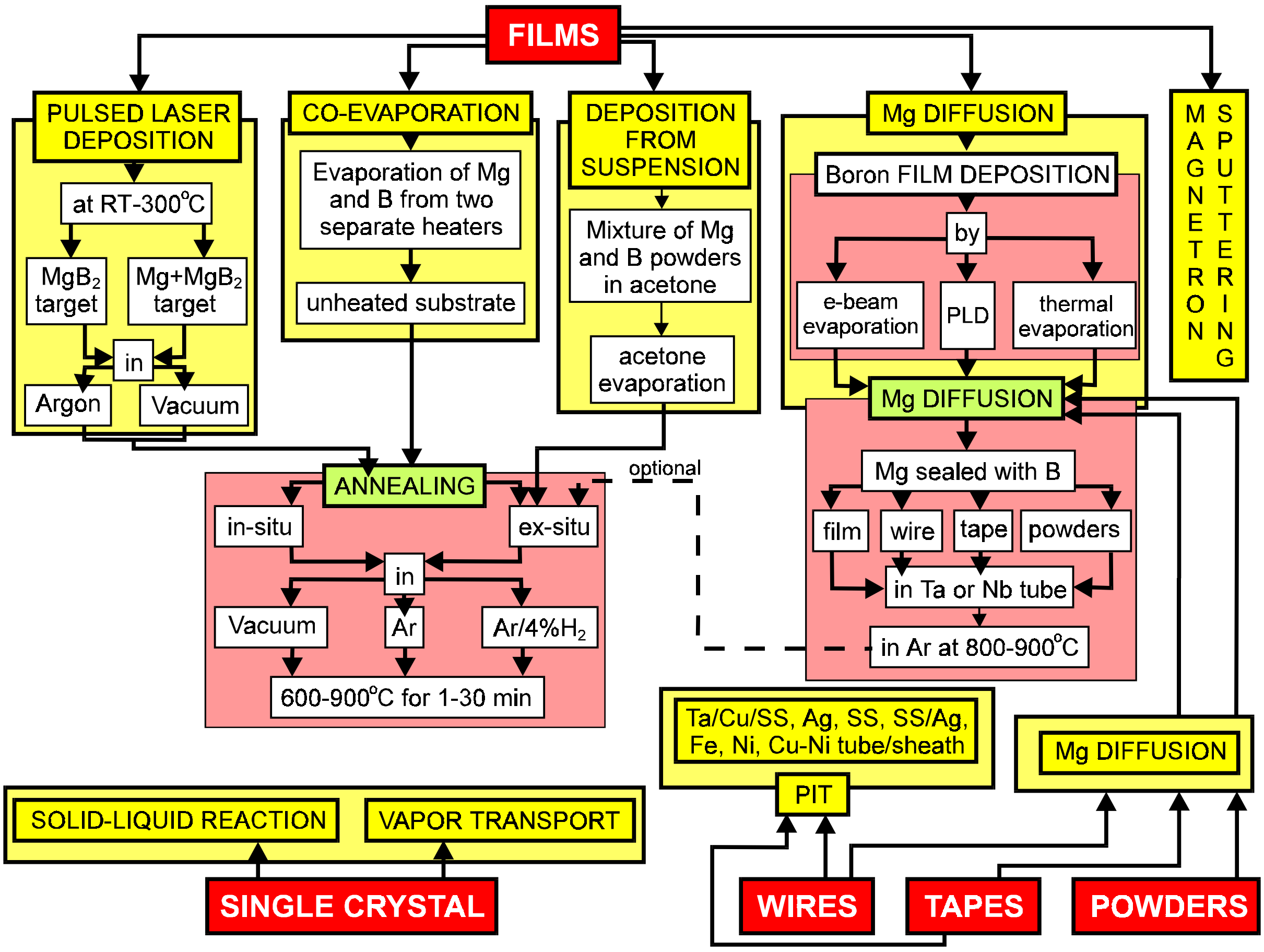

**Figure 5.** Schematics of the methods of preparation used for magnesium diboride

A recent report on PLD shows that the temperature of the film varies during pulsed laser deposition, the variation depending on the deposition parameters: substrate temperature, pressure in the ablation chamber, deposition rate [Buzea]. This may be an important factor due to Mg volatility.

In addition to sapphire, good quality films can be prepared on $SrTiO_3$ [Eom], Si [Plecenik (b)], and SS [Li (a)]. However, the films prepared on SS have poor adhesion on the substrate [Li (a)].

What is important to note from Fig. 6 is that the most important factor is the deposition method and not the type of substrate. The best method for $MgB_2$ thin film fabrication has proven to be Mg diffusion method. Why the type of substrate is not so important? Probably because the hexagonal structure of $MgB_2$ can accommodate substrates with different lattice parameters. However, we expect that further experiments to show a dependence of the critical temperature of $MgB_2$ with the type of substrate, as the critical temperature varies with the B-B bond length.

For electronic applications it is desirable that films with a high $T_c$ of 39K be made by a single step in-situ process. Usually, the magnesium diboride films with high superconducting temperatures made up to date are fabricated in a two-step process, film deposition followed by annealing.

What is interesting to note is that the Mg diffusion method is used not only for the fabrication of thin films, but also bulk, powders, wires and tapes. This method consists in Mg diffusion into B with different geometries. Due to the fact that Mg is highly volatile, Mg together with B are sealed in Nb or Ta tubes and heated up to 800-900 C. During this procedure, magnesium diffuses into the boron, increasing the size of the final reacted material.

For practical applications of $MgB_2$ (such as magnets and cables) it is necessary to develop tapes and wires.

Various research groups have already reported the fabrication of tapes and wires. Several critical issues relevant for practical fabrication of bulk wires remain unresolved. One of them is that $MgB_2$ is mechanically hard and brittle, therefore the drawing into fine-wire geometry is not possible. The wire and tape fabrication is achieved by two methods: the Mg diffusion and powder-in-tube (PIT) method.

Mg diffusion into B wires is a relatively easy method which can rapidly convert already commercially existent B wires into superconductive $MgB_2$ wires [Canfield], [Cunningham]. The magnesium diffusion has also attempted for fabricating tapes [Che].

However, the powder-in-tube (PIT) method is the most popular for achieving good quality wires [Glowacki], [Goldacker], [Jin (b)], [Wang (b)] and tapes [Grasso], [Sumption], [Liu (b)], [Song], [Kumakura], [Soltanian].

**Table.2.** Critical temperature and preparation conditions of films deposited by pulsed laser ablation PLD on SiC, Si, MgO, $SrTiO_3$ and $Al_2O_3$ substrates.

| Reference | Substrate | $T_{con}$ (K)<br>$T_{c0}$ (K) | Preparation conditions | post annealing | Obs. |
|---|---|---|---|---|---|
| Blank | SiC | 25<br>21 | Mg enrich target, 0.17mbar Ar, preablation of Mg, 4J/cm$^2$, 10Hz, RT | in-situ annealing, 0.2mbar in Ar at 600°C in Mg -plasma | target sintered in N-flow for 3h at 640°C, 10h at 500°C |
| Brinkman (a) | Si (100) | 27<br>15.5 | $MgB_2$ un-sintered target, KrF laser, 248nm, 0.17 mbar Ar, 10Hz, | ex-situ annealing in 0.2mbar Ar, increase of T to 600°C in 4 min, cooling to RT at a rate of 50°C/min | |
| Brinkman (a) | Si (100) | 27<br>11 | $MgB_2$ un-sintered target, KrF laser, 248nm, 0.17 mbar Ar, 10Hz, | ex-situ annealing in 0.2mbar Ar, increase of T to 600°C in 4 min, cooling to RT at a rate of 50°C/min | |
| Zhai (a) | Si (100) | 25.5<br>24 | stoichiometric targets, 0.1mTorr Ar, 248nm, 1.7-3.3J/cm$^2$, 15Hz | in-situ annealed in 0.2mTorr Ar/4%$H_2$, heated to 630°C at a rate of 100°C/min, held 20 min, cooled below 200°C at 50°C/min in 1atm Ar/4%$H_2$ | films contained micron-size PLD droplets |
| Zhai (b) | Si | 25<br>21.4 | $MgB_2$ target at 0.2 mTorr Ar/$H_2$ | in-situ annealing in vacuum at 630°C for 20 min | |
| Zhai (a) | Si (100) | 25<br>18 | stoichiometric targets, 0.1mTorr Ar, 248nm, 1.7-3.3J/cm$^2$, 15Hz | in-situ annealed in 0.2mTorr Ar/4%$H_2$, heated to 600°C at a rate of 100°C/min, held 20 min, cooled below 200°C at 50°C/min in 1atm Ar/4%$H_2$ | |
| Blank | Si | 24<br>16 | Mg enrich target, 0.17mbar Ar, preablation of Mg, 4J/cm$^2$, 10Hz, RT | in-situ annealing, 0.2mbar in Ar at 600°C | target sintered in N-flow for 3h at 640°C, 10h at 500°C |
| Blank | MgO | 25<br>23 | Mg enrich target, 0.17mbar Ar, preablation of Mg, 4J/cm$^2$, 10Hz, RT | in-situ annealing, 0.2mbar in Ar at 600°C in Mg -plasma | target sintered in N-flow for 3h at 640°C, 10h at 500°C |
| Eom | $SrTiO_3$ (111) | 36<br>34 | sintered $MgB_2$ targets, deposition at RT with a KrF laser, 248nm, 4J/cm$^2$, at 10Hz, in 0.3 Pa Ar | ex-situ annealing of films with Mg pellets in an evacuated small Nb tube at 850°C for 15 min, quenched to RT | Mg:B:O ratio 1.0:1.0:0.4; |
| Eom | $SrTiO_3$ (111) | 34<br>30 | sintered $MgB_2$ targets, deposition at RT with a KrF laser, 248nm, 4J/cm$^2$, at 10Hz, in 0.3 Pa Ar | ex-situ annealing of films with Mg pellets wrapped in a Ta envelope, in an evacuated small Nb tube at 750°C for 30 min, quenched to RT | Mg:B:O ratio 1.0:1.2:0.3; $J_c$(4.2K,1T) =$3\times10^6$A/cm$^2$, |
| Eom | $SrTiO_3$ (111) | 34<br>29 | sintered $MgB_2$ targets, deposition at RT with a KrF laser, 248nm, 4J/cm$^2$, at 10Hz, in 0.3 Pa Ar | ex-situ annealing of films with Mg pellets in an evacuated large quartz tube at 750°C for 30 min, quenched to RT | large amount of O in the film due to large volume of quartz tube; Mg:B:O ratio 1.0:0.9:0.7; large Jc |
| Blank | $SrTiO_3$ | 23<br>21 | Mg enrich target, 0.17mbar Ar, preablation of Mg, 4J/cm$^2$, 10Hz, RT | in-situ annealing, 0.2mbar in Ar at 600°C in Mg -plasma | target sintered in N-flow for 3h at 640°C, 10h at 500°C |
| Zeng | $Al_2O_3$ (0001) | 38<br>34 | films deposited in 120mTorr Ar at 250-300°C from unsintered targets of Mg:$MgB_2$ molar ratio of 4:1; 5J/cm$^2$, 5Hz | in-situ annealed, heated to 600°C at a rate of 40°C/min, held 10 min, cooled to RT in 20Torr Ar | large Jc; in the annealing step the film thickness decreases indicating evaporation of Mg |
| Grassano | $Al_2O_3$ | 28.6<br>23.4 | $MgB_2$ sintered target, $10^{-8}$-$10^{-9}$ mbar vacuum, deposition at RT | ex-situ annealing in Mg vapors at 650°C, in an evacuated quartz tube for 30 min, quenched to RT | |
| Zhai (b) | $Al_2O_3$ | 28<br>25 | $MgB_2$ target at 0.2 mTorr Ar/$H_2$ | C ex-situ annealed for 1h at 900°C in excess Mg | |
| Christen | $Al_2O_3$ | 26.5<br>22.5 | $MgB_2$/Mg segmented target, deposition at RT, $10^{-4}$Torr Ar/4%$H_2$, 1.7-3.3J/cm$^2$, 15Hz | in-situ annealed of Mg rich $MgB_2$ films with Mg cap layer at 0.7 atm Ar/4%$H_2$, heated to 550-600°C at a rate of 100°C/min, held 20 min, quenched to RT | |
| Zhai (b) | $Al_2O_3$ | 25<br>24 | $MgB_2$ target at 0.2 mTorr Ar/$H_2$ | in-situ annealing in vacuum at 600°C for 20 min | |
| Grassano | $Al_2O_3$ | 25<br>22.9 | Mg rich targets, non-sintered, 0.02mbar Ar, 450°C, 30Hz | | |
| Grassano | $Al_2O_3$ | 9.4<br>6.8 | $MgB_2$ sintered target, $10^{-8}$-$10^{-9}$ mbar vacuum, deposition at T>RT and T<750°C | ex-situ annealing in Mg vapors at 650°C, in an evacuated quartz tube for 30 min, quenched to RT | |

**Table.3.** Critical temperature and preparation conditions of films deposited by co-deposition (CD) on Si and $Al_2O_3$ substrates.

| Reference | Substrate | $T_{con}$ (K)<br>$T_{c0}$ (K) | Preparation conditions | post annealing | Obs. |
|---|---|---|---|---|---|
| Plecenik (b) | Si<br>(100) | 35<br>27 | evaporation of Mg and B from two separate resistive heaters on unheated substrates | in-situ annealed in vacuum, at 900°C for 30s, quenched to RT<br>ex-situ annealing in 1 atm Ar, for 15 min at 600°C | the Tc dependence on ex-situ annealing time was studied |
| Plecenik (b) | Si<br>(100) | 27<br>17 | evaporation of Mg and B from two separate resistive heaters on unheated substrates | in-situ annealed in vacuum, at 900°C for 30s, quenched to RT | films contained cracks |
| Plecenik (b) | $Al_2O_3$ | 33.3<br>32 | evaporation of Mg and B from two separate resistive heaters on unheated substrates | ex-situ annealing in Ar, at 600°C for 15 min, quenched to RT | smooth surface |
| Plecenik (b) | $Al_2O_3$ | 26<br>16 | evaporation of Mg and B from two separate resistive heaters on unheated substrates | in-situ annealing in vacuum, at 900°C for 30s, quenched to RT | smooth surface |

**Table 4.** Critical temperature and preparation conditions of films deposited by Mg diffusion method on $Al_2O_3$ substrates.

| Reference | Substrate | $T_{con}$ (K)<br>$T_{c0}$ (K) | Preparation conditions | Post annealing | Obs. |
|---|---|---|---|---|---|
| Zhai (b) | $Al_2O_3$ | 39<br>38.8 | e-beam evaporated B, reacted with $MgB_2$ and Mg at 900°C in a Ta tube | B ex-situ annealed for 1 h at 900°C | |
| Kim (a) | $Al_2O_3$<br>(1$\underline{1}$02) | 39<br>38.8 | B film deposited by PLD at RT, sealed with Mg in a Nb tube in Ar, 900°C for 10-30 min in a evacuated quartz tube, quenched to RT | | highly c-axis oriented structure with no impurity phase, RRR=2.3; $J_c$(5K,0T)=4×10$^7$A/cm$^2$, $J_c$(15K,5T)=10$^5$A/cm$^2$ |
| Paranthaman (a) | $Al_2O_3$<br>(102) | 39 38.3<br>38.6 38 | e-beam evaporated B films at RT in 10$^{-6}$Torr; B films with $MgB_2$ pellets and excess Mg sealed in a Ta tube, heated in an evacuated quartz tube to 600°C, held 5 min, T increased to 890°C, held for 10-20 min, cooled to RT | | c-axis and random grains are observed, $J_c$(20K,0T)=2×10$^6$A/cm$^2$, $J_c$(20K,1T)=2.5×10$^5$A/cm$^2$ |
| Kang (d) | $Al_2O_3$<br>(1$\underline{1}$02) | 39<br>37.6 | B film deposited by PLD at RT, sealed with Mg in a Ta tube in Ar, heated in a evacuated quartz tube to 900°C in 5 min, held for 10-30 min, quenched to RT | | c-axis oriented films, $J_c$(5K,0T)=6×10$^6$A/cm$^2$, $J_c$(35K,0T)=3×10$^5$A/cm$^2$ |
| Plecenik (b) | $Al_2O_3$ | 39<br>37 | B films thermally evaporated, sealed in Nb tube with Mg, 3kPa Ar, RT to 800°C in 60 min, kept 30 min, and quenched to RT | | |
| Wang (a) | $Al_2O_3$<br>(0001) | 39<br>36 | PLD deposition of B at 900°C at 6×10$^{-4}$Pa with a XeCl excimer laser; after deposition B films dipped in alcohol to remove $B_2O_3$ | B films with $MgB_2$ pellets wrapped in Ta foil, annealed in evacuated quartz tube, at 900°C for 60 min, cooled slowly to RT | most grains have c-axis orientation |
| Zhai (b) | $Al_2O_3$ | 38.3<br>38 | e-beam evaporated B, reacted with $MgB_2$ and Mg at 900°C in a Ta tube | Aex-situ annealed for 20 min at 900°C | |

PIT approach has been used to fabricate metal-clad $MgB_2$ wires/ribbons using various metals, such as: stainless steel SS [Song], [Kumakura], Cu [Glowacki], Ag [Glowacki], Ag/SS [Glowacki], Ni [Suo], Cu-Ni [Kumakura], Nb, Ta/Cu/SS [Goldacker], Fe [Jin (b)], [Wang (b)], [Soltanian], [Suo]

Usually, the PIT method consists in the following procudure. $MgB_2$ reacted powder or a mixture of Mg and B powders with stoichiometric composition is packed in various metal tubes or sheaths. These tubes are drwan into wires, cold-worked into ribbons, followed optionally by a heat treatment at 900-1000 °C.

For fabricating metal-clad $MgB_2$ wires/ribbons, hard but ductile and malleable metals are essential. These metals have to play a role of diffusion barrier for the volatile and reactive Mg. Also, it is important to find a suitable sheath material which does not degrade the superconductivity. Mg and $MgB_2$ tends to react and combine with many metals, such as Cu, Ag, forming solid solutions or intermetallics with low melting points, which renders the metal cladding useless during sintering of $MgB_2$ at 900-1000°C. One can see that there are only a small number of metals which are not soluble or do not form intermetallic compounds with Mg [Jin (b)]. These are Fe, Mo, Nb, V, Ta, Hf, W. Of these, the refractory metals (Mo, Nb, V, Ta, Hf, W) have inferior ductility compared to Fe, which makes iron the best candidate material as a practical cladding metal or diffusion barrier for $MgB_2$ wire and tape fabrication which includes annealing.

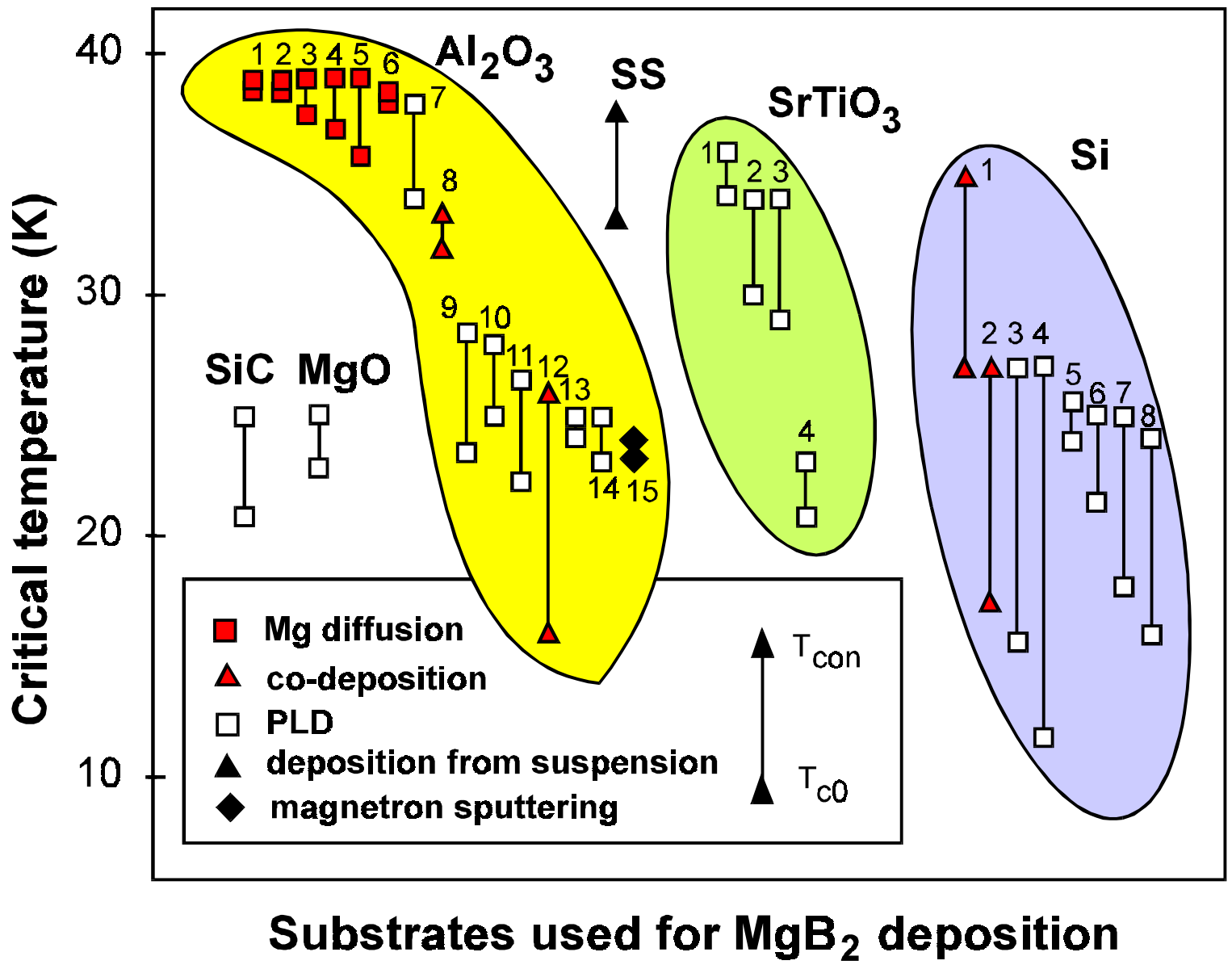

**Figure 6.** Critical temperature and critical temperature width for $MgB_2$ films deposited on different substrates. The data were taken from the references: $Al_2O_3$ - 1 [Zhai (b)], [Kim (a)], 2 [Paranthaman (a)], 3 [Kang (d)], 4, 8, 12 [Plecenik (b)], 5 [Wang (a)], 6, 10, 13 [Zhai (b)], 7 [Zeng], 9, 14 [Grassano], 11 [Christen], 14 [Ermolov]; $SrTiO_3$ - 1, 2, 3 [Eom], 4 [Blank]; Si - 1, 2 [Plecenik (b)], 3, 4 [Brinkman (b)], 5, 6, 7 [Zhai (b)], 8 [Blank]; MgO - [Blank]; SiC - [Blank]; Stainess Steel (SS) - [Li (a)].

If the annealing process is skipped, more metals could be used as sheaths, their reactivity with Mg being on a secondary plane. A fabrication process with no heat treatment would also reduce the fabrication costs.

In order to improve the superconducting properties of bulk $MgB_2$, two methods have been used: hot deformation [Handstein], [Frederick], [Indrakanti], [Shields], and high-pressure sintering [Jung (b)], [Takano], [Tsvyashchenko], [Jung (a)].

Single crystals are currently obtained by solid-liquid reaction method from Mg-rich precursor [Jung (c)], under high pressure in Mg-B-N system [Lee (a)], and vapor transport method [Xu].

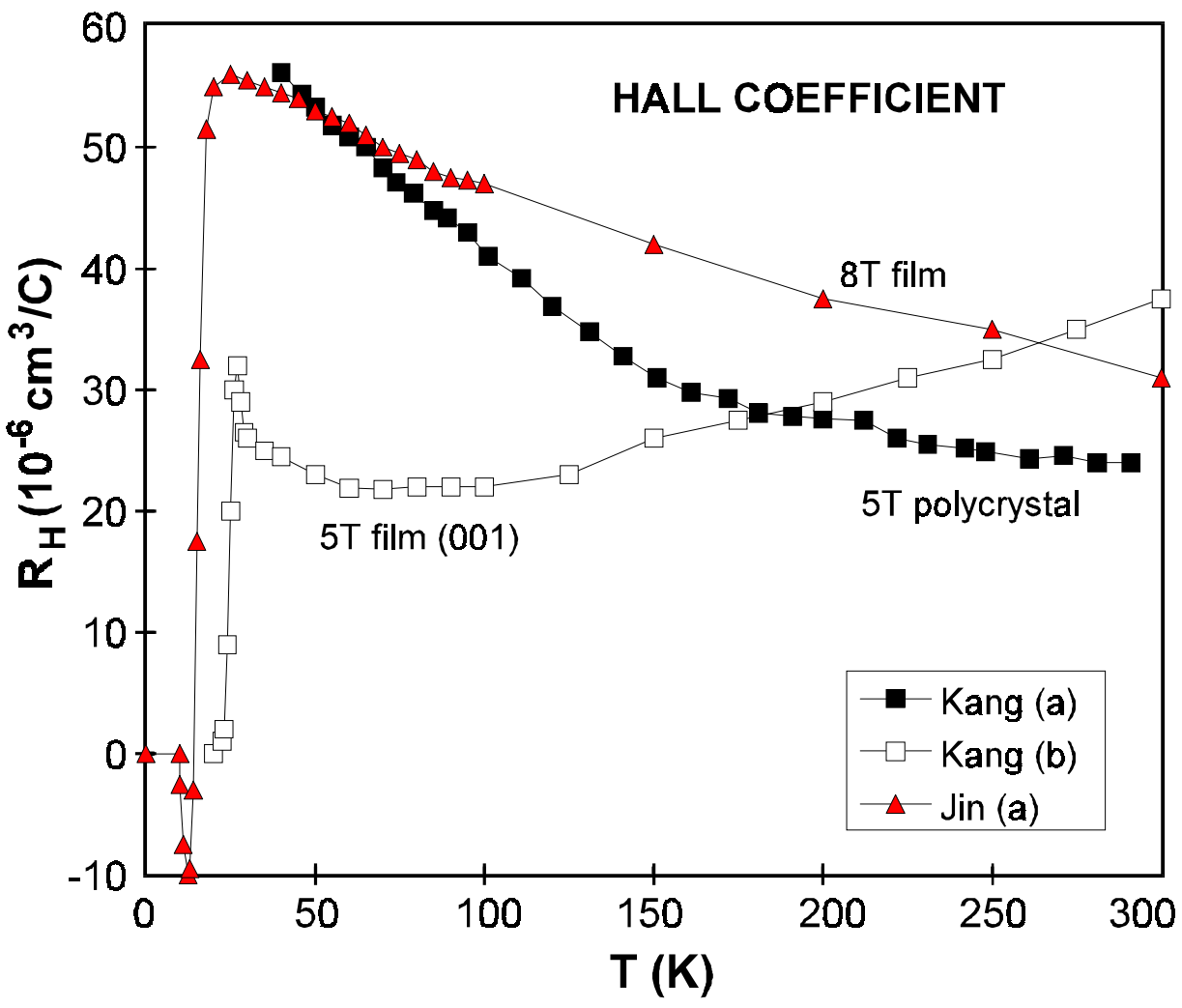

**Figure 7.** Hall coefficient versus temperature.

## 4. Hall coefficient

There are only three reports about Hall effect in $MgB_2$ until now. These are for polycrystals [Kang (a)], c-axis oriented films [Kang (b)], and films without preferential orientation [Jin (a)]. All reports agree with the fact that the normal state Hall coefficient $R_H$ is positive (Fig. 7), therefore the charge carriers in magnesium diboride are holes with a density at 300K of between $1.7 \div 2.8 \times 10^{23}$ holes/cm$^3$, about two orders of magnitude higher than the charge carrier density for $Nb_3Sn$ and YBCO [Kang (a)]. The three reports disagree weather the Hall coefficient in normal state increases or decreases in temperature. For the Hall coefficient measured on the c-axis oriented film [Kang (b)] one can notice a peak just above the transition. In the case of the non-oriented film [Jin (a)] one can notice a sign reversal of $R_H$ in the mixed state.

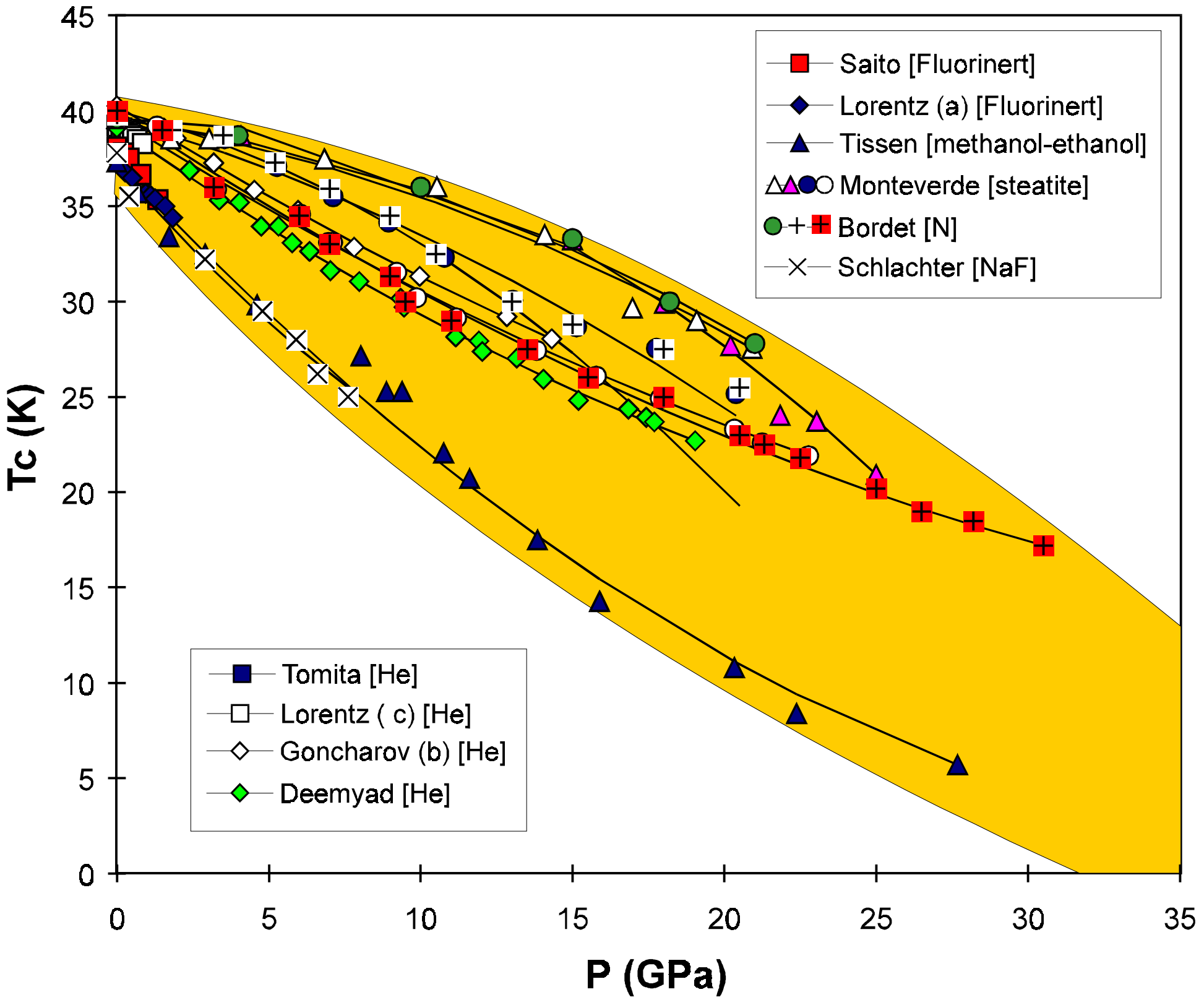

**Figure 8.** The critical temperature of MgB2 versus applied pressure. The legends indicates the pressure medium used by each author.

## 5. Pressure dependent properties

### *5.1. Critical temperature versus pressure*

The response of $MgB_2$ crystal structure to pressure is important for testing the predictions of competing theoretical models, but it also might give valuable clues for guiding chemical substitutions. For example, in simple metals BSC-like superconductors, as Aluminium, critical temperature $T_c$ decreases under pressure due to the reduced electron-phonon coupling energy from lattice stiffening [Gubser]. Also, a large magnitude of the pressure derivative $dT_c/dP$ is a good indication that higher values of $T_c$ may be obtained through chemical means.

The pressure effect on the superconducting transition of $MgB_2$ is negative to the highest pressure studied. Fig. 8 shows the evolution of the critical temperature with pressure from several references [Bordet], [Deemyad], [Goncharov (b)], [Lorenz (a)], [Lorenz (c)], [Monteverde], [Saito], [Schlachter], [Tissen], [Tomita]. All reports agree with the fact that the critical temperature of $MgB_2$ is shifted to lower values, giving different rates of decrease $-dT_c/dP$.

$T_c$ follows a quadratic or linear dependence on applied pressure, decreasing monotonically. Despite the fact that $T_c(P)$ data from different authors differs considerably, in Fig. 8 one can notice a pattern. Samples with lower $T_c$ at zero pressure have a much steeper $T_c(P)$ dependence than the samples with higher $T_c$. More exactly, the initial slope rate of the samples with lower $T_c$ is about -2 K/GPa, while that of samples with higher $T_c$ is about -0.2 K/GPa, as can be seen in Fig. 9 inset.

The initial rate of the derivative $-dT_c/dP$ is invers proportional to pressure, most of the data falling in the quadratic dependence depicted in Fig. 9 inset, in the shadowed region. Several data do not fit this dependence [Monteverde], but taking into account the solid pressure medium (steatite) they used, the quasi-hydrostatic nature of their experiment makes this explainable.

Also, from Fig. 8 can be seen that samples with higher $T_c$ have a negative curvature of $T_c(P)$ dependence, changing to positive for samples with lower $T_c$. The change in the sign of curvature is illustrated by the derivative $-dT_c/dP(P)$ from Fig. 9.

Taking into account the strong compressibility anisotropy [Bordet], [Prassides], [Goncharov (b)], [Vogt], [Schlachter] which will be described in the next paragraph, it is likely that shear stress of sufficient magnitude will cause important changes in the $T_c(P)$ dependence.

Large shear stresses are generated by changing the pressure on a solid medium, such as steatite [Tomita]. The shear stresses generated in cooling Fluorinert or other liquids with similar melting curve are much smaller, depending on the experimental procedure (cooling rate, change in applied force).

One report shows the existence of a cusp in the $T_c(P)$ dependence at about 9 GPa, the authors attributing it to a pressure-induced electronic transition [Tissen]. However, the data from other reports do not show any cusp.

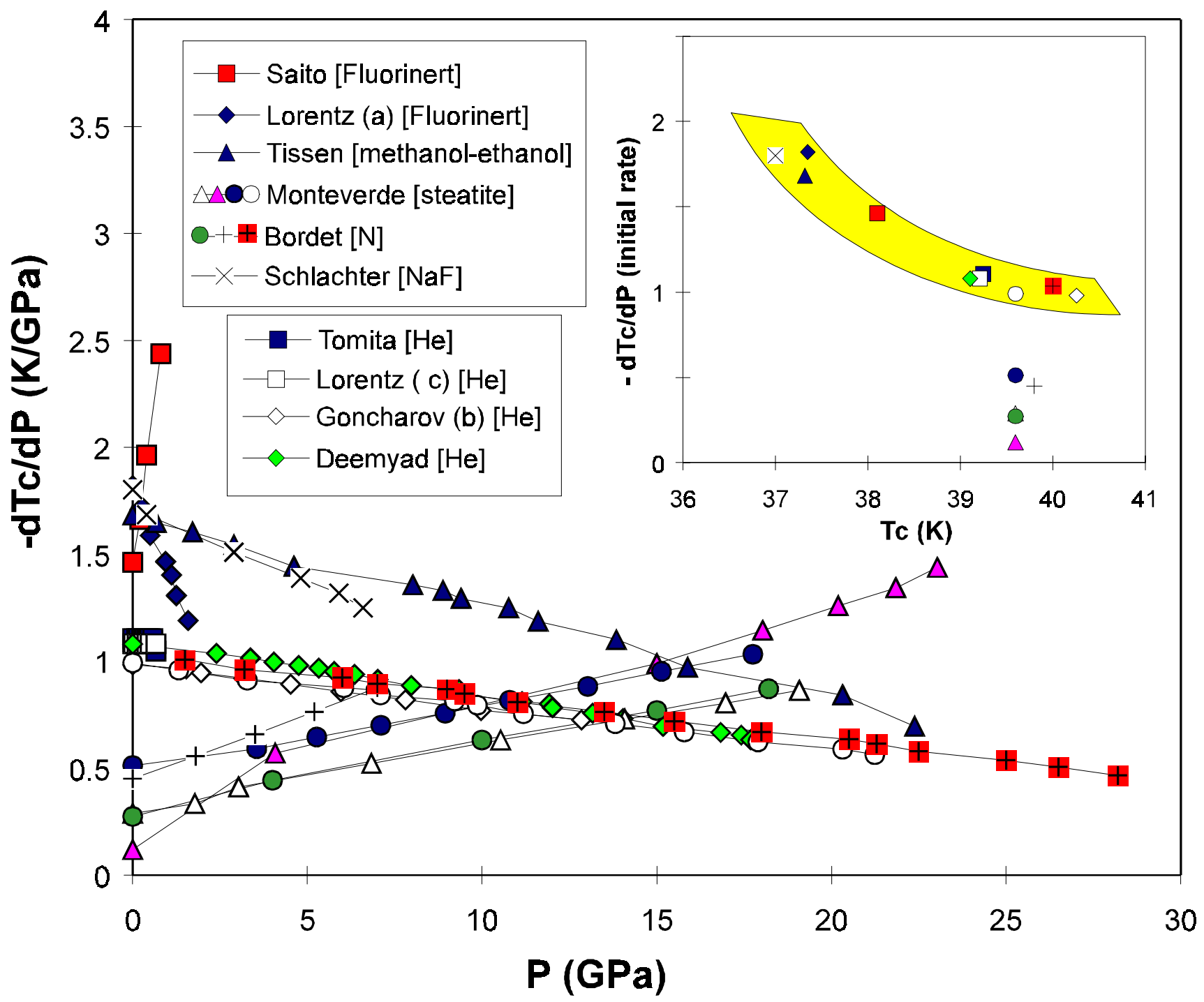

**Figure 9.** The derivative of critical temperature in pressure of MgB2 versus applied pressure. Inset shows the initial rate of variation of the derivative versus $T_c$ at zero pressure. The legend indicates the pressure medium used by each author.

Schlachter et al. noticed that the pressure effect is fully reversible in a He gas pressure cell while in the diamond-anvil cell (DAC) the application of pressure leads to a stronger $T_c$-decrease than in the He pressure cell, and after the release of pressure a degradation effect with lower $T_c$ and a broader transition compared to the first measurement at ambient pressure occurs [Schlachter]. Such degradation may be explained by shear stresses and uniaxial pressure components, which cannot be avoided in a DAC at low temperatures.

The discrepancy in the $T_c$ dependence by various groups may arise partially due to different pressure transmitting media used in experiments, pointed out in the legend of Figs. 8 and 9. Considering its anisotropic structure, $MgB_2$ may be sensitive to non-hydrostatic pressure components, which would explain the spread of $dTc/dp$ values reported in the literature.

But more interesting, several authors report different $T_c(p)$ dependencies for different $MgB_2$ samples measured in the same experimental setup (see for example [Monteverde], [Bordet]), which points toward the Mg nonstoichiometry as an important factor in determining the pressure dependent behavior of the critical temperature.

The reduction of $T_c$ under pressure is consistent with a BCS-type pairing interaction mediated by high-frequency boron phonon modes. This indicates that the reduction of the density of states at the Fermi energy, due to the contraction of B-B and B-Mg bonds, dominates the hardening phonon frequencies that could cause increase of $T_c$ as an external pressure is applied.

A hole-based theoretical scenario for explaining the superconductivity of $MgB_2$ predicted a positive pressure coefficient on $T_c$, as a result of decreasing in-plane B-B distance with increasing pressure [Hirsch (a)], [Hirsch (b)], [Hirsch (c)]. This contradicts all experimental data. However, the situation is more complex if pressure also affects the charge transfer between Mg-B, resulting in different responses of the system in the underdoped and overdoped regimes.

## *5.2. Anisotropic compressibility*

Diffraction studies at room temperature under pressure have been performed by a series of authors. Most of the reports study the lattice compression up to 6 GPa [Prassides], [Goncharov (a)], [Vogt], [Schlachter], while there is a report which goes up to 30 GPa [Bordet].

$MgB_2$ remains strictly hexagonal until the highest pressure, no sign of structural transition being seen.
This is illustrated in the pressure variation of the normalized hexagonal lattice constants a and c from Fig. 10. One notice a clear anisotropy in the bonding of the $MgB_2$ structure. All reports show that the lattice parameter along c-axis decreases faster with pressure than along a-axis (Fig. 10), demonstrating that the out-of-plane Mg-B bonds are much weaker than in-plane Mg-Mg bonds. This fact is emphasized also by the lattice parameters variation versus temperature (Fig. 13).

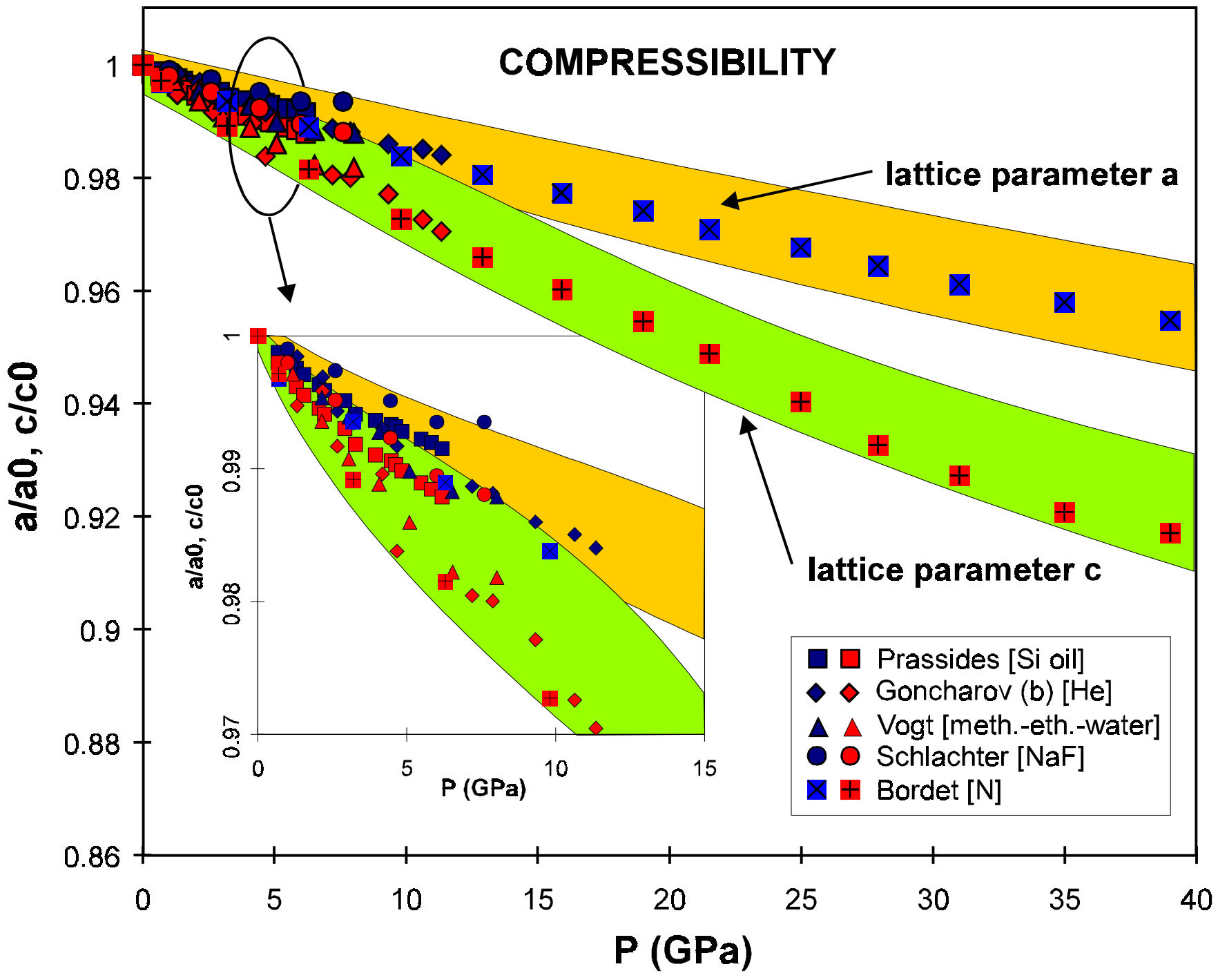

**Figure 10.** The normalized lattice parameters to the zero pressure value versus applied pressure. Inset shows the same data at lower pressures at an enlarged scale. The legend indicates the pressure medium used by each author.

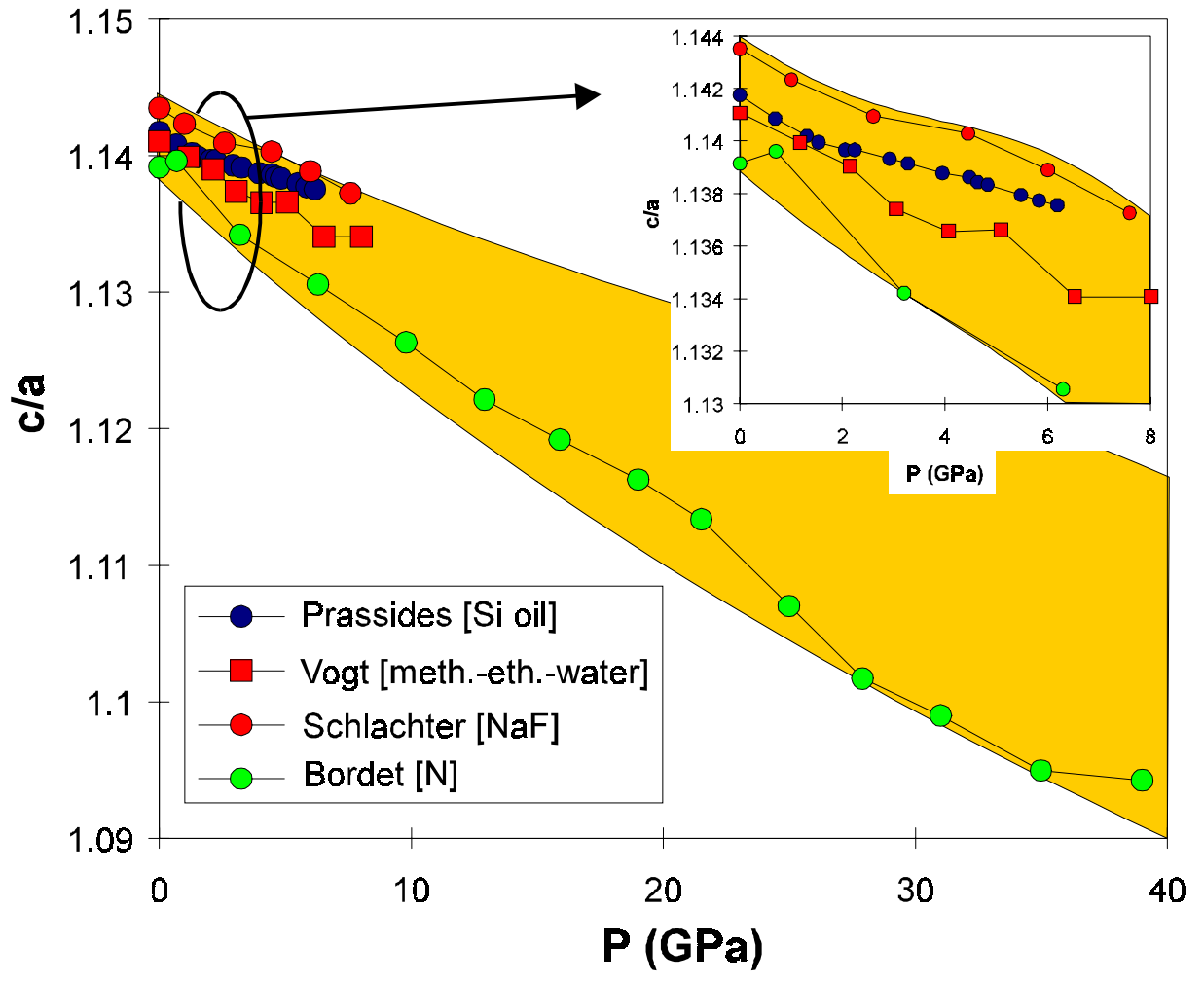

**Figure 11.** The ratio between the lattice parameters along c- and a-axis versus pressure. Inset shows the same data at lower P. The legend indicates the pressure medium used by each author.

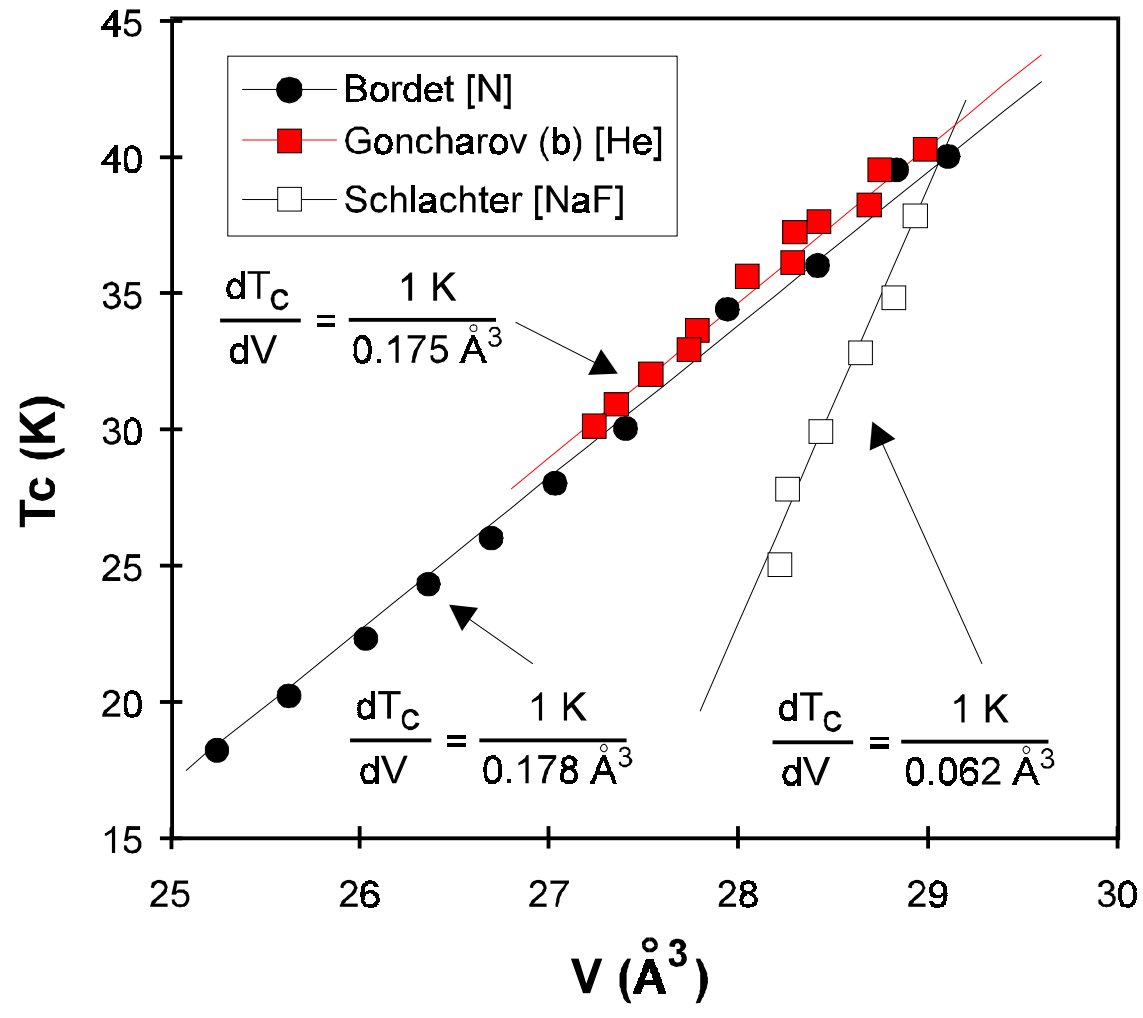

**Figure 12.** The critical temperature of $MgB_2$ versus the volume of the unit cell. The data were calculated from Figs. 8 and 10. The legend indicates the pressure medium used by each author.

The difference in compressibility values obtained in different reports may arise from the fact of using different pressure-transmitting media: helium [Lorenz (c)], [Tomita], [Goncharov (b)], [Deemyad]; Fluorinert [Lorenz (a)], [Saito]; methanol-ethanol [Tissen]; methanol-ethanol-water [Vogt]; solid pressure medium steatite [Monteverde]; NaF [Schlachter]; silicon oil [Prassides]; and nitrogen [Bordet].

The compressibility anisotropy decreases linearly with pressure, as illustrated in Fig. 11.

From the critical temperature dependence on applied pressure corroborated with the data of compressibility we calculated the dependence of $T_c$ on the unit cell volume. We plotted $T_c(V)$ in Fig. 12.

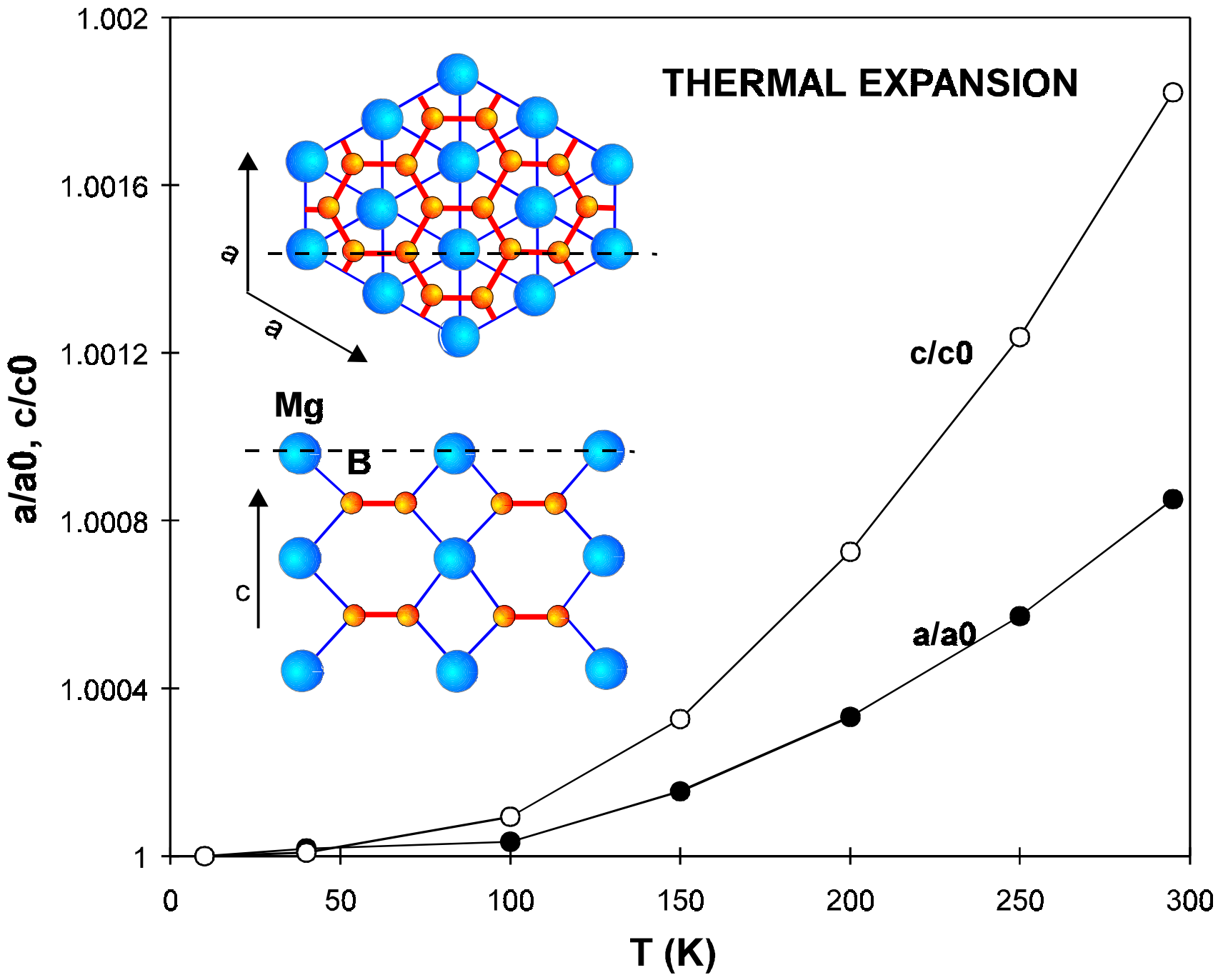

**Figure 13.** The normalized thermal expansion along a and c-axis. Inset shows the boron-boron and magnesium-boron bonds. The data thermal expansion data are taken from [Jorgensen].

The large value of critical temperature variation with small modification in the unit cell volume demonstrate that Mg-B and B-B bonding distances are crucial in the superconductivity of $MgB_2$ at such a high $T_c$ compared to other materials. The reduction of critical temperature with 1 K is achieved by lowering the unit cell volume with only 0.17 Å$^3$ deduced from the data of [Bordet] and [Goncharov (b)]. This implies a very sensitive dependence of the superconducting properties to the interatomic distances.

## 6. Thermal expansion

Thermal expansion, analogous to compressibility, exhibits a pronounced anisotropy, with the c-axis responses substantially higher than a-axis, as illustrated in Fig. 13. The lattice parameter along c-axis increases twice compared to the lattice parameter along a-axis at the same temperature [Jorgensen].

This fact demonstrates that the out-of-plane Mg-B bonds are much weaker than in-plane Mg-Mg bonds.

Band structure calculations clearly reveal that, while strong B-B covalent bonding is retained, Mg is ionized and its two electrons are fully donated to the B-derived conduction band [Kortus]. Then it may be assumed that the superconductivity in $MgB_2$ is essentially due to the metallic nature of the 2D sheets of boron and high vibrational frequencies of the light boron atoms lead to the high $T_c$ of this compound.

## 7. Effect of substitutions on critical temperature

The substitutions are important from several points of view. First, it may increase the critical temperature of one compound. Secondly, it may suggest the existence of a related compound with higher $T_c$. And last but not least, the doped elements which do not lower the $T_c$ considerably may act as pinning centers and increase the critical current density.

In the case of $MgB_2$, several substitutions have been tried up to date: carbon [Ahn], [Mehl], [Paranthaman (b)], [Takenobu], [Zhang (a)]; aluminium [Bianconi (b)], [Cimberle], [Li (b)], [Lorenz (b)], [Slusky], [Xiang], [Ogita], [Postorino]; lithium, silicon [Cimberle], [Zhao (a)]; beryllium [Felner], [Mehl]; zinc [Kazakov], [Moritomo]; copper [Mehl], [Kazakov]; manganese [Ogita], [Moritomo]; niobium, titanium [Ogita]; iron, cobalt, nickel [Moritomo].

In Fig. 14 is shown $T_c$ versus the doping content, $0<x<0.2$, for substitutions with Al, C, Co, Fe, Li, Mn, Ni, Si, and Zn.

The critical temperature decreases at various rates for different substitutions, as can be seen in Figs. 14 and 15. The largest reduction is given by Mn [Moritomo], followed by Co [Moritomo], C [Takenobu], Al [Li (b)], Ni, Fe [Moritomo]. The elements which do not reduce the critical temperature of $MgB_2$ considerable are Si and Li [Cimberle].

Up to date, all the substitutions alter the critical temperature of magnesium diboride with an exception: Zn, which increases $T_c$ slightly, with less than one degree [Moritomo], [Kazakov]. There are only two reports regarding Zn doping. Both agree with the fact that at a certain doping level $T_c$ increases, but disagree with the doping level for which this fact occur. This may be due to the incorporation of a smaller amount of Zn than the doping content. Anyway, Zn doping deserves further attention.

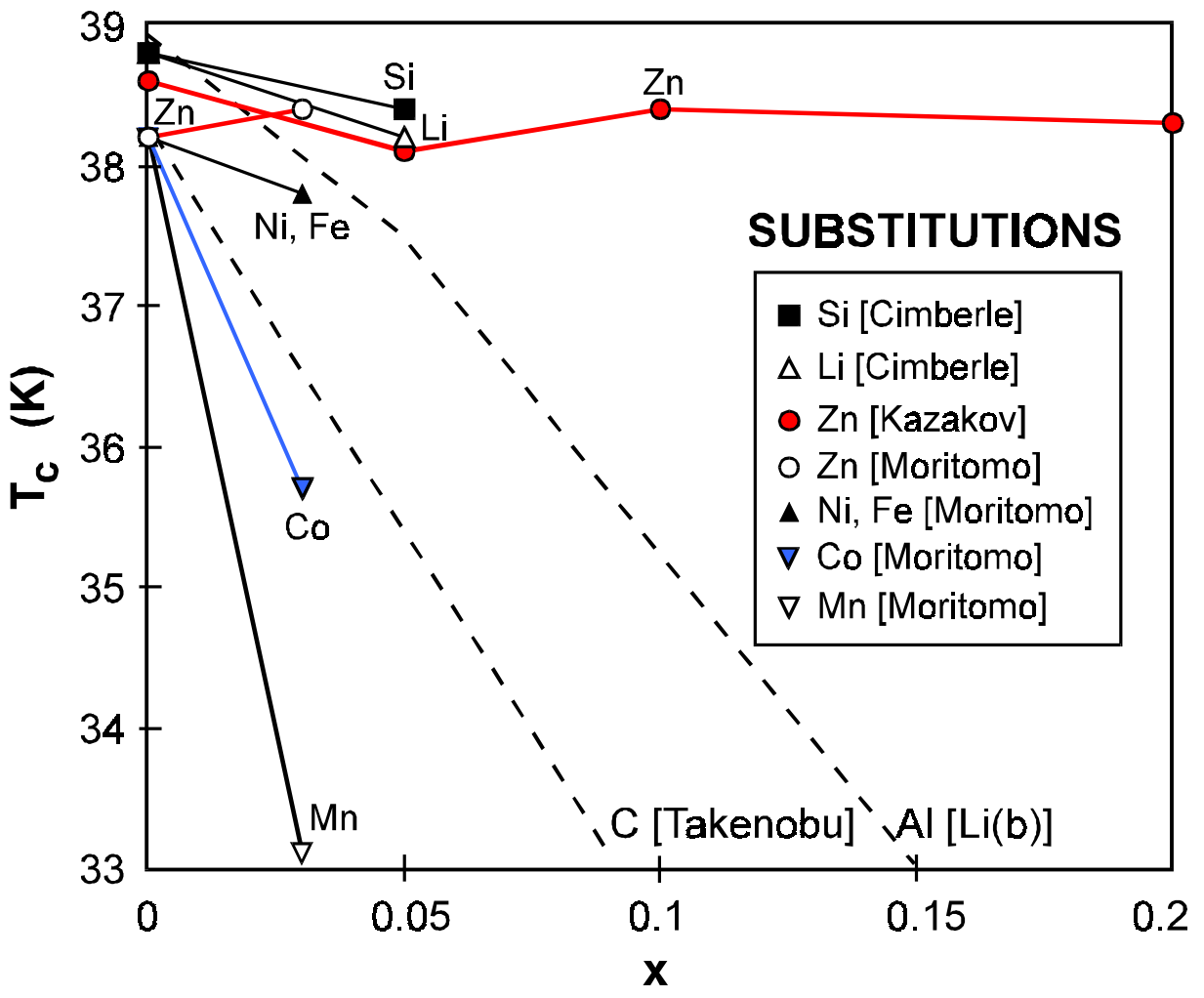

**Figure 14.** Critical temperature dependence on doping content x for substitutions with Zn, Si, Li, Ni, Fe, Al, C, Co, Mn (0<x<0.2).

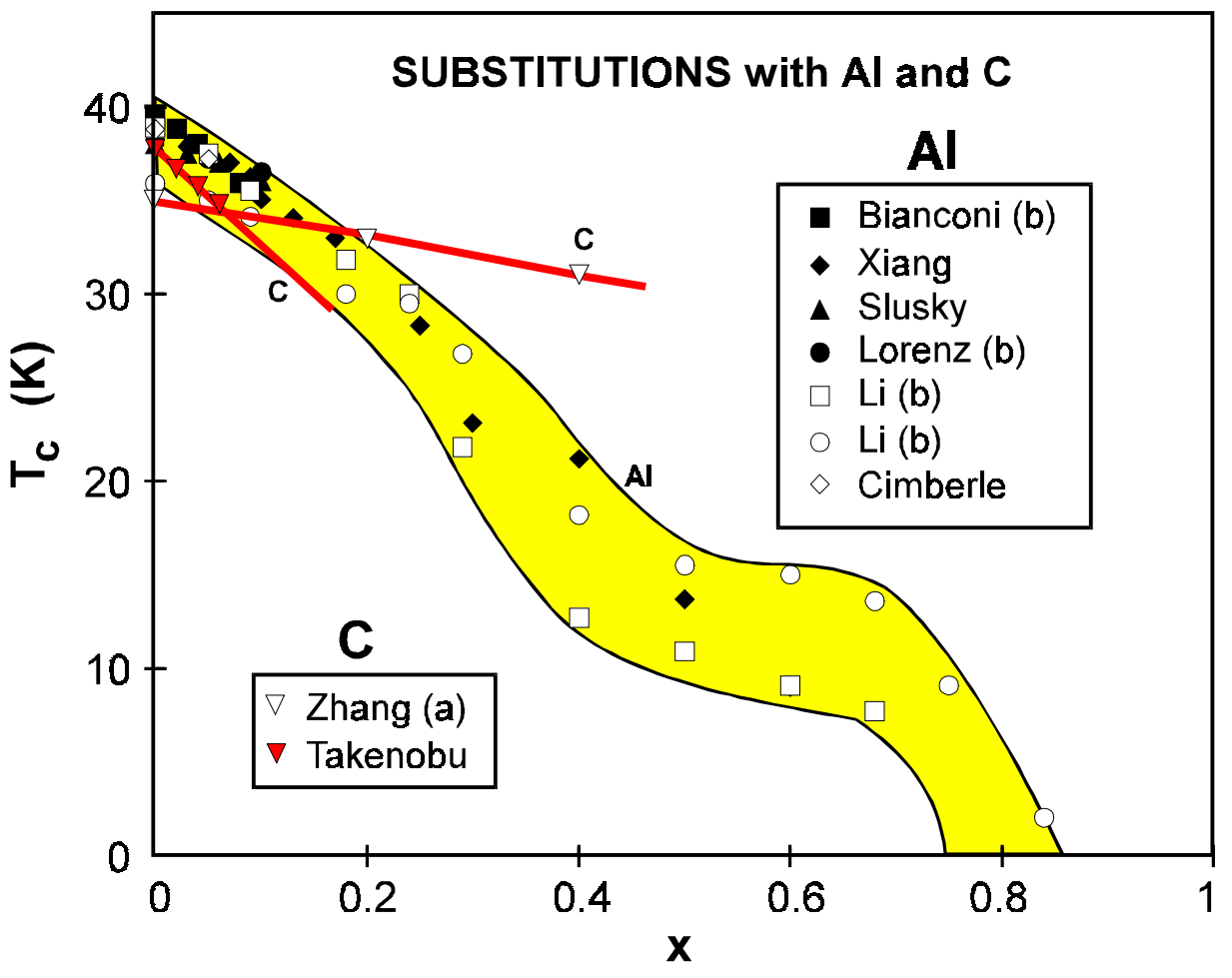

**Figure 15.** Critical temperature dependence on doping content x for substitutions with Al and C.

In Fig. 15 is shown $T_c$ versus doping level 0<x<0.82 for substitutions with C [Zhang], [Takenobu] and Al [Bianconi], [Xiang], [Slusky], [Lorenz (b)], [Li (b)], [Cimberle]. The critical temperature variation versus x for Al reflects the existence of structural transitions at different doping levels, the slopes $dT_c/dx$ from different reports being in agreement with each other.

The investigation of $T_c$ and lattice parameters with Al substitution in $Mg_{1-x}Al_xB_2$, lead to the conclusion that $MgB_2$ is near a structural instability that can destroy superconductivity [Slusky]. Critical temperature decreases smoothly with increasing x from 0<x<0.1, accompanied by a slight decrease of the c-axis parameter. At x≈1 there is an abrupt transition to a non-superconducting isostructural compound which has a c-axis shortened by about 0.1 Å. The loss of superconductivity associated with decreasing the c-axis length with no change in the cell symmetry suggests that the structure parameters of $MgB_2$ are particularly important in its superconductivity at high $T_c$.

In the case of C doping the two reports [Zhang], [Takenobu] disagree with the value of the critical temperature at different doping levels. This may be due to the fact that carbon was not completely incorporated into the $MgB_2$ structure in the report of Zhang [Zhang (a)].
Also, the existence of different critical temperatures for the starting $MgB_2$ at zero doping levels may give different $T_c(x)$ behaviours. As it was pointed out previously, we believe Mg nonstoichiometry leads to different critical temperature dependencies versus the applied pressure, therefore we may expect different $T_c(x)$ behaviours as a function of small Mg nonstoichiometry.

However, in order to have a clear picture about the effect of substitutions on $MgB_2$ more data in a wider range of doping levels are necessary.

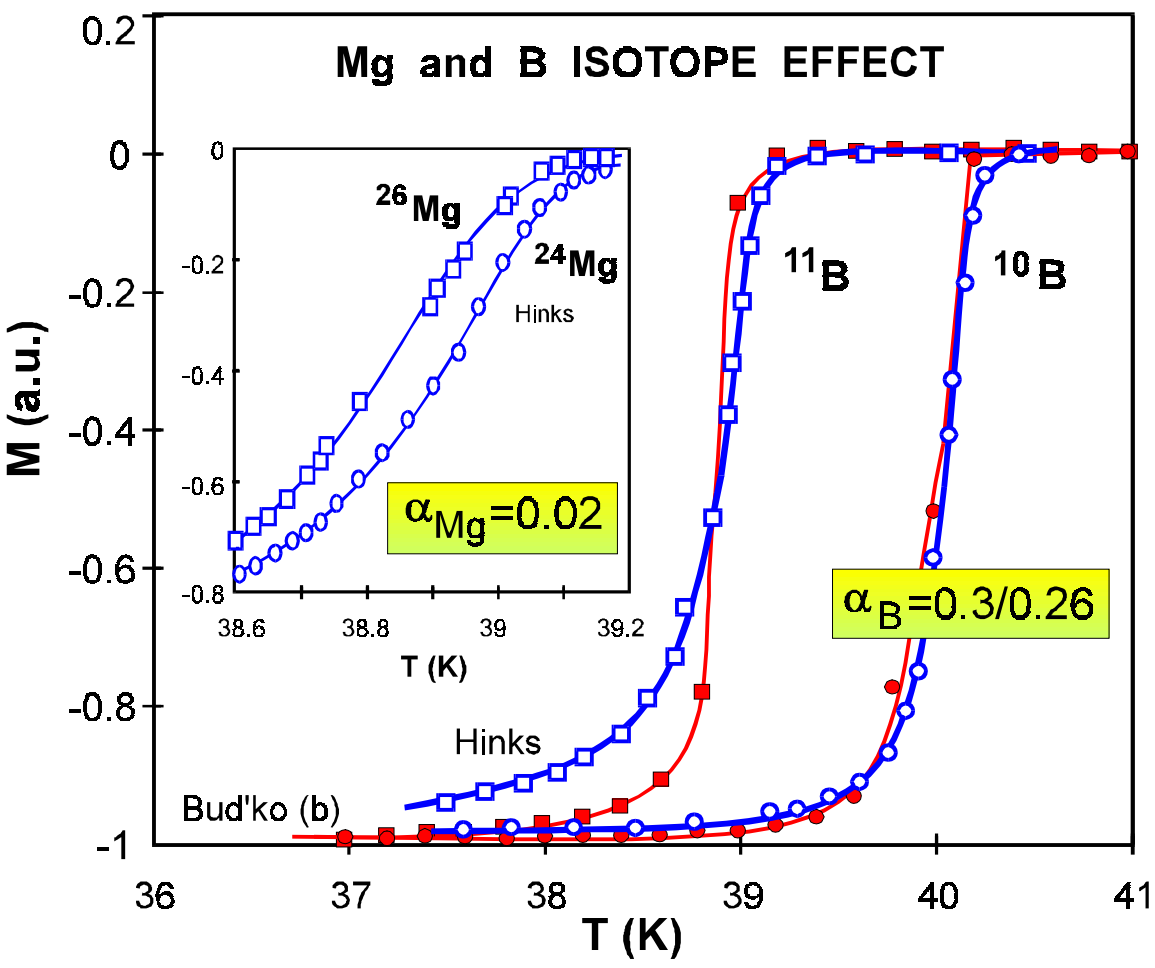

**Figure 16.** The relative magnetization versus temperature for B isotopically substituted samples. Inset shows Mg isotope effect.

## 8. Total isotope effect

In Fig. 16 is illustrated the critical temperature of $MgB_2$ at isotopic substitutions of Mg and B. The large value of the partial boron isotope exponent, $\alpha_B$, of 0.26 [Bud'ko (b)], 0.3 [Hinks] shows that phonons associated with B vibration play a significant role in $MgB_2$ superconductivity.

On the other hand, the magnesium isotope effect, $\alpha_{Mg}$, is very small, 0.02 [Hinks], as can be seen in Fig. 16 inset. This means that the vibrational frequencies of Mg have a low contribution on $T_c$. The B isotope substitution shifts $T_c$ with about 1K, while the Mg isotope substitution changes $T_c$ ten times less. Overall, the presence of an isotope effect clearly indicates a phonon coupling contribution to $T_c$. The difference between the value of the total isotope effect $\alpha_T = \alpha_B + \alpha_{Mg} \approx 0.3$ in $MgB_2$ and the 0.5 BCS-value may be related to the high $T_c$ of this material.

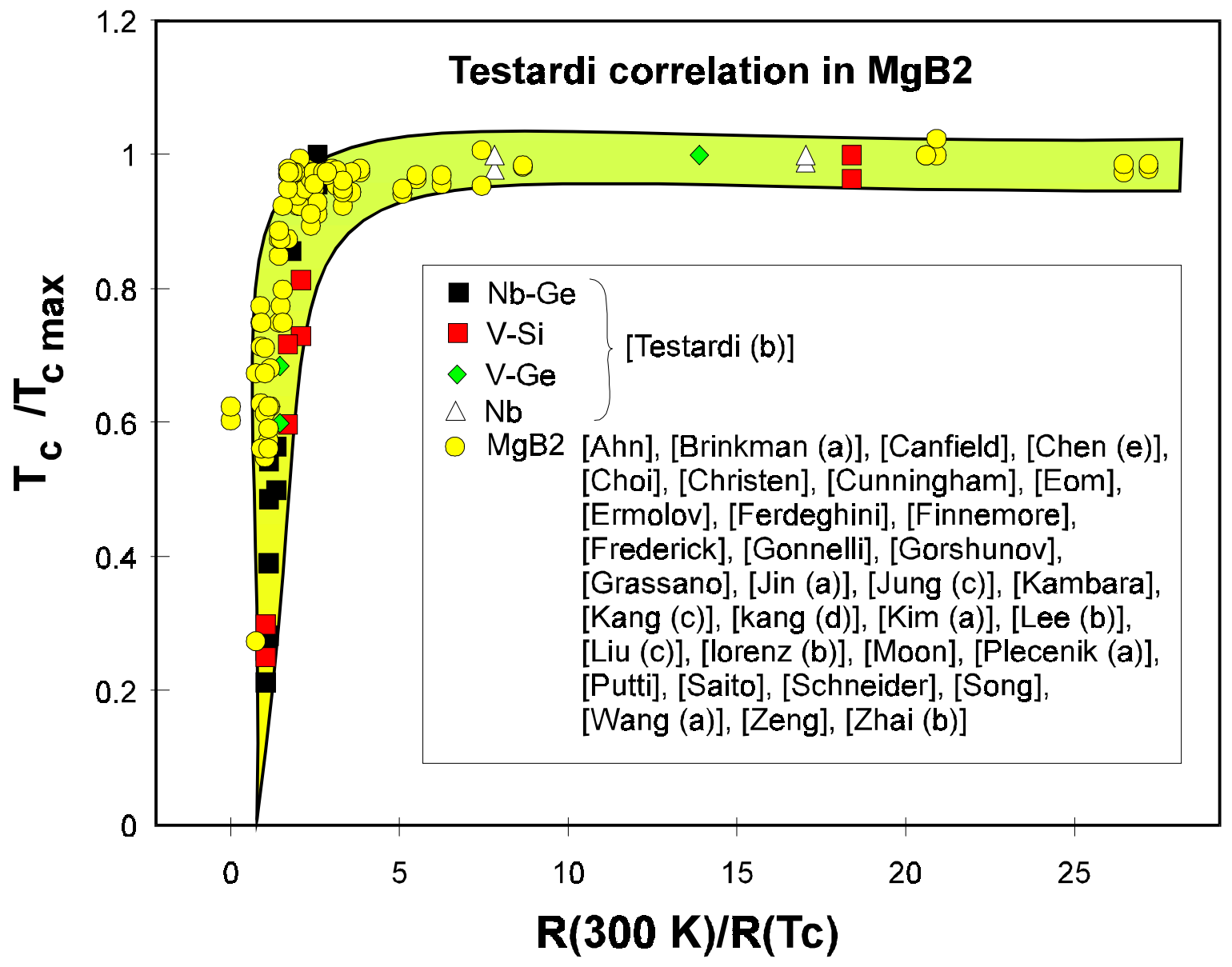

**Figure 17.** The correlation between the critical temperature of zero resistivity normalized to the onset critical temperature versus the ratio of resistance at 300K to the resistance near $T_c$.

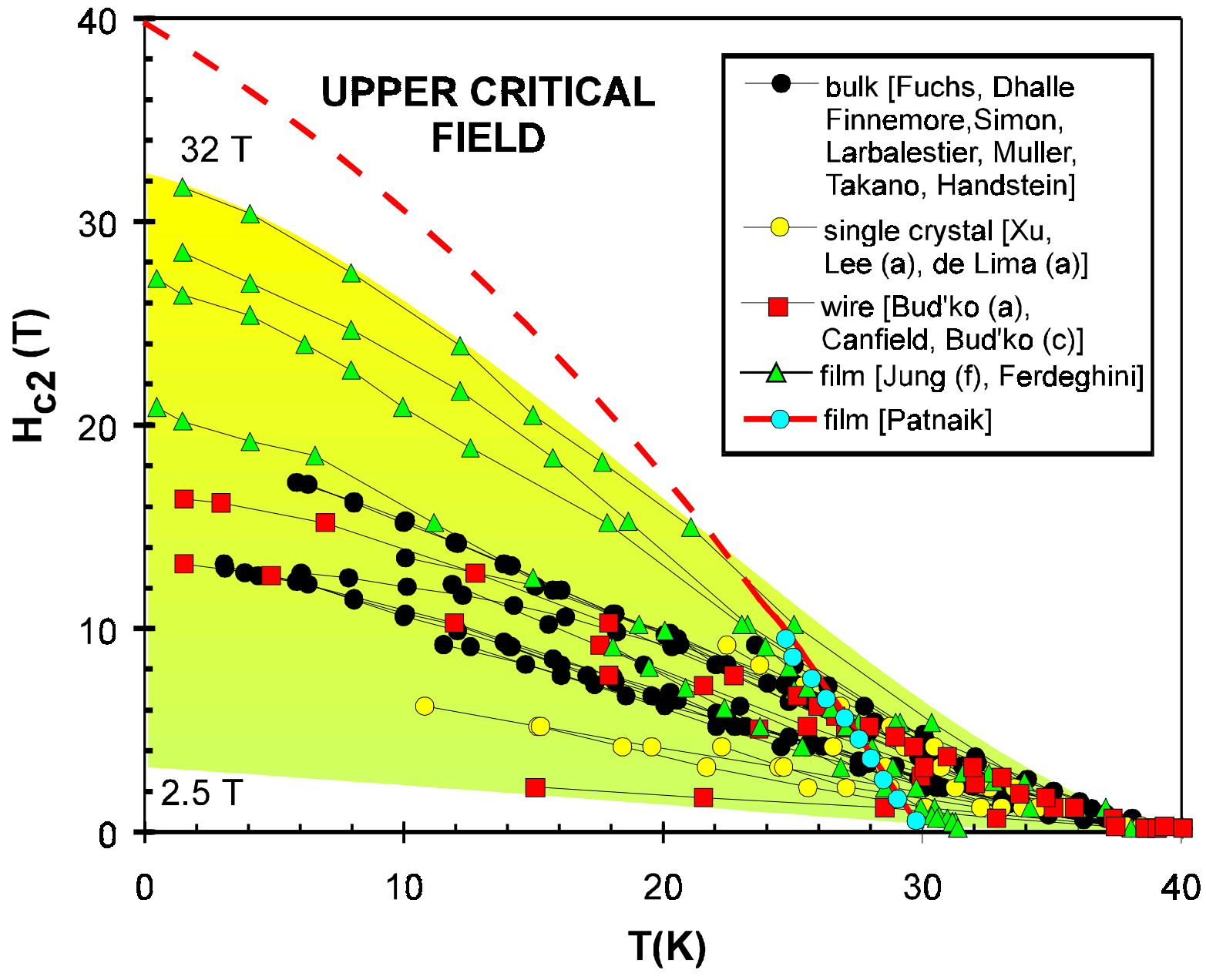

**Figure 18.** Upper critical field $H_{c2}$ versus temperature T for MgB2 in different configurations: bulk, single crystals, wires, and films.

## 9. Testardi correlation between $T_c$ and RR

One more proof in favour of a dominant phonon mechanism in $MgB_2$ superconductivity is the correlation between $T_c$ and the ratio of resistivity at room temperature and near $T_c$, RR=R(300K)/R($T_c$), also known as the Testardi correlation [Testardi (a)], [Testardi (b)], [Poate], [Park].

In 1975 Testardi showed that disorder decreases both $\lambda$ - the McMillan electron-phonon coupling constant and $\lambda_{tr}$ - in phonon-limited resistivity of normal transport phenomena, leading to the universal correlation between $T_c$ and RR [Testardi (a)]. Decreasing $T_c$, no matter how is achieved, is accompanied by the loss of thermal resistivity (electron-phonon interaction) [Testardi (b)]. The Testardi correlation translates into: samples with metallic behavior will have higher $T_c$ than samples with higher resistivity near $T_c$.

In Fig. 17 is shown the critical temperature of zero resistivity normalized to the onset critical temperature versus the ratio of resistance at 300K to the resistance near $T_c$, i.e. the Testardi correlation, for A15 compounds [Testardi (b)] and for $MgB_2$.

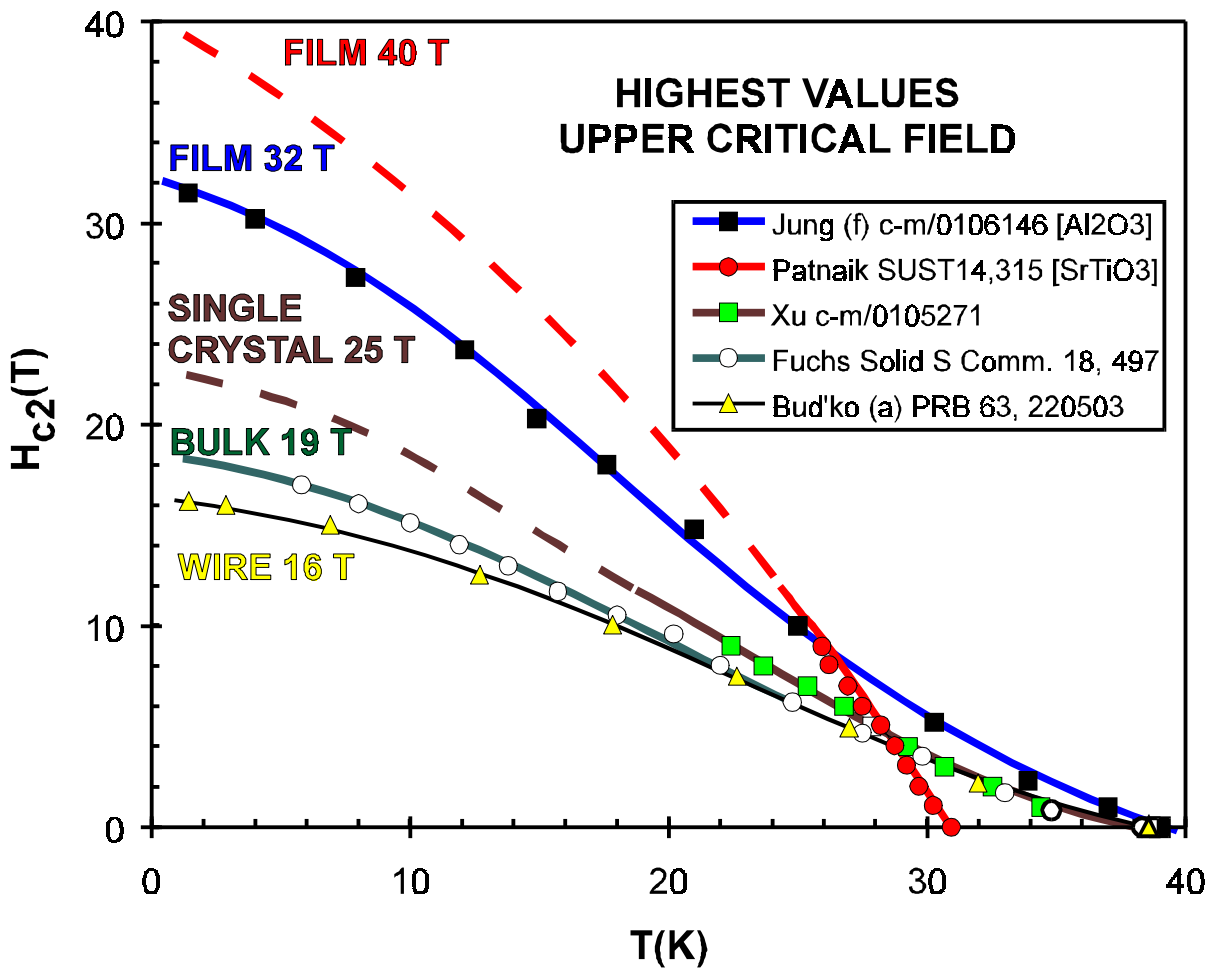

**Figure 19.** Highest values of $H_{c2}(T)$ for $MgB_2$ in different geometries (bulk, single crystals, wires, and films).

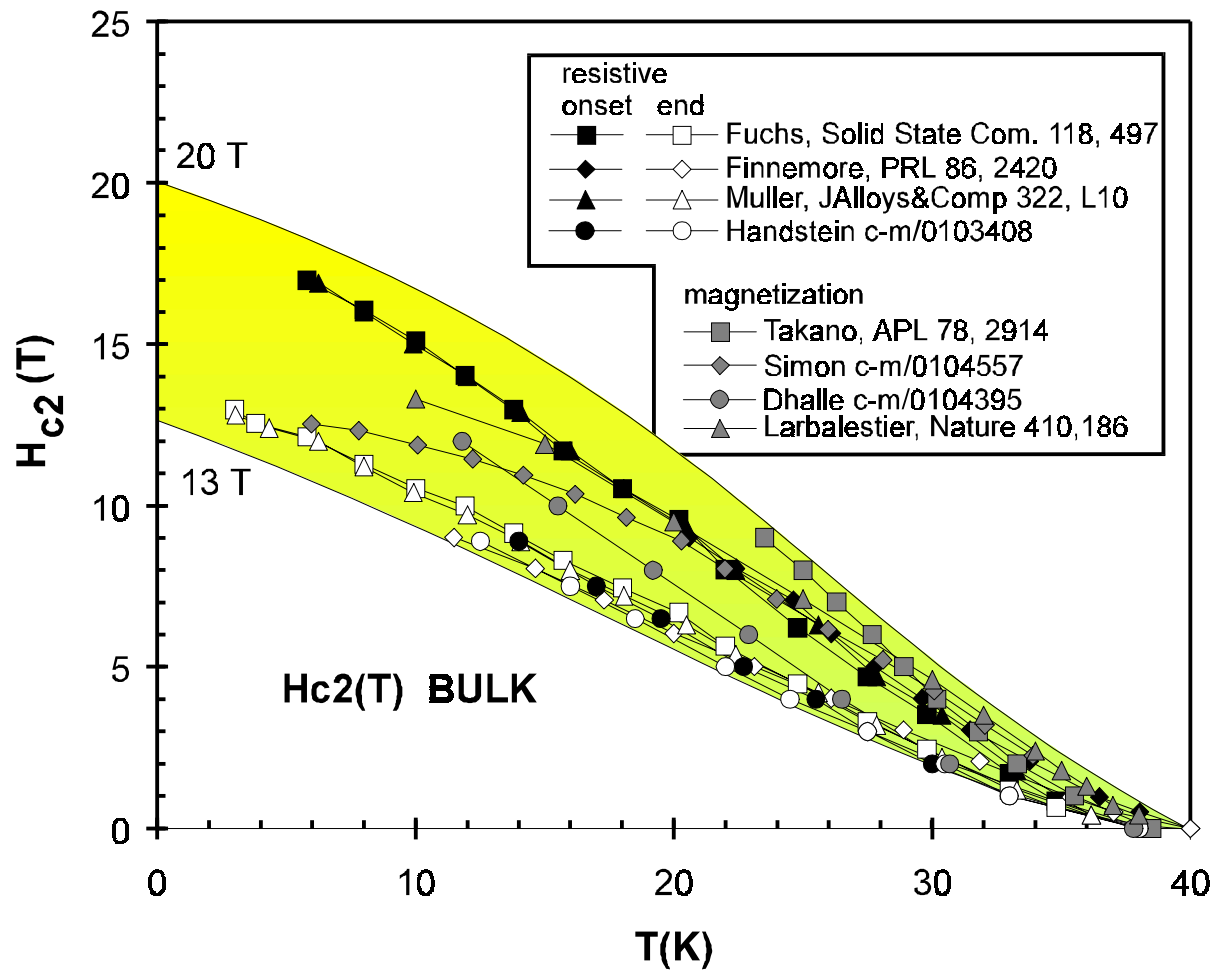

**Figure 20.** Upper critical field versus temperature for $MgB_2$ bulk.

The data for magnesium diboride were taken from [Ahn], [Brinkman (a)], [Canfield], [Chen (e)], [Choi], [Christen], [Cunningham], [Eom], [Ermolov], [Ferdeghini], [Finnemore], [Frederick], [Gonnelli], [Gorshunov], [Grassano], [Jin (a)], [Jung (c)], [Kambara], [Kang (c)], [Kang (d)], [Kim (a)], [Lee (b)], [Liu (c)], [Lorenz (b)], [Moon], [Plecenik (a)], [Putti], [Saito], [Schneider], [Song], [Wang (a)], [Zeng], and [Zhai (b)].

From Fig. 17 one notices that $MgB_2$ shows Testardi corelation between the critical temperature and resistivity ratio in normal state and near $T_c$, being one more proof in favour of a phonon-mediated mechanism in the superconductivity of this compound.

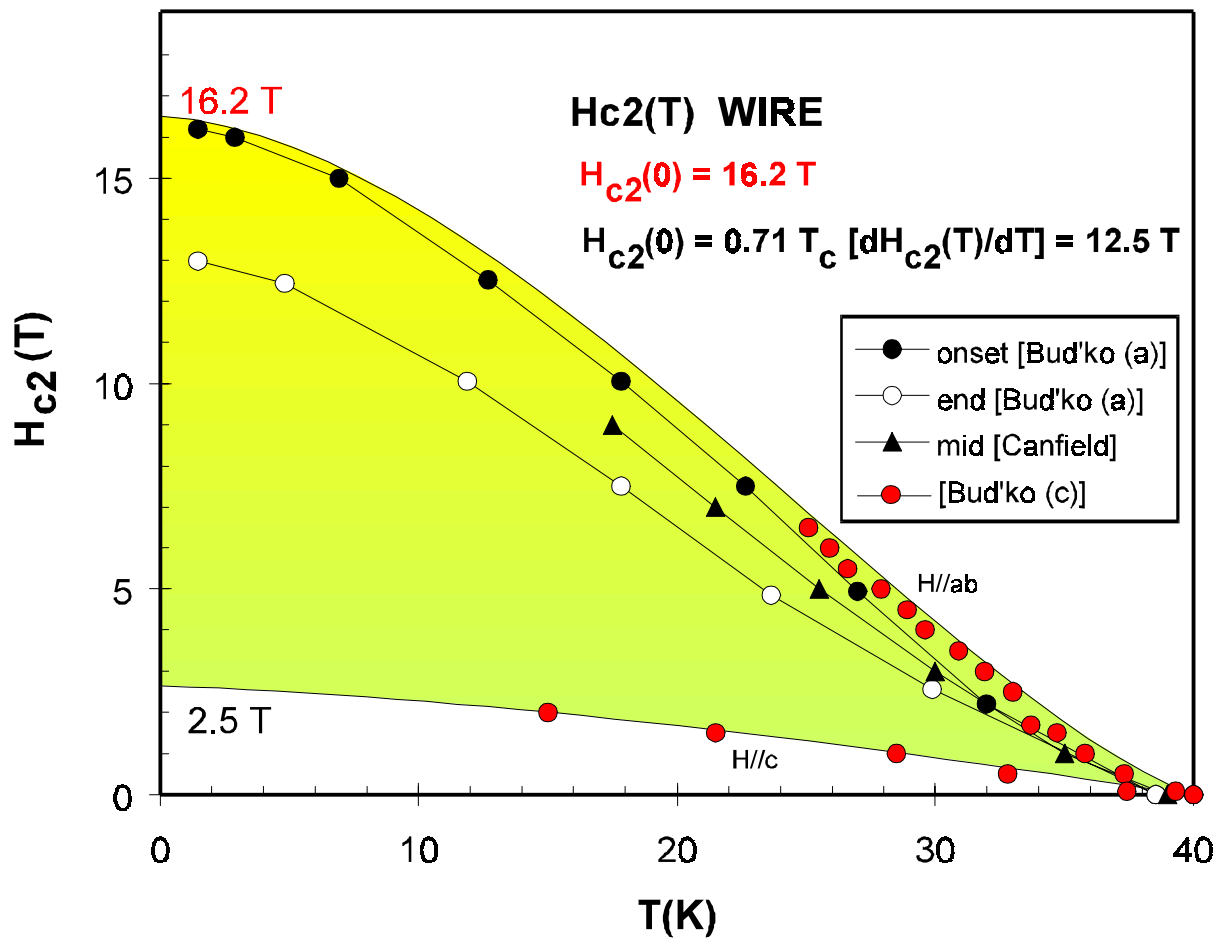

**Figure 21.** Upper critical field versus temperature for $MgB_2$ wire.

## 10. Critical fields

### *10.1. $H_{c2}(T)$ highest values*

Measurements of the upper critical field in temperature show a wide range of values for the $H_{c2}(0)$, from 2.5T up to 32T, as depicted in Fig. 18. However, even higher upper critical fields (40T) may be obtained for films with oxygen incorporated [Patnaik]. Unfortunately, due to oxygen alloying, these films have a lower $T_c$, of about 31K. Although, shortening the coherence length of $MgB_2$ (Table 1) is the basis of improving high field performances, the ability to maintain high $\xi$ is very advantageous for electronic applications. Understanding and controlling the superconducting properties of $MgB_2$ by alloying will be crucial in the future applications of this material.

In Fig. 19 are shown the curves $H_{c2}(T)$ with the highest values at low temperatures for $MgB_2$ in different configurations. The highest values of the upper critical field are achieved for films. The films with the usual critical temperature of 39 K have upper critical fields of $H_{c2}(0)$ = 32 T [Jung (f)]. However, films with lower $T_c$'s can reach higher upper critical fields up to 40 T [Patnaik]. The second best values for the upper critical fieldsw are attained by single crystals with $H_{c2}(0)$ = 25 T [Xu], followed by bulk with $H_{c2}(0)$ = 19 T [Takano], [Fuchs] and wires 16 T [Bud'ko (a)].

In Figs. 20-23 are the temperature dependencies of the upper critical field obtained for $MgB_2$ in different configurations: bulk, wires, films and single crystals, respectively.

In the plot from Fig. 21 can be easily observed that the $H_{c2}(T)$ dependence is linear in a large T-range, saturating at low temperatures.

A particular feature of $H_{c2}(T)$ curve for $MgB_2$ is the pronounced positive curvature near $T_c$, similar to the one observed in borocarbides $YNi_2B_2C$ and $LuNi_2B_2C$, considered superconductors in the clean limit [Shulga (b)].

### *10.2. $H_{c2}(T)$ anisotropy*

Anisotropy is very important both for basic understanding of this material and practical applications, strongly affecting the pinning and critical currents. The question related to the anisotropy degree of $MgB_2$ is still unresolved, reports giving values ranging between 1.1 and 9.

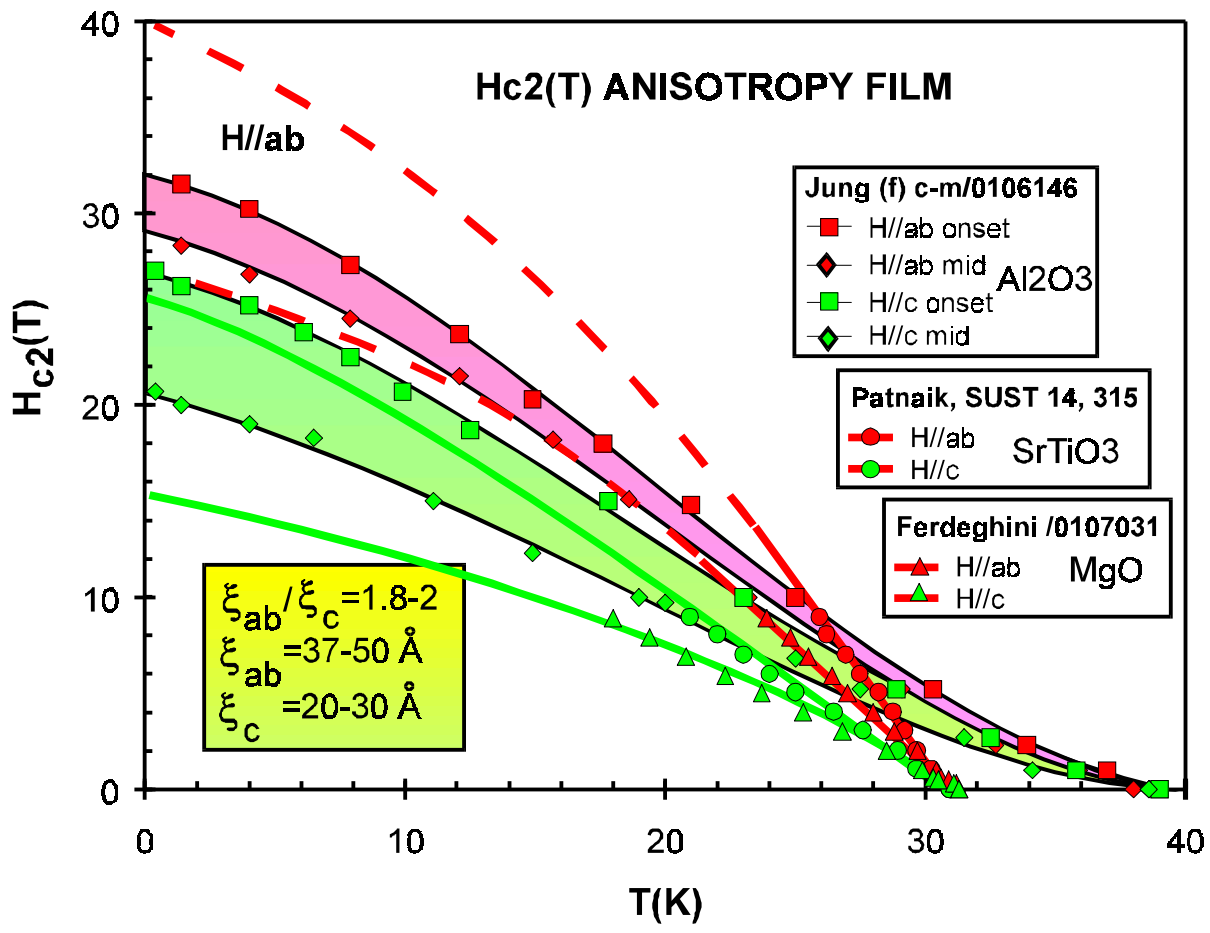

**Figure 22.** Anisotropic data of $H_{c2}(T)$ for $MgB_2$ films.

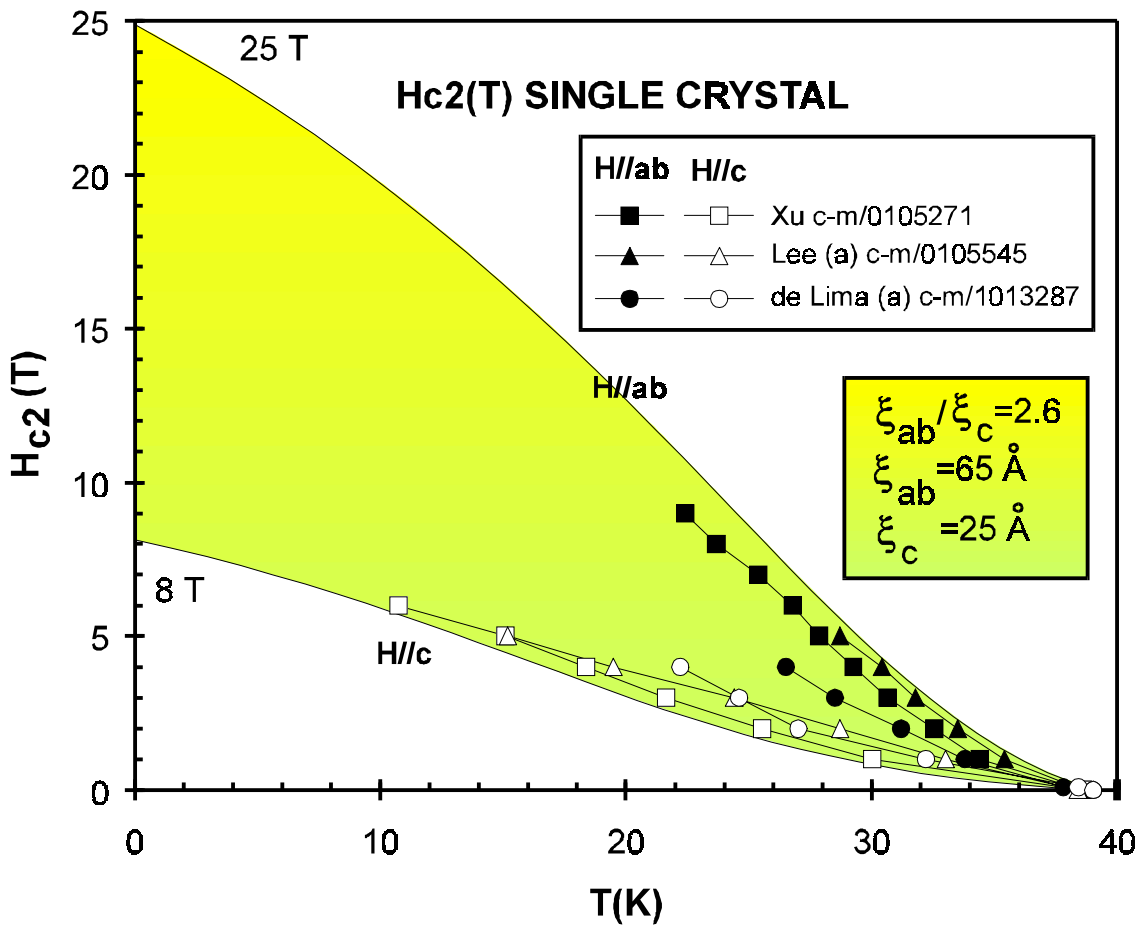

**Figure 23.** $H_{c2}(T)$ dependence for $MgB_2$ single crystals.

For textured bulk and partially oriented crystallites, the anisotropy ratio $\gamma = H_{c2}^{//ab}/H_{c2}^{//c}$ is reported to be between 1.1 and 1.7 [Handstein], [de Lima (a)], [de Lima (b)]; for c-axis oriented films 1.2 ÷ 2 [Jung (f)], [Ferdeghini], [Patnaik]; in single crystals slightly larger values than in aligned powders or films, between 1.7 ÷ 2.7 [Jung (a)], [Xu], [Lee (a)]; and finally for powders unexpectedly large values, ranging from 5 to 9 [Bud'ko (c)], [Simon].

Generally, the anisotropy of one material can be estimated on aligned powders, epitaxial films or/and single crystals. The method using aligned powders consists in mixing the superconducting powders with epoxy, followed by the alignment in magnetic fields made permanent by curing the epoxy. In order this method to give reliable results, the powders must consist of single crystalline grains with a considerable normal state magnetic anisotropy. Regularly, this method gives underestimates of anisotropy coefficient $\gamma$ due to uncertainties in the degree of alignment. The c-axis oriented films may also have a certain degree of misorientation, therefore the anisotropy coefficient will be smaller than the real value. Usually, the most reliable values are for single crystals.

Recently, Bud'ko et al. proposed a method of extracting the anisotropy parameter $\gamma = H_{c2}^{max}/H_{c2}^{min}$ from the magnetization M(H,T) of randomly oriented powders [Buk'ko (c)]. Their method is based on two features in $(\partial M/\partial T)_H$. The maxim upper critical field $H_{c2}^{max}$ is associated with the onset of diamagnetism at $T_c^{max}$ and the minimum upper critical field $H_{c2}^{min}$ is associated with a kink $\partial M/\partial T$ at lower temperatures, $T_c^{min}$. In order to prove this method is reliable, they measured the anisotropy coefficient for $LuNi_2B_2C$ and $YNBi_2B_2C$ powders, the data they obtained being in agreement with the previous values reported in the literature. For $MgB_2$ powders they obtained a very large anisotropy factor $\gamma \approx 6 \div 7$ [Bud'ko (c)].

Still, even higher $H_{c2}$ anisotropy $\gamma = 6 \div 9$ was inferred from conduction electron spin resonance measurements on high purity and high residual resistance samples [Simon].

In Fig. 24 are plotted the anisotropic upper critical fields measured for single crystals [Xu], [Lee (a)], powders and wires [Bud'ko (c)] together with the data for bulk [Fucks], [Dhalle], [Finnemore], [Simon], [Handstein], [Larbalestier], [Muller], [Takano].

One can notice that the values for bulk are situated between the anisotropic upper critical field curves for H//ab and H//c. The anisotropic upper critical field value $H_{c2}^{//ab}$ for both single crystals and powders is closed to highest values for bulk, suggesting the upper limit of $H_{c2}$ determined from anisotropy measurements may be closed to the real value for $MgB_2$.

On the other side, the upper critical field for fields parallel to c-axis $H_{c2}^{//c}$ inferred from non-aligned powders by the new method of Bud'ko et al. [Bud'ko (c)] is much lower than the lowest values obtained for bulk and single crystals, implying an underestimation of $H_{c2}^{//c}$. In order to determine with certainty the anisotropy of $MgB_2$ more experiments on larger single crystals are necessary.

## *10.3. Coherence lengths*

A comparison between the values of the coherence lengths, the anisotropy parameter $\gamma$, and upper critical field determined from experiments performed on aligned powders, thin films, single crystals and randomly aligned powders can be seen in Table 5. In order to deduce the values of the anisotropic coherence lengths from the upper critical fields we used the anisotropic Ginzburg-Landau theory equations: for the magnetic field applied along c-axis $H_{c2}^{//c} = \phi_0/2\pi\xi_{ab}^2$, and for the magnetic field applied in the ab-plane $H_{c2}^{//ab} = \phi_0/2\pi\xi_{ab}\xi_c$, where $\phi_0$ is the flux quantum, $\xi_{ab}$, $\xi_c$ are the coherence lengths along ab plane and c-axis. The previous formulae are in CGS system.

Overall, the coherence lengths values along the ab-plane range between $\xi_{ab}(0) = 3.7 \div 12.8$ nm and along c-axis between $\xi_c(0) = 1.6 \div 5.0$ nm.

Probably the most reliable data are for single crystals, with $\xi_{ab}(0) = 6.1 \div 6.5$nm and $\xi_c(0) = 2.5 \div 3.7$nm.

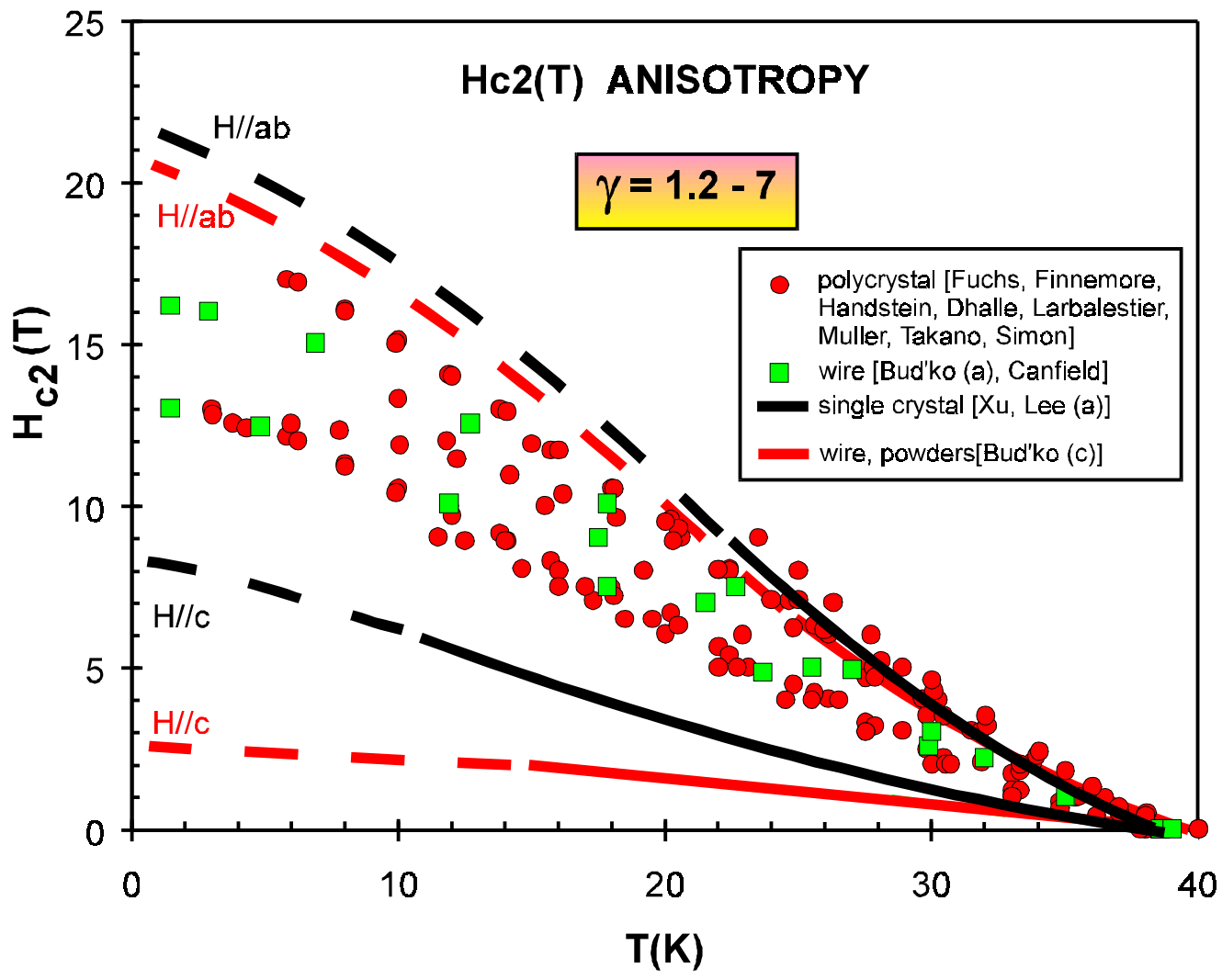

**Figure 24.** Upper critical field anisotropy versus temperature for $MgB_2$ single crystals, wire, powders. Notice that the $H_{c2}(T)$ data for $MgB_2$ bulk falls between the anisotropic dependencies of $H_{c2}(T)$ for H//c and H//ab.

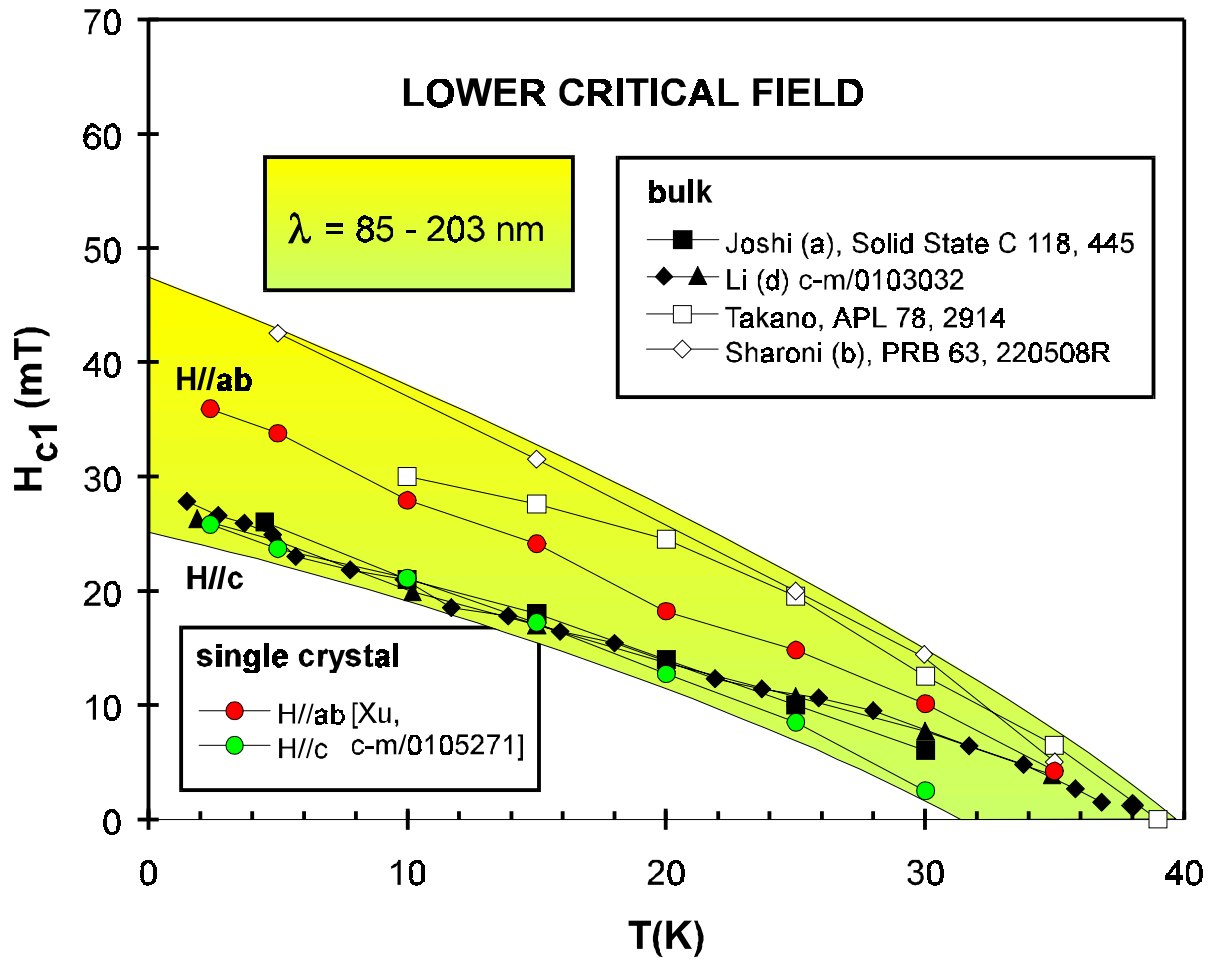

**Figure 25.** Lower critical field versus temperature.

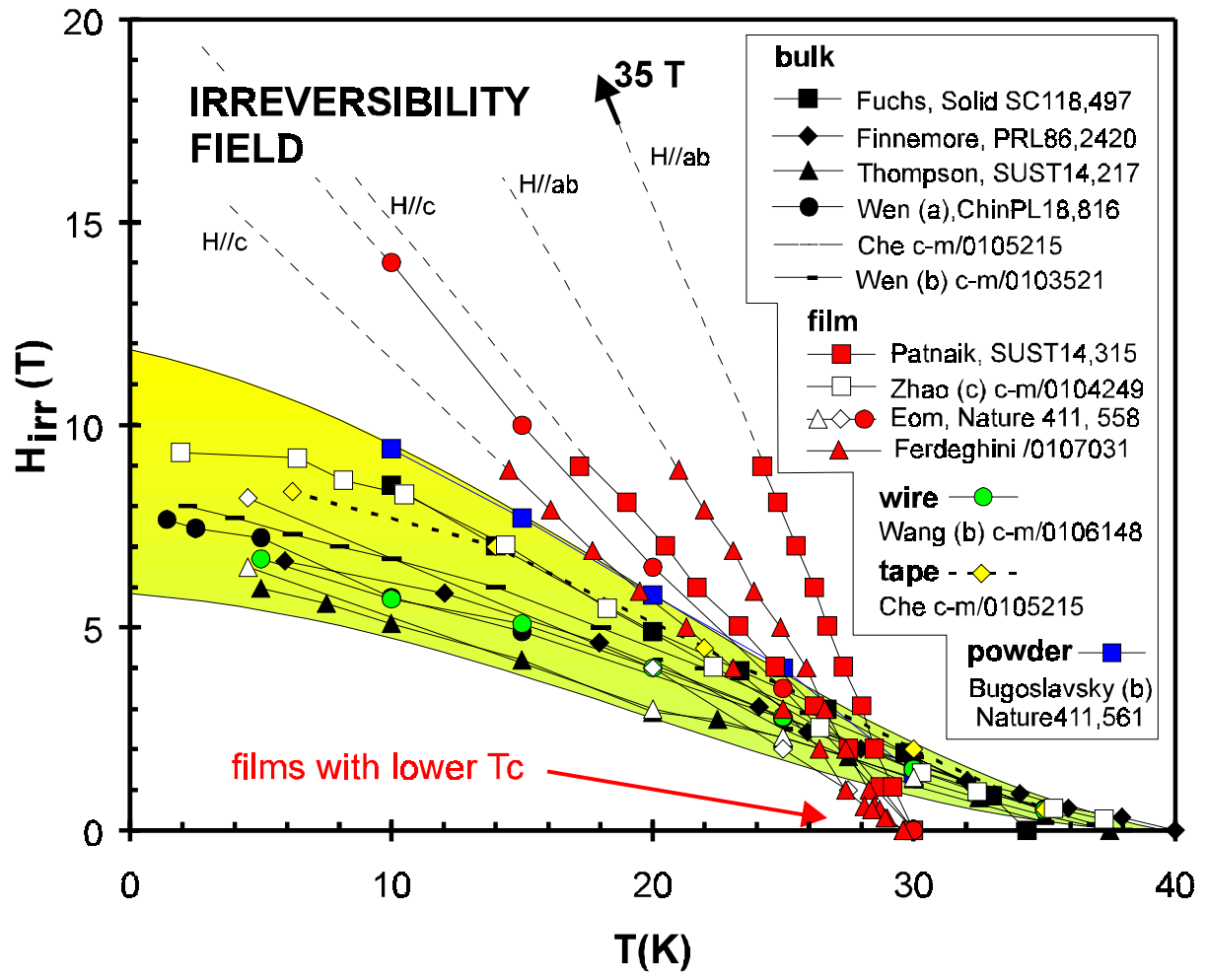

**Figure 26.** Irreversibility field versus temperature for different geometries of $MgB_2$ (bulk, film, wire and powder).

**Table 5**. Anisotropy of the upper critical field and coherence lengths inferred from experiments on aligned powders, thin films, single crystals and randomly aligned powders.

| Form | Reference | $H_{c2}^{//ab}(0)$ [T] | $H_{c2}^{//c}(0)$[T] | $\xi_{ab}$ (0)[nm] | $\xi_c(0)$ [nm] | γ |
|---|---|---|---|---|---|---|
| textured bulk | [Handstein] | 12 | 11 | 5.5 | 5.0 | 1.1 |
| aligned | [de Lima (a)] | 11 | 6.5 | 7.0 | 4.1 | 1.7 |
| crystallites | [de Lima (b)] | 12.5 | 7.8 | 6.5 | 4.0 | 1.6 |
| films | [Jung (f)] | 30 | 24 | 3.7 | 3.0 | 1.25 |
| | [Ferdeghini] | 26.4 | 14.6 | 4.7 | 2.6 | 1.8 |
| | [Patnaik] | 22.5 | 12.5 | 5.0 | 2.8 | 1.8 |
| | [Patnaik] | 24.1 | 12.7 | 5.0 | 2.6 | 1.9 |
| | [Patnaik] | 39 | 19.5 | 4.0 | 2.0 | 2 |
| single | [Jung (a)] | 14.5 | 8.6 | 6.1 | 3.7 | 1.7 |
| crystals | [Xu] | 25.5 | 9.2 | 6.5 | 2.5 | 2.6 |
| | [Lee (a)] | | | | | 2.7 |
| powders | [Bud'ko (c)] | 20 | 2.5 | 11.4 | 1.7 | 5÷8 |
| | [Simon] ESR | 16 | 2 | 12.8 | 1.6 | 6÷9 |

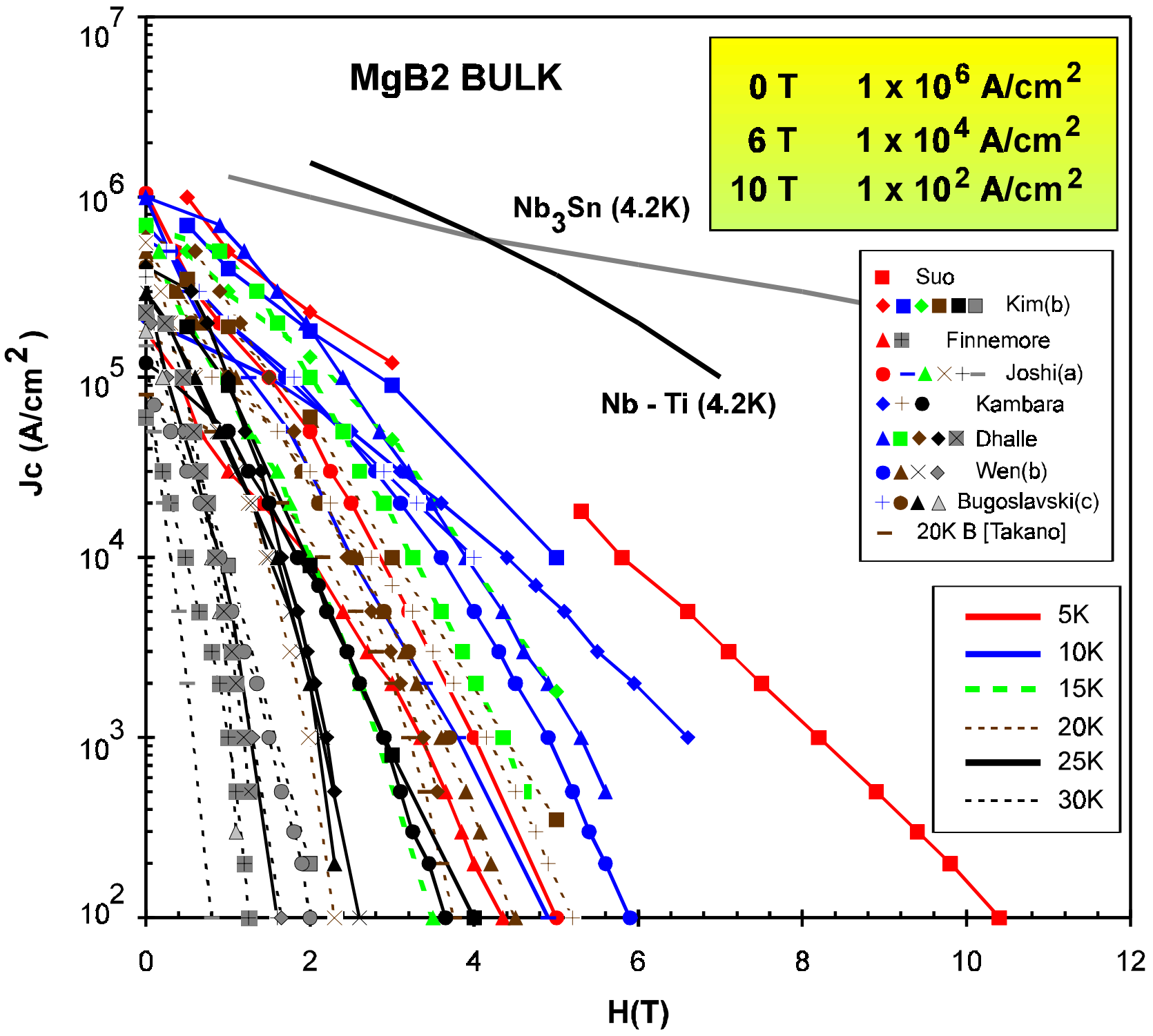

**Figure 27.** Critical current densities versus magnetic field for $MgB_2$ bulk samples [Suo], [Kim (b)], [Finnemore], [Joshi (a)], [Kambara], [Dhalle], [Wen (b)], [Bugoslavski (c)], [Takano]. The data for Nb-Ti [Heussner] and $Nb_3Sn$ [Kim (e)] at 4.2 K are shown for comparison.

### *10.4. Lower critical field $H_{c1}(T)$*

The lower critical field data versus temperature is shown in Fig. 24. Most of the values are situated between 25 and 48mT.

The data of anisotropic $H_{c1}^{//ab}$ and $H_{c1}^{//c}$ measured using single crystals [Xu]does not encompass the values for bulk [Joshi (a)], [Li (d)], [Takano], [Sharoni (b)], suggesting that the data for single crystal is not accurate.

The values of the penetration depth deduced from the lower critical field data range between 85 and 203 nm.

### *10.5. Irreversibility field $H_{irr}(T)$*

The knowledge of the irreversibility line is important in potential applications, as non-zero critical currents are confined to magnetic fields below this line.

The irreversibility fields extrapolated at zero temperature range between 6 and 12 T for $MgB_2$ bulk, films, wires, tapes and powders, as illustrated in Fig. 26. A substantial enhancement of the irreversibility line accompanied by a significantly large $J_c$ between $10^6$ to $10^7$ A/cm$^2$ at 4.2 K and 1 T and have been reported in $MgB_2$ thin films with lower $T_c$ [Patnaik], [Eom], [Ferdeghini]. These results give further encouragement to the development of $MgB_2$ for high current applications.

## 11. Critical current density versus applied magnetic field $J_c(H)$

### *11.1. $J_c(H)$ in bulk*

Many groups have measured the critical current density and its temperature and magnetic field dependence for different geometrical configurations of $MgB_2$: powders [Bugoslavsky (b)], [Dhalle], - bulk [Bugoslavsky (c)], [Dhalle], [Finnemore], [Frederick], [Joshi (a)], [Kambara], [Takano], [Wen (b)], - films [Eom], [Johansen], [Kim (a)], [Kim (b)], [Li (a)], [Moon], [Paranthaman (a)], - tapes [Che], [Grasso], [Kumakura], [Soltanian], [Song], [Sumption], - wires [Canfield], [Glowacki], [Goldacker], [Jin (b)], [Wang (b)],.

The consensus which seems to emerge is that, unlike in HTSC, $J_c(T,H)$ in $MgB_2$ is determined by its pinning properties and not by weak link effects. These pinning properties are strongly field dependent, becoming rather poor in modest magnetic fields. The inductive measurements indicate that in dense bulk samples, the microscopic current density is practically identical to the intra-granular $J_c$ measured in dispersed powders [Dhalle], therefore the current is not limited by grain boundaries [Kawano].

In Fig. 27 are shown data of critical current versus applied magnetic field, $J_c(H)$, for bulk $MgB_2$ samples, taken at different temperatures, 5, 10, 15, 20, 25 and 30K. We have to mention that most of the $J_c$ data we will present in the followings were inferred from magnetization measurements.

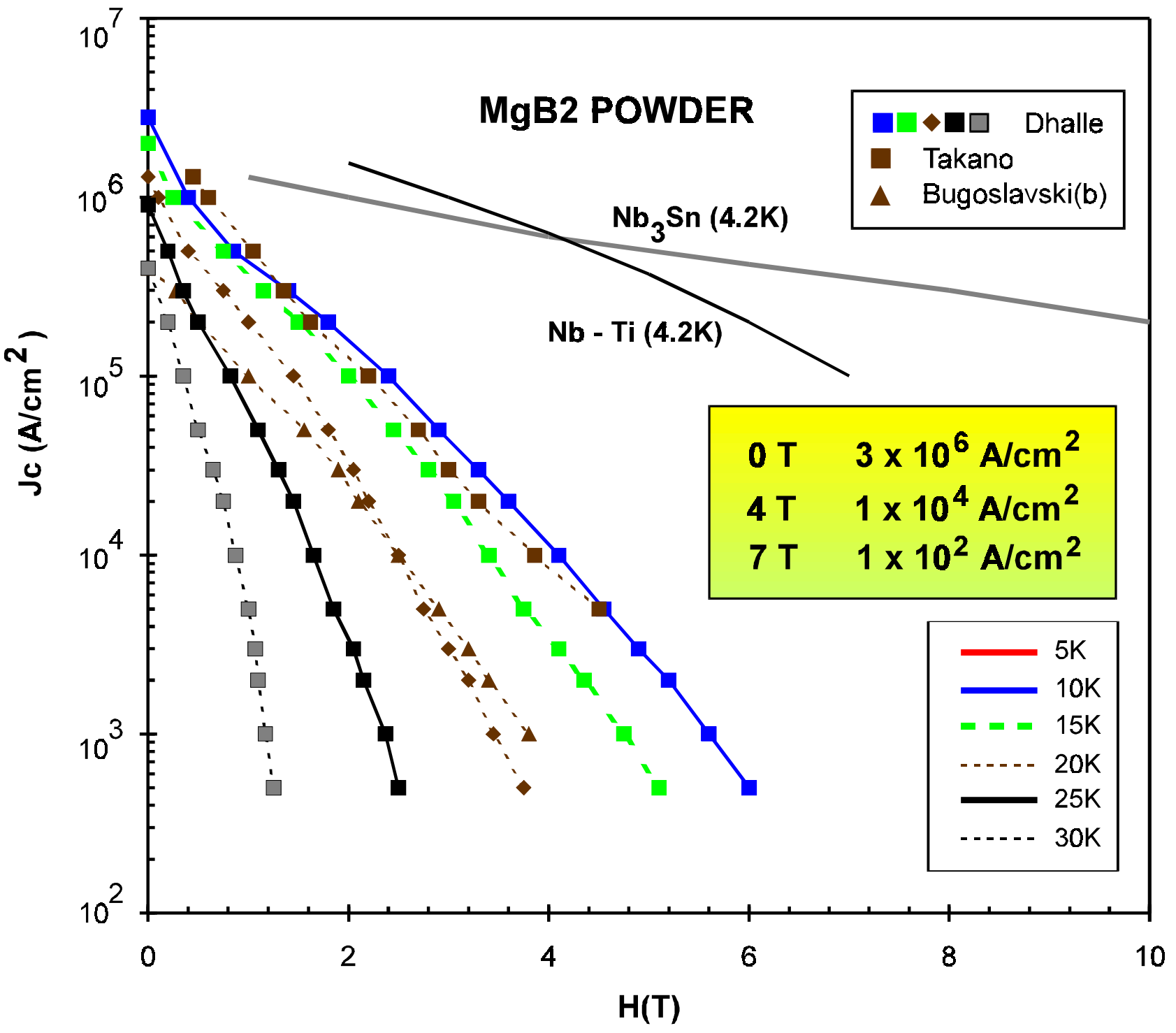

**Figure 28.** Critical current densities versus magnetic field for $MgB_2$ powders [Dhalle], [Takano], [Bugoslavski (b)]. The data for Nb-Ti [Heussner] and $Nb_3Sn$ [Kim (e)] at 4.2 K are shown for comparison.

For comparison are shown the $J_c(H)$ data for Nb-Ti [Heussner] and $Nb_3Sn$ [Kim (e)] at 4.2 K. In self fields bulk $MgB_2$ achieve moderate values of critical current density, up to $10^6$ A/cm$^2$. In applied magnetic fields of 6 T $J_c$ maintains above $10^4$ A/cm$^2$, while in 10 T $J_c$ is about $10^2$ A/cm$^2$.

## *11.2. $J_c(H)$ in powders*

Fig. 28 illustrates the critical current density versus field for $MgB_2$ powders [Dhalle], [Takano], [Bugoslavski (b)]. Very high current densities can be achieved in low fields, of up to $3\times10^6$ A/cm$^2$. However, magnetic fields of 7 T quenches the current density to low values - $10^2$ A/cm$^2$, $J_c(H)$ having a steeper dependence in field than bulk $MgB_2$.

## *11.3. $J_c(H)$ in wires and tapes*

In Figs. 29 and 30 are shown the critical current density dependence in magnetic field for $MgB_2$ wires and tapes, respectively. The data for are taken from references [Canfield], [Che], [Goldacker], [Glowacki], [Jin (b)], [Kumakura], [Soltanian], [Song], [Suo], [Wang (b)].

Compared to bulk and powders $MgB_2$, the wires and tapes have lower values of $J_c$ in low fields, of about $6\times10^5$ A/cm$^2$. However, the $J_c(H)$ dependence becomes more gradual in field, allowing larger current density values in higher fields, $J_c(5T) > 10^5$ A/cm$^2$. Due to geometrical shielding properties, the tapes can achieve superior currents in relatively high magnetic fields than the wires.

Suo et al. found that annealing of the tapes increases core density and sharpened the superconducting transition, raising $J_c$ by more than a factor of 10 [Suo].

Wang et al. studied the effect of sintering time on the critical current density of $MgB_2$ wires [Wang (b)]. They found that there is no need for prolonged heat treatment in the fabrication of Fe-clad wires. Several minutes sintering gives the same performances as longer sintering time. Therefore, these findings substantially simplify the fabrication process and reduce the cost for large-scale production of $MgB_2$ wires.

Jin et al. showed that alloying $MgB_2$ with Ti, Ag, Cu, Mo, Y, has an important effect upon $J_c$, despite the fact that $T_c$ remains unaffected or slightly reduced by these elements [Jin (b)]. Iron addition seems to be least damaging, whereas Cu addition causes $J_c$ to be significantly reduced by 2-3 orders of magnitude.

Iron is also beneficial as metal-cladd, as it shields the core from external fields, the shielding being less effective for fields parallel to the tape plane [Soltanian]. When there is no external field, the transport current will generate a self-field surrounding the tape. Because Fe is ferromagnetic, the flux lines will suck into the Fe sheath, particularly at the edges of the tape. Therefore, the sheath will reduce the effect of self-field on $I_c$. When external fields are applied, the Fe sheath acts as a shield, reducing the effect of external field. Therefore, using Fe-clad tapes may be beneficial for power transmission lines.

In order to increase $J_c$ in wires and tapes, the fabrication process must be optimised by using finer starting powders or by incorporating nanoscale chemically inert particles that would inhibit the grain growth.

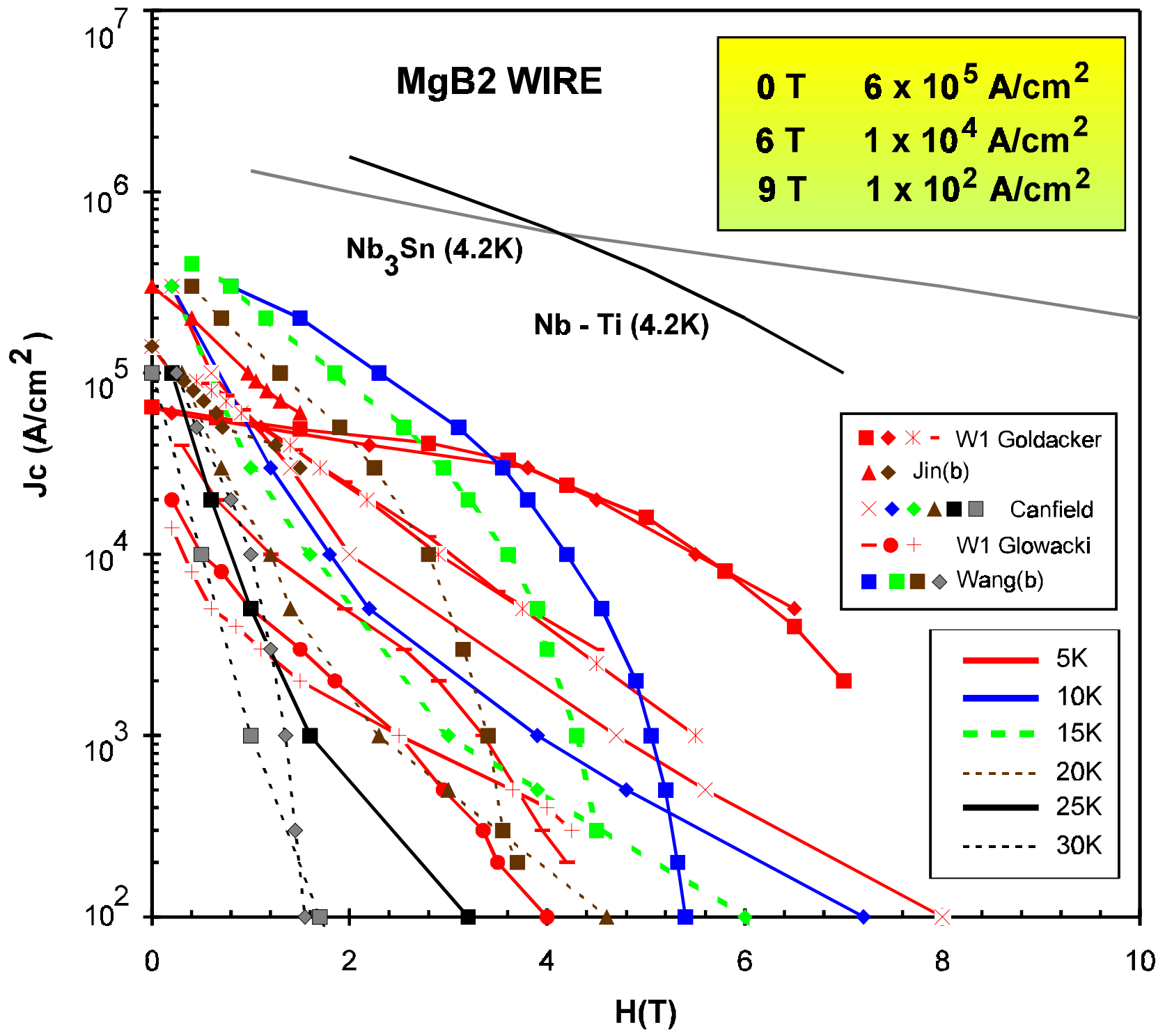

**Figure 29.** Critical current densities versus magnetic field for $MgB_2$ wires [Goldacker], [Jin (b)], [Canfield], [Glolwacki], [Wang (b)]. The data for Nb-Ti [Heussner] and $Nb_3Sn$ [Kim (e)] at 4.2 K are shown for comparison.

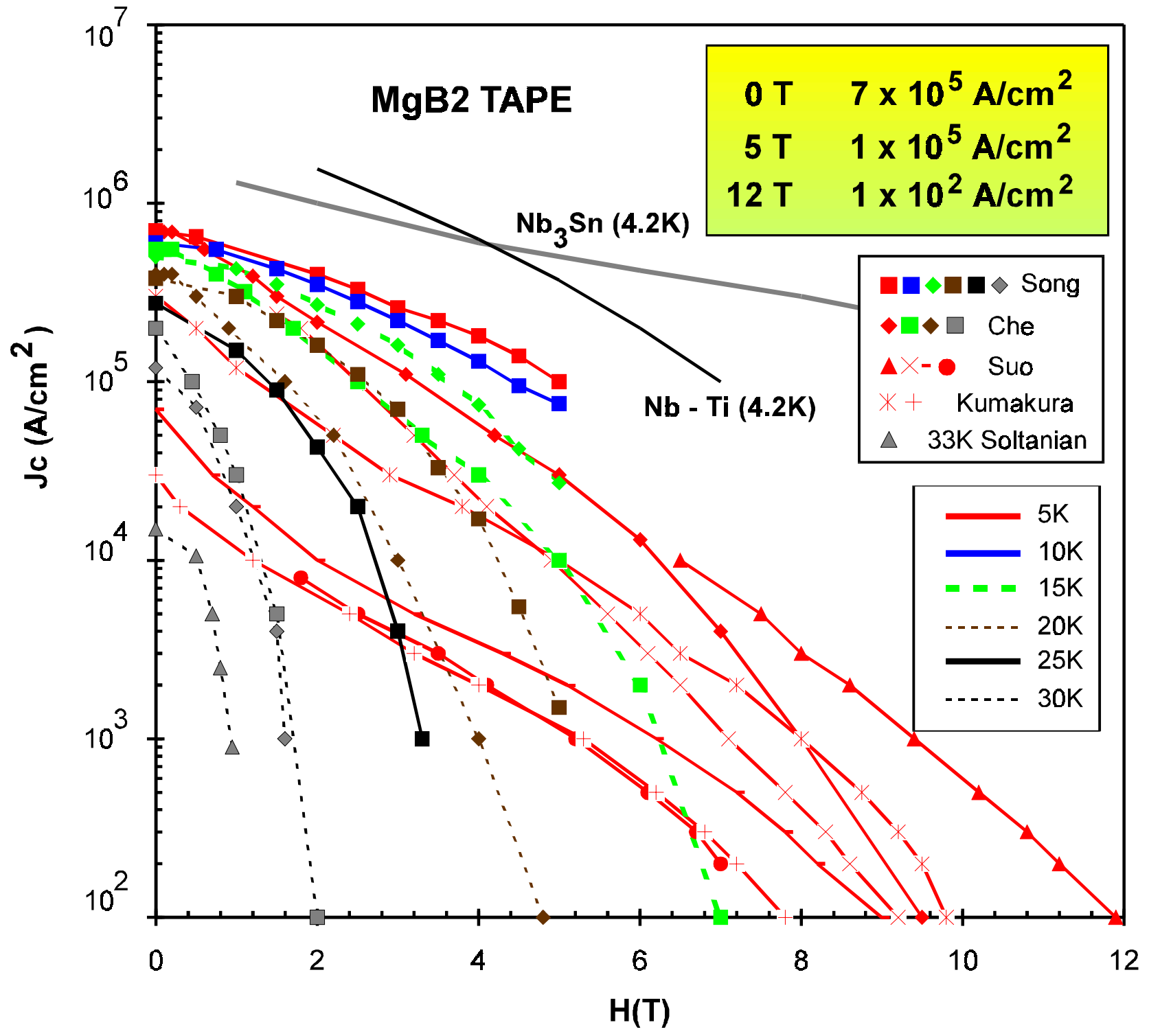

**Figure 30.** Critical current densities versus magnetic field for $MgB_2$ tapes [Song], [Che], [Suo], [Kumakura], [Soltanian]. The data for Nb-Ti [Heussner] and $Nb_3Sn$ [Kim (e)] at 4.2 K are shown for comparison.

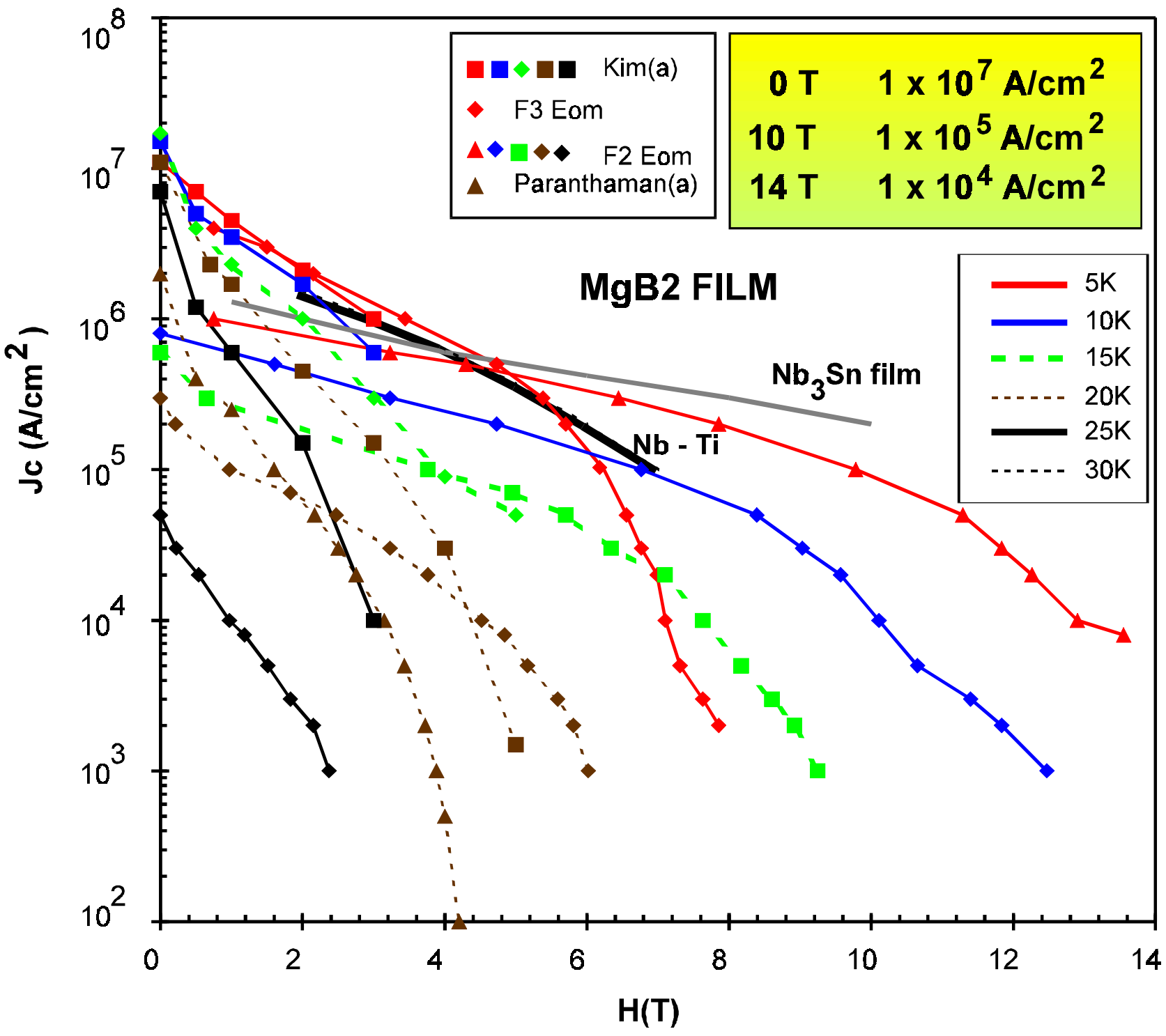

**Figure 31.** Critical current densities versus magnetic field for $MgB_2$ films [Kim (a)], [Eom], [Paranthaman (a)]. The data for Nb-Ti [Heussner] and $Nb_3Sn$ [Kim (e)] at 4.2 K are shown for comparison.

### *11.4. $J_c(H)$ in thin films*

In Fig. 31 are shown the values of critical current density versus magnetic field in $MgB_2$ films [Kim (a)], [Eom], [Paranthaman (a)]. To our great surprise, the data for thin films have given us the proof that the performances of $MgB_2$ can rival and perhaps eventually exceed that of existing superconducting wires. One can see in Fig. 31 that in low fields, the current density in $MgB_2$ is higher [kim (a)], [Eom] than the current in $Nb_3Sn$ films [Kim (e)] and Nb-Ti [Heussner]. In larger magnetic fields Jc in $MgB_2$ decreases faster than for Nb-Sn and Nb-Ti superconductors. However, a Jc of $10^4$ A/cm$^2$ can be attained in 14T for films with oxygen and MgO incorporated [Eom].

These high current densities, exceeding 1 MA/cm$^2$, measured in films [Kim (a)], [Eom], demonstrate the potential for further improving the current carrying capabilities of wires and tapes.

### *11.5. Highest $J_c(H)$ at different temperatures*

As we can see in Figs. 27-31, $MgB_2$ has a great potential for high-current and high-field applications, as well as microelectronics. Josephson junctions may be much easier to fabricate that those made from HTSC, having the performances of conventional superconductors (Nb, NbN), but operate at much higher temperatures.

In particular, as illustrated in Fig. 32, $MgB_2$ has similar performances regarding critical current density in low temperatures with best existing superconductors.

Up to date several authors succeeded in improving $J_c$ of $MgB_2$ by: oxygen alloying [Eom], proton irradiation [Bugoslavsky (b)], while other studied the influence of doping [Jin (b)] or sample preparation [Dhalee] on $J_c$.

To take advantage of the relatively high $T_c$ of 39 K of $MgB_2$, it is important to have high $J_c$ values at temperatures above 20 K. The boiling point of H at atmospheric pressure is 20.13 K, so that is possible to use liquid hydrogen as cryogen for cooling $MgB_2$. In Fig. 33 are shown the best values of Jc(H) for temperatures of 25 K and 30 K, respectively. For applications at above 20 K it is necessary to improve the flux-pinning properties through structural and microstructural modifications. For example, chemical doping, introduction of precipitates, atomic-scale control of defects such as vacancies, dislocations, grain boundaries.

### *11.6. Absence of weak links*

Many magnetization and transport measurements show that $MgB_2$ does not exhibit weak-link electromagnetic behavior at grain boundaries [Larbalestier] or fast flux creep [Thompson], phenomena which limit the performances of high-$T_c$ superconducting cuprates.

As stated previously, high critical current densities have been observed in bulk samples, regardless of the degree of grain alignment [Kim (b)], [Suo]. This would be an advantage for making wires or tapes with no degradation of $J_c$, in contrast to the degradation due to grain boundary induced weak-links which is a common and serious problem in cuprate high temperature superconductors.

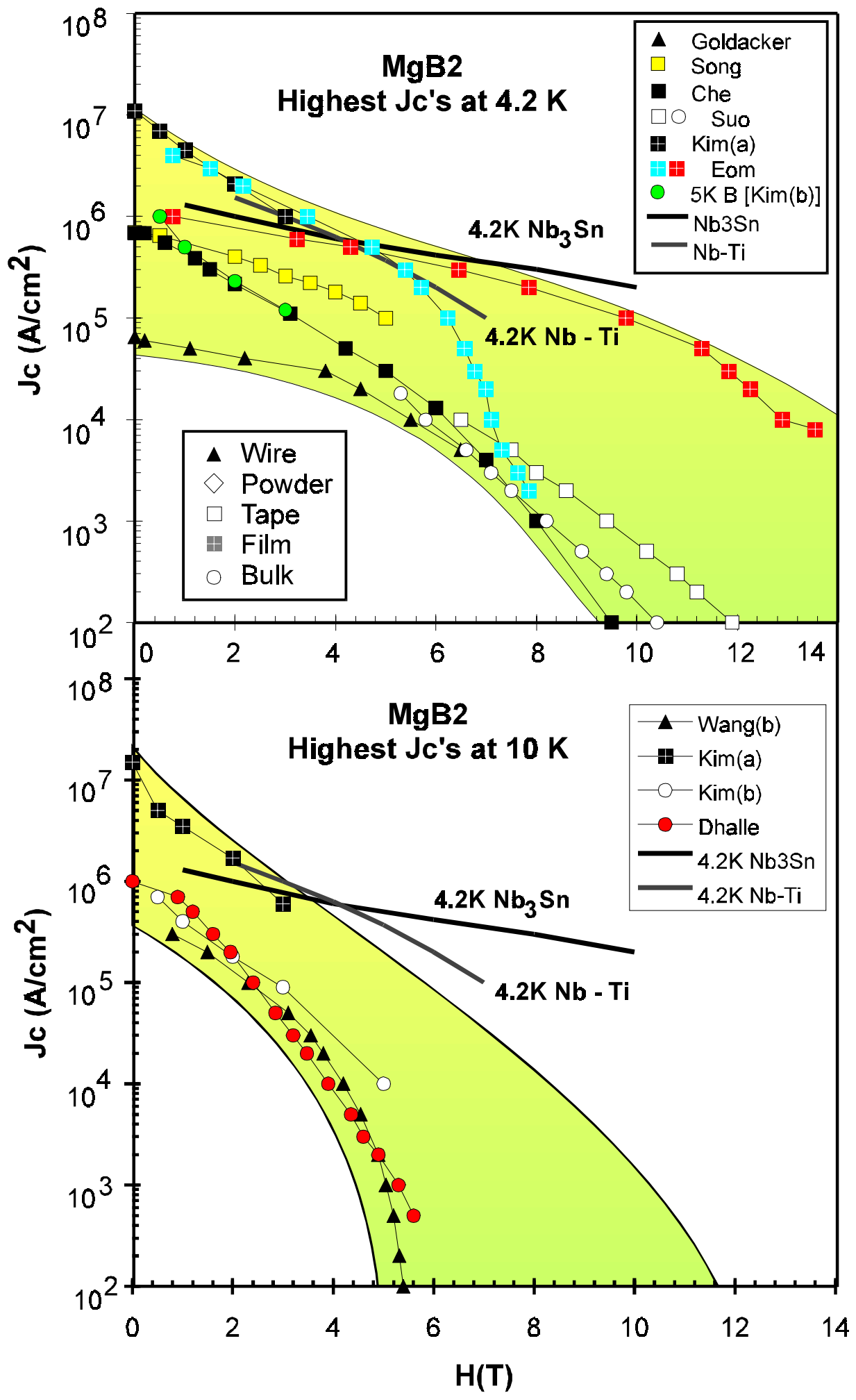

**Figure 32.** Highest critical current densities versus magnetic field for $MgB_2$ at 4.2K and 10 K. Data at 4.2 K are taken from [Goldacker], [Song], [Che], [Suo], [Kim (a)], [Eom], [Kim (b)]; data at 10K are taken from [Wang (b)], [Kim (a)], [Kim (b)], [Dhalle]. The data for Nb-Ti [Heussner] and $Nb_3Sn$ [Kim (e)] at 4.2 K are shown for comparison.

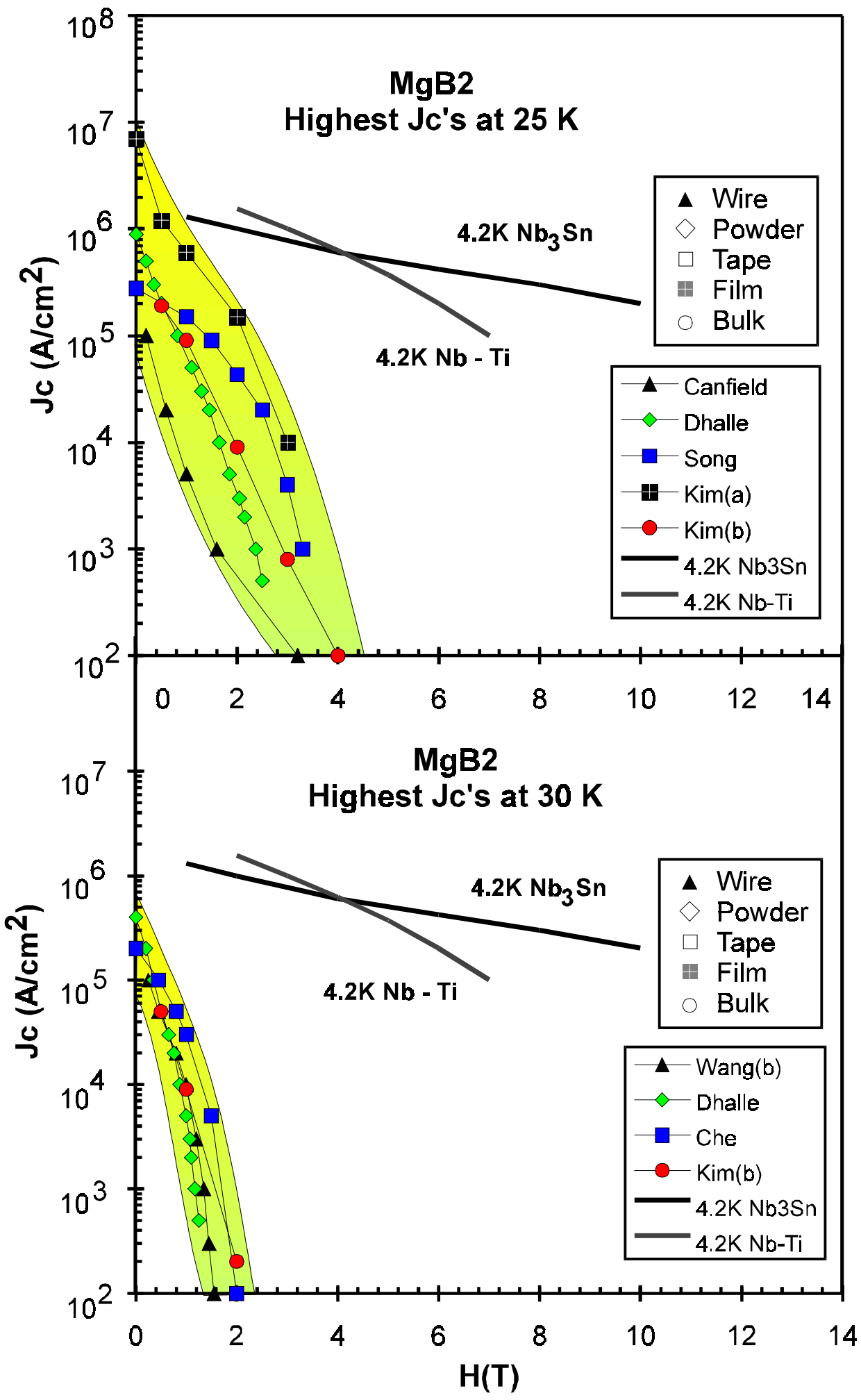

**Figure 33.** Highest critical current densities versus magnetic field for $MgB_2$ at 25K and 30K. Data at 25K are taken from [Canfield], [Dhalle], [Song], [Kim (a)], [Kim (b)]; data at 30K are taken from [Wang (b)], [Dhalle], [Che], [Kim (b)]. The data for Nb-Ti [Heussner] and $Nb_3Sn$ [Kim (e)] at 4.2 K are shown for comparison.

In Fig. 34 is illustrated the absence of weak links in $MgB_2$. The transport measurements in high magnetic fields of dense bulk samples yields very similar $J_c$ values as the inductive measurements [Dhalle], [Kim (b)]. This confirms that the inductive current flows coherently throughout the sample, unaffected by grain boundaries. Therefore the flux motion will determine $J_c$ dependence in field and temperature.

Jin et al. [Jin (b)] found that some materials used as tubes or sheaths in the PIT method dramatically reduce the critical current of $MgB_2$. Although magnesium diboride itself does not show the weak-link effect, contamination does result in weak-link -like behaviour.

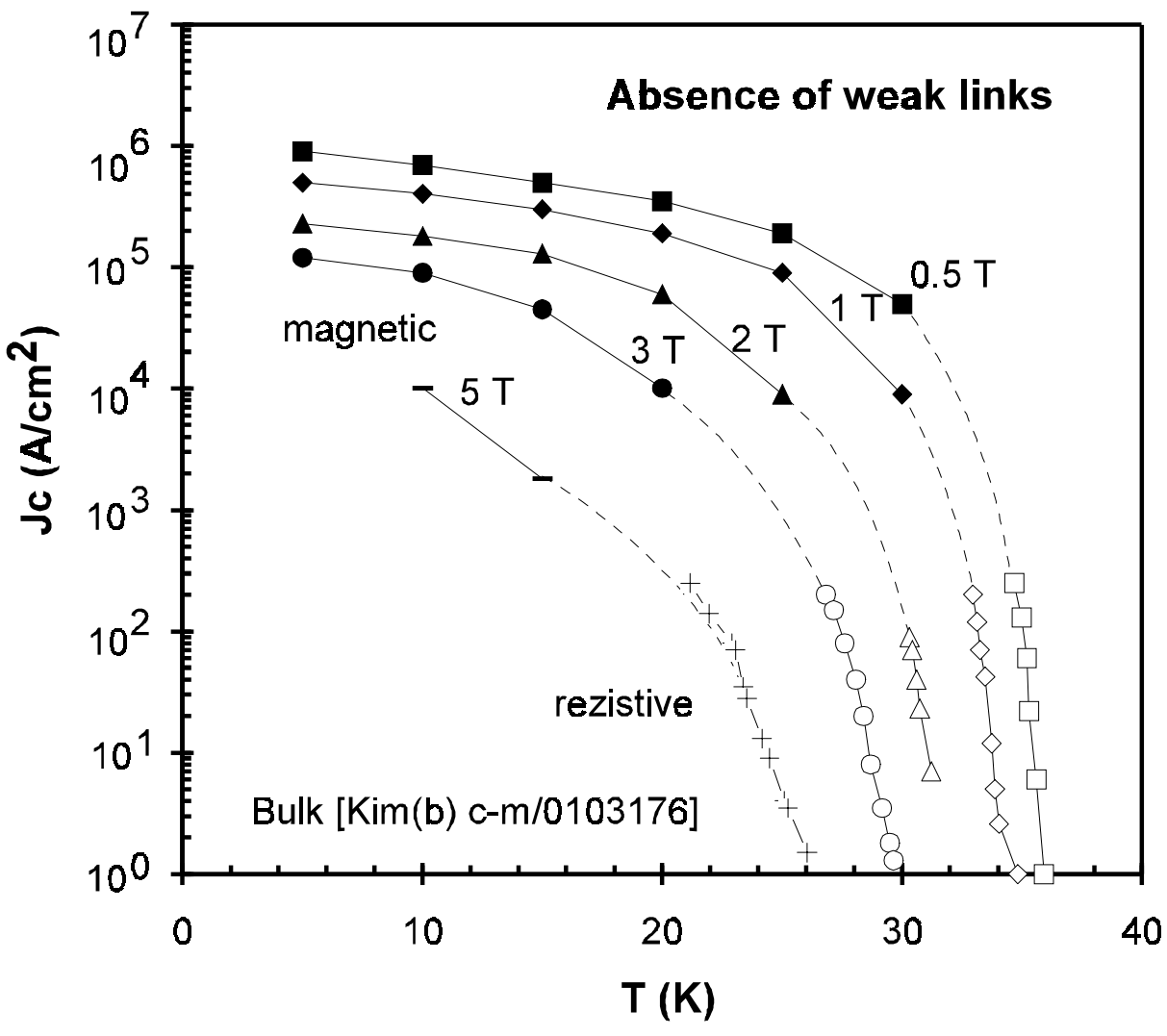

**Figure 34.** Critical current density dependence in magnetic field. data taken from resistive and magnetic measurements [Kim (b)].

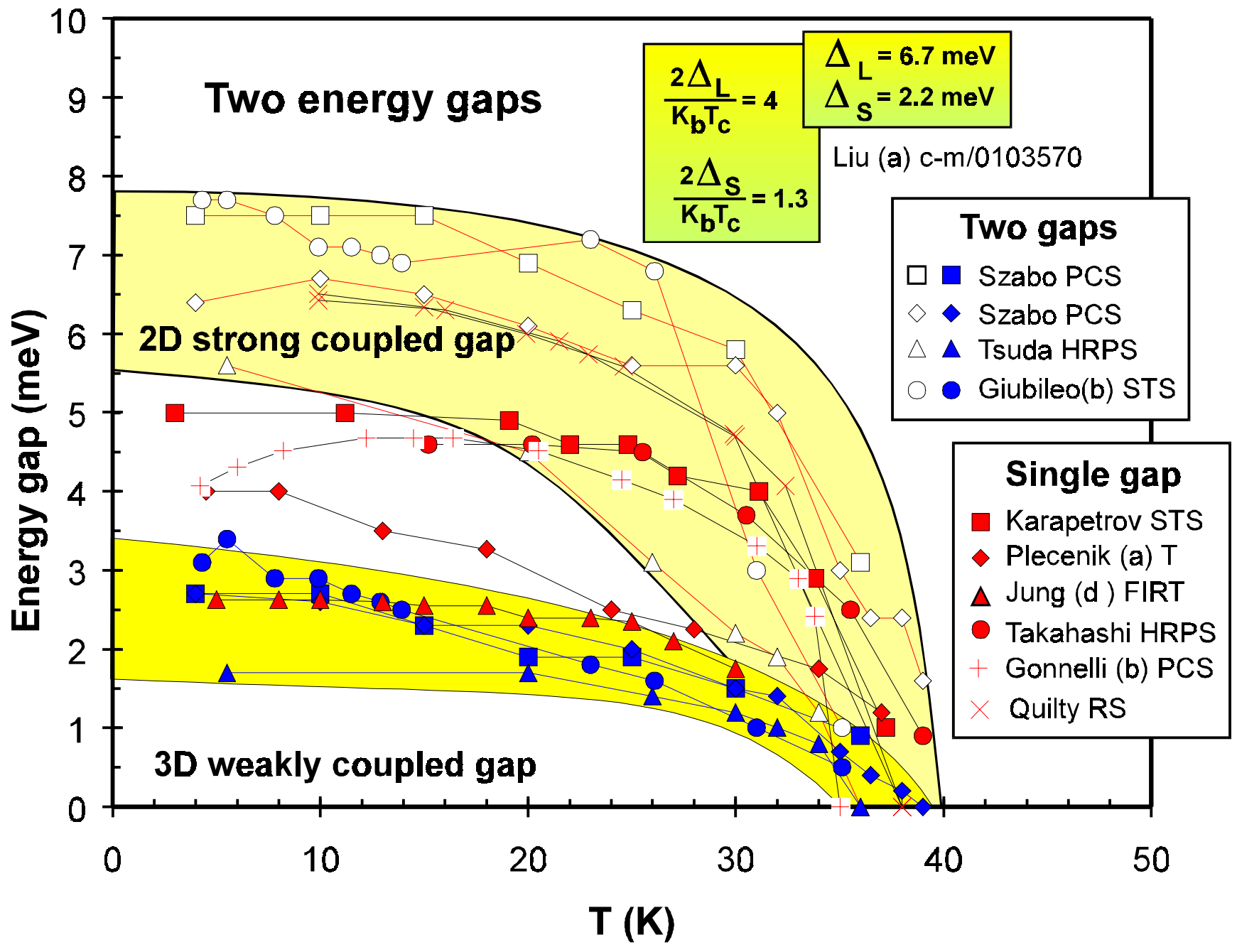

**Figure 35.** Energy gap dependence on temperature obtained from point contact spectroscopy (PCS), high-resolution photoemission spectroscopy (HRPS), scanning tunneling spectroscopy (STS), tunneling (T), far-infrared transmission (FIRT), Raman spectroscopy (RS) experiments. Data are taken from references [Szabo], [Tsuda], [Giubileo (b)], [Karapetrov], [Plecenik (a)], [Jung (d)], [Takahashi], [Laube], [Gonnelli (b)], [Quilty].

## 12. Energy gap

There is no consensus yet about the gap values in $MgB_2$ and weather or not this material has a single anisotropic gap or a double gap, as shown in Fig. 34.

Energy gap values have been inferred by using tunneling spectroscopy [Karapetrov], [Sharoni (b)], [Sharoni (a)], [Chen (a)], [Giubileo (a)], [Giubileo (b)], [Rubio-Bollinger], point contact tunneling [Schmidt], [Szabo], [Laube], [Zhang (b)], [Gonnelli (b)], specific heat studies [Kremer], [Walti], [Wang (c)], [Bauer], [Junod], [Fisher], [Bouquet (b)], high-resolution photoemission spectroscopy (HRPS) [Takahashi], [Tsuda], far-infrared transmission studies (FIRT) [Gorshunov], [Jung (d)], [Kaindl], Raman spectroscopy [Chen (c)], [Quilty], tunneling junctions [Plecenik (a)]. Energy gaps in superconductors are usually investigated by spectroscopic techniques, which are subject to errors associated with surface impurities or non-uniformity. In the case of $MgB_2$ the gap structure is so pronounced that specific heat measurements can be used to infer its values.

As shown in Fig. 35, several experiments measured a single gap, with values between 2.5÷5 meV, while latest experiments claim to have brought some clarification about the gap features in $MgB_2$. According to tunneling spectroscopy [see Giubileo (a)], point contact spectroscopy [Szabo], and Raman scattering [Chen (c)], there is evidence, suggested earlier by Liu et al. [Liu (a)], of two distinct gaps associated with the two separate segments of the Fermi surface [Belashchenko]. The width values of these two gaps were determined to be between 1.8÷3 meV for the small 3D weakly coupled gap, and between 5.8÷7.7 meV for the large strongly coupled gap.

Specific heat measurements [Wang (c)], [Bouquet (a)], [Bouquet (b)] show that it is necessary to involve either two gaps or a single anisotropic gap [Haas] to describe the data.

Microwave measurements results can be explain by the existence of an anisotropic superconducting gap or the presence of a secondary phase, with lower gap width, in some of the $MgB_2$ samples [Zhukov (c)].

## 13. Conclusions

To summarise, in this article we presented a review of the main normal and superconducting properties of magnesium diboride. $MgB_2$ has an unusual high critical temperature of about 40K among binary compounds, with an $AlB_2$-type structure with graphite-type boron layers separated by hexagonal close-packed layers of Mg. The presence of the light boron as well as its layered structure may be important factors which contribute to superconductivity at such a high temperature for a binary compound.

According to initial findings, $MgB_2$ seemed to be a low-$T_c$ superconductor with a remarkably high critical temperature, its properties resembling that of conventional superconductors rather than of high-$T_c$ cuprates. This include isotope effect, a linear T- dependence of the upper critical field with a positive curvature near $T_c$ (similar to borocarbides), a shift to lower temperatures of both $T_c$(onset) and $T_c$(end) at increasing magnetic fields as observed in resistivity R(T) measurements. On the other hand, the quadratic T-dependence of the penetration depth $\lambda(T)$, as well as the sign reversal of the Hall coefficient near $T_c$, indicates unconventional superconductivity similar to cuprates.

Several other related materials are known to be superconductive, but $MgB_2$ holds the record of $T_c$ in its class. The hope that the critical temperature could be raised above 40K initiated a search for superconductivity in similar compounds, up to now several materials being discovered to superconduct: $TaB_2$ ($T_c$=9.5K), $BeB_{2.75}$ ($T_c$=0.7K), C-S composites ($T_c$=35K), and the elemental boron under pressure ($T_c$=11.2K).

As a guide in the search for new related superconducting materials, we suggest several issues to be taken into account. First, one should try several compositions, as the superconductivity may arise only in nonstoichiometric compounds. Secondly, the contamination by non-reacted simple elements or other phases has to be ruled out by comparing the critical temperature of the new compound with the $T_c$ of the simple elements contained in the composition, and with other possible phases. In order to make easier the search for new superconducting borides, we gave updated information on $T_c$ of binary and ternary borides, as well as for simple chemical elements.

In Table 6 is presented a list with the most important parameters of $MgB_2$. In the followings we will sumarise this review. Up to date, $MgB_2$ has been synthesised as bulk, single crystals, thin films, tapes and wires. Thin films are fabricated by PLD, co-evaporation, deposition from suspension, magnetron sputtering and Mg diffusion. The highest critical temperatures and sharpest transitions are achieved by Mg diffusion method. This method is also used for fabrication of powders, wires and tapes. The most popular method for wires and tapes fabrication is the powder-in-tube PIT method. Several metal-cladding have been tried, the best results being achieved by iron. Other metals are reacting with Mg during a post-annealing process. High enough current densities can be achieved by skipping the sintering, which makes cheaper the fabrication process and expands the range of metals used in cladding. Single crystals are currently obtained by solid-liquid method, under high pressure, and by vapor-transport method. The charge carriers in $MgB_2$ are holes with a hole density at 300K between $1.7 \div 2.8 \times 10^{23}$ holes/cm$^3$. The critical temperature of $MgB_2$ decreases under pressure, the compound remaining hexagonal until the highest pressure studied. $T_c(P)$ data differs considerably for different authors. However, a pattern emerges in the $T_c(P)$ dependence: samples with lower $T_c$ at zero pressure have a positive curvature and a much steeper dependence than samples with higher $T_c$, which show a negative curvature. The initial rate of the critical temperature derivative in pressure $-dT_c/dp$ range between -1.1 and -2, being inverse proportional to pressure. The observed $T_c(P)$ may correlate with Mg nonstoichiometry in this compound. In order to clarify this subject, data which specify the correlation between $T_c$ and the Mg nonstoichiometry are necessary. $MgB_2$ shows anisotropic compressibility and thermal expansion, with the c-axis responses substantially higher than a-axis. This fact demonstrates that out-of-plane Mg-B bonds are much weaker than in-plane bonds.

**Table 6.** List of superconducting parameters of $MgB_2$

| Parameter | Values |
|---|---|
| critical temperature | $T_c = 39 \div 40$ K |
| hexagonal lattice | a = 0.3086 nm, |
| parameters | b = 0.3524 nm |
| theoretical density | $\rho = 2.55$ g/cm$^3$ |
| pressure coefficient | $dT_c/dP = -1.1 \div 2$ K/GPa |
| carrier density | $n_s = 1.7 \div 2.8 \times 10^{23}$ holes/cm$^3$ |
| isotope effect | $\alpha_T = \alpha_B + \alpha_{Mg} = 0.3 + 0.02$ |
| resistivity near $T_c$ | $\rho(40K) = 0.4 \div 16$ $\mu\Omega$cm |
| resistivity ratio | RR= $\rho(40K)/\rho(300K) = 1 \div 27$ |
| upper critical field | $H_{c2}$//ab(0) =14 ÷ 39 T |
| | $H_{c2}$//c(0) = 2 ÷ 24 T |
| lower critical field | $H_{c1}(0) = 27 \div 48$ mT |
| irreversibility field | $H_{irr}(0) = 6 \div 35$ T |
| coherence lengths | $\xi_{ab}(0) = 3.7 \div 12$ nm |
| | $\xi_c(0) = 1.6 \div 3.6$ nm |
| penetration depths | $\lambda(0) = 85 \div 180$ nm |
| energy gap | $\Delta(0) = 1.8 \div 7.5$ meV |
| Debye temperature | $\Theta_D = 750 \div 880$ K |
| critical current | $J_c(4.2K,0T) > 10^7$ A/cm$^2$ |
| densities | $J_c(4.2K,4T) = 10^6$ A/cm$^2$ |
| | $J_c(4.2K,10T) > 10^5$ A/cm$^2$ |
| | $J_c(25K,0T) > 5 \times 10^6$ A/cm$^2$ |
| | $J_c(25K,2T) > 10^5$ A/cm$^2$ |

Critical temperature decreases at various rates for substitutions with Si, Li, Ni, Fe, Co, Mn, while Zn doping seems to slightly increase $T_c$ at a certain doping level (less than 1K). The total isotope effect $\alpha = 0.32$ and Testardi correlation between $T_c$ and resistivity ratio RR seems to point out towards a phonon-mediated mechanims. High upper critical fields values of $H_{c2}(0)$=39T may be attained for films with lower $T_c$ (31K). Single crystals give second best values for the upper critical fields $H_{c2}(0)$=25T, followed by bulk $H_{c2}(0)$=19T and wires $H_{c2}(0)$=16T. For textured bulk and partially oriented crystallites, the anisotropy ratio $\gamma = H_{c2}^{//ab}/H_{c2}^{//c}$ is reported to be between 1.1 and 1.7; thin films give values of $1.2 \div 2$; single crystals show slightly higher values than in aligned powders or films, between $1.7 \div 2.7$; while measurements on non-aligned powders give unexpectedly large values, ranging from 5 to 9. Lower critical field data range between 25 and 48 mT, with penetration depths in the range 85-203 nm. The highest values of current density are obtained in $MgB_2$ thin films with incorporated impurities (O, MgO), showing similar or higher performances than the best existing superconducting wires. The high critical current densities attained in thin films give hopes for improving the current carrying capabilities of wires and tapes.

Altogether, relative low costs of fabrication, high critical currents and fields, large coherence lengths, its high critical temperature of 39K, absence of weak-links, makes $MgB_2$ a promising material for applications at above 20.13K, the temperature of boiling hydrogen at normal pressure.

In conclusion, in this article we have presented a review on $MgB_2$ normal and superconducting properties from studies appeared during the last seven month, from January until July. Since the progress in this field has been so wide and fast, it is possible that we may have unintentionally omitted some of the data. Also, despite the fact some issues have been studied in the literature, we did not cover it in this review, from special reasons. This may include microwave properties [Hakim], [Joshi (b)], [Lee (b)], [Zhukov (a)], [Nefyodov], [Klein], irradiation-induced properties [Karkin], [Bugoslavsky (b)], Josephson properties [Brinkman (b)], [Gonnelli (a)], [Burnell], [Zhang (b)]. These issues will be discussed in a later review to be included as a special chapter in our book [Yamashita] Nevertheless, we tried to update this review with the latest information in the field, hoping the reader will be provided with the current situation and trends that are to be pursued in the near future. For orientation purpose, in the reference list we cite all the $MgB_2$ studies appeared to our knowledge in printed or electronic format.

**Acknowledgments**

This work was supported by CREST (Core Research for Evolutional Science and Technology) of Japan Science and Technology Corporation (JST) and JSPS (Japan Society for the Promotion of Science).